\documentclass[reqno]{amsart}

\usepackage{booktabs} % For formal tables
\usepackage{IEEEtrantools}
\usepackage{amssymb,latexsym,amsfonts,amsmath}
\usepackage{graphicx}
\usepackage{mathrsfs}
\usepackage{dsfont}

\usepackage{tikz}
\usetikzlibrary{calc,shapes,arrows}

\usepackage{algorithmic}

\usepackage{xspace}

\usepackage{booktabs} % For formal tables
\usepackage{latexsym,amsfonts,amsmath}
\usepackage{hyperref}
\usepackage{subfigure}
\usepackage{dsfont}
\usepackage{extarrows}
\usepackage[ruled,vlined,linesnumbered]{algorithm2e}

\usepackage{paralist}
\usepackage{capt-of}
\usepackage{xcolor}

\newcommand{\R}{{\mathbb{R}}}

\newcommand{\N}{{\mathbb{N}}}

\newcommand{\ie}{{\it i.e.}}

\newcommand{\argmax}{\textrm{arg}\max}

\newcommand{\q}{\mathsf{q}}

\newcommand{\ul}{\underline}

\definecolor{myco}{rgb}{0.55, 0.0, 0.63}

\newcommand\blfootnote[1]{%
	\begingroup
	\renewcommand\thefootnote{}\footnote{#1}%
	\addtocounter{footnote}{-1}%
	\endgroup
}

\newtheorem{theorem}{Theorem}[section]
\newtheorem{lemma}[theorem]{Lemma}
\newtheorem{problem}[theorem]{Problem}
\newtheorem{proposition}[theorem]{Proposition}

\newtheorem{definition}[theorem]{Definition}
\newtheorem{example}{Example}

\newtheorem{remark}[theorem]{Remark}
\newtheorem{assumption}[theorem]{Assumption}
\numberwithin{equation}{section}

\begin{document}
	
\begin{abstract}
High performance but unverified controllers, \emph{e.g.,} artificial intelligence-based (\emph{a.k.a.} AI-based) controllers, are widely employed in cyber-physical systems (CPSs) to accomplish complex control missions.
However, guaranteeing the safety and reliability of CPSs with this kind of controllers is currently very challenging, which is of vital importance in many real-life safety-critical applications. 
To cope with this difficulty, we propose in this work a Safe-visor architecture for sandboxing unverified controllers in CPSs operating in noisy environments (\emph{a.k.a.} stochastic CPSs). 
The proposed architecture contains a history-based supervisor, which checks inputs from the unverified controller and makes a compromise between functionality and safety of the system, and a safety advisor that provides fallback when the unverified controller endangers the safety of the system. 
Both the history-based supervisor and the safety advisor are designed based on an approximate probabilistic relation between the original system and its finite abstraction.
By employing this architecture, we provide formal probabilistic guarantees on preserving the safety specifications expressed by accepting languages of deterministic finite automata (DFA).
Meanwhile, the unverified controllers can still be employed in the control loop even though they are not reliable. 
We demonstrate the effectiveness of our proposed results by applying them to two (physical) case studies.
\end{abstract}

\title[Safe-visor Architecture for Sandboxing (AI-based) Unverified Controllers in Stochastic CPSs]{Safe-visor Architecture for Sandboxing (AI-based) Unverified Controllers in Stochastic Cyber-Physical Systems}

\author{Bingzhuo Zhong$^{1*}$}
\author{Abolfazl Lavaei$^{2}$}
\author{Hongpeng Cao$^{1}$}
\author{Majid Zamani$^{3,4}$}
\author{Marco Caccamo$^{1}$}
\blfootnote{*Corresponding Author.}
\address{$^1$Department of Mechanical Engineering, Technical University of Munich, Germany.}
\email{bingzhuo.zhong@tum.de}
\email{cao.hongpeng@tum.de}
\email{mcaccamo@tum.de}
\address{$^2$Institute for Dynamic Systems and Control, ETH Zurich, Switzerland}
\email{alavaei@ethz.ch}
\address{$^3$Department of Computer Science, University of Colorado Boulder, USA.}
\email{majid.zamani@colorado.edu}
\address{$^4$Department of Computer Science, LMU Munich, Germany.}
\maketitle

\section{Introduction}\label{sec:introd}
{\bf Motivations.}
Cyber-physical systems (CPSs) are complex heterogeneous systems combining both cyber (computation and communication) and physical components, which tightly interact with each other in a feedback loop.
In the past few decades, \emph{stochastic} CPSs have received significant attention since they provide a powerful modeling framework for describing many engineering systems operating in noisy environments, including traffic networks, transportation systems, power grids, and so on.
Nowadays, many high performance but unverified controllers (\emph{e.g.,} deep neural network or black-box controllers from third parties) are widely employed to
accomplish complex missions for stochastic CPSs.
Nevertheless, the application of such controllers makes it increasingly challenging to ensure the overall safety of CPSs. 
In particular, ensuring safety is essential for safety-critical applications, in which system's failures (\emph{e.g.,} collision) are not acceptable.

\begin{figure}
	\centering
	\includegraphics[width=8cm]{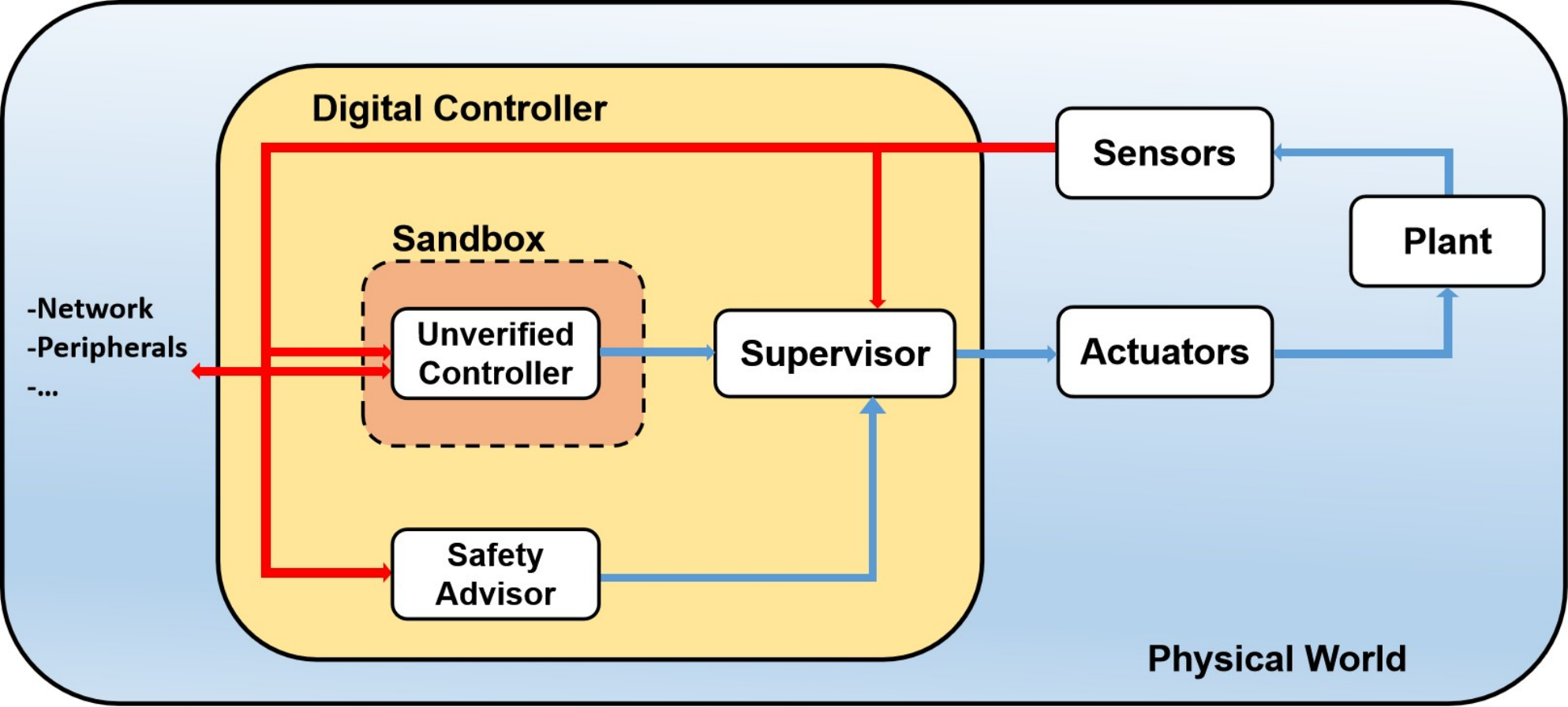}
	\caption{Safety advisor-supervisor (Safe-visor architecture) for sandboxing unverified \\ controllers.} \label{fig1:Svisor_arc}
\end{figure}

To cope with this issue, we propose a correct-by-construction architecture, namely Safe-visor architecture~\cite{Zhong2019Sandboxing}, for sandboxing (AI-based) unverified controllers at runtime to ensure the safety of physical systems.
The proposed architecture employs a safety advisor and supervisor (cf. Figure~\ref{fig1:Svisor_arc}).
At runtime, the supervisor checks control inputs provided by the unverified controllers.
These inputs would only be accepted when they are not disobeying the safety rules defined in the sandboxing mechanism. 
Otherwise, inputs from the safety advisor would be applied to the system to ensure the overall safety.
It is worth mentioning that the term \emph{sandbox}~\cite{Reis2009Browser} here is an idea borrowed from the computer security community.
In general, the sandbox is a common mechanism that ensures the security of CPSs containing untested and untrusted software components.
A verified sandbox is designed to isolate these components from the critical part of a host machine in a sandbox control environment, including its operating system. 
Consequently, an untrusted controller can only actuate the plant through a tightly-monitored interface so that the plant would not be endangered by accidental or malicious harm caused by the unverified controllers.

In this paper, we focus on stochastic CPSs that can be modeled as general Markov decision processes (gMDPs)~\cite{Haesaert2017Verification} with continuous/uncountable state and input sets. 
Moreover, we are interested in those safety specifications modeled by the accepting languages of deterministic finite automata (DFA)~\cite{Baier2008Principles}, which is a powerful tool for characterizing complex safety properties, \emph{e.g.,} those expressed as linear temporal logic formulae over finite traces (\emph{a.k.a.} LTL$_F$ formulae~\cite{DeGiacomo2013Linear}).
To provide safety guarantees for physical systems, the Safe-visor architecture specifies verifiable safety rules for the unverified controller. 
At runtime, the safety advisor is responsible for providing an advisory input that maximizes the probability of satisfying the safety specification of interest. 
Meanwhile, the supervisor checks the input given by the unverified controller at every time instant according to the same safety specification. 
The unverified inputs would only be accepted when they follow their corresponding rules; otherwise, the supervisor would accept the advisory input from the safety advisor.
Note that the unverified controller is designed to realize some tasks which are much more complex than just keeping the system safe.
Accordingly, the safety advisor would only be used if the unverified controller tries to perform some harmful actions. 
Therefore, using our proposed architecture, one can exploit the functionalities offered by the unverified controller while preventing the system from being unsafe. 

Although we focus on enforcing safety specifications for systems with arbitrary types of unverified controllers, our proposed results can be potentially leveraged to provide safety guarantees over unverified \emph{AI-based controllers}, which are becoming increasingly important to tackle complex control problems for CPSs~\cite{Bojarski2016End,Julian2019Deep}. 
Recent results~\cite{Papernot2016limitations,NTSB2018Preliminary,Kwiatkowska2019Safety} show that the application of AI-based controllers gives rise to many reliability problems, which could lead to catastrophic consequences with unexpected actions.
Our work here provides a system level~\cite{Claviere2020Safety}, correct-by-construction architecture to ensure the overall safety of the AI-based controlled systems.

{\bf Related Works.}
In the setting of discrete-space systems, \emph{e.g.,} reactive systems, the work in~\cite{Bloem2015Shield,Humphrey2016Synthesis,Bharadwaj2019Synthesis,Alshiekh2018Safe} proposes a shield to enforce some safety properties at runtime.
For systems with continuous state and input sets, plenty of results are applicable to deterministic systems concerning simple invariance properties, in which systems are expected to stay within a fixed safety set.
A Simplex architecture is proposed in~\cite{Sha2001Using,Crenshaw2007Simplex} that enables the application of unverified, high-performance controllers in the control loop by using a Lyapunov-function-based elliptic recovery region (\emph{a.k.a.} safety invariant set).
This region is associated with a verified, linear, state-feedback controller, which serves as a high-assurance controller. 
Later, this architecture is further developed in~\cite{Wang2013L1Simplex,Wang2018RSimplex,Yao2013NetSimplex,Abdi2018Preserving} to handle uncertainty and bounded-time delay in the system dynamics, as well as undetectable cyber attacks.

Note that in the Simplex architecture, the Lyapunov-function-based elliptic recovery region is usually very conservative and, therefore, unnecessarily restricts the use of high-performance controllers~\cite[Figure 7]{Bak2014Real}.
However, high-performance controllers are expected to be applied as often as possible.
To provide more flexibility for high-performance controllers, results in~\cite{Bak2011Sandboxing,Bak2014Real,Abdi2017Application} employ reachability analysis to enlarge the recovery region.
Although some of these results are still called ``Simplex design", they are quite different from the basic idea of Simplex architecture as in~\cite{Sha2001Using}.
In Simplex architecture, high-assurance and high-performance controllers are both designed for the same tasks and are different in terms of the performance of finishing those tasks.
By enlarging the recoverable region as in~\cite{Bak2011Sandboxing,Bak2014Real,Abdi2017Application}, the high-assurance controller is no longer able to finish the same tasks as the high-performance one but only ensures that the system would not leave the desired safety set.
Thus, high-assurance controllers in these results work similarly to the safety advisor of the Safe-visor architecture in our paper.
As mentioned, however, these results only work for deterministic systems and simple invariance properties.

Results in~\cite{Kenton2019Generalizing,Ma2018Improved,Thananjeyan2020Safety,Fisac2019Bridging,Shyamsundar2016Reinforcement} improve the robustness of systems concerning safety in which AI-based controllers are applied, but they do not provide any formal safety guarantee.
Meanwhile, results in~\cite{Cheng2019End,Deshmukh2019Learning,Yang2019Safety,Huh2020Safe,Larsen2017Safe,Fisac2018general,Wabersich2018Linear,Wabersich2018Safe,Muntwiler2019Distributed,Hewing2020Learning} provides formal safety guarantees for AI-based controllers regarding simple invariance properties; 
a few recent results also handle complex logical properties for systems with continuous state and input sets, e.g.,~\cite{Li2019Temporal} for deterministic systems, and~\cite{Lavaei2020Formal,Kazemi2020Formal} for stochastic systems.
In these results, formal guarantees are achieved by adequately cooperating the desired properties in the reward functions and guiding the learning process. 

{\bf Contributions.} 
In this paper, we propose abstraction-based approaches for designing the Safe-visor architecture as in Figure~\ref{fig1:Svisor_arc} with respect to safety properties modeled by DFAs, which are more general than simple invariance properties.
A limited subset of the provided results in this paper has been recently presented in~\cite{Zhong2019Sandboxing}. 
Our proposed approach here differs from the one in~\cite{Zhong2019Sandboxing} in three main directions:
\begin{enumerate}[(i)]
	\item We propose new history-based supervisors in the Safe-visor architecture by enlarging the class of safety specifications that can be expressed by the accepting languages of deterministic finite automata (DFA), whereas~\cite{Zhong2019Sandboxing} handles only simple invariance properties;
	\item By employing ($\epsilon$,$\delta$)-approximate probabilistic relations proposed in~\cite{Haesaert2017Verification} while designing the history-based supervisor, we provide formal guarantees for probabilities of satisfying desired specifications.
	Meanwhile, the results in~\cite{Zhong2019Sandboxing} do not consider the effect of state set's discretization on the provided safety guarantee so that the guarantee can only be validated via experiments;
	\item Our proposed method allows constructing a reduced-order model for the original system when synthesizing the safety advisor and supervisor. 
	As a result, our method applies to large-scale systems, which was not the case in~\cite{Zhong2019Sandboxing}. 
	This issue is crucial since the complexity of synthesizing a safety advisor via a finite abstraction grows exponentially with respect to the dimension of the system's state set.
\end{enumerate}
	Additionally, we also compare our approaches with related results in the literature.
	In comparison with those results that adapt shielding approach in the learning process to ensure safety  (e.g.~\cite{Alshiekh2018Safe,Cheng2019End,Kazemi2020Formal}, as discussed above in the related works), our approaches decouple the desired safety properties from the reward functions in the learning process so that the provided safety guarantee is valid for AI-based controllers trained with arbitrary learning methods.
	Particularly, our methods are applicable when the reward functions for the control tasks are challenging to be designed (e.g.,~\cite{Biyik2020Learning}), while satisfying some safety properties are still required.
	Compared with those results that also decouple the desired safety properties from the synthesis of the unverified controllers (e.g., Simplex architecture in~\cite{Wang2018RSimplex,Abdi2017Application}, see detailed discussion above in the related works), our proposed results are the first to deal with stochastic CPSs with continuous state and input sets while ensuring safety properties modeled by DFAs. 	
	Although~\cite{Bloem2015Shield,Humphrey2016Synthesis,Bharadwaj2019Synthesis} also enforce safety specifications that are modeled by automata, they are only applicable to discrete-space systems.  

It is also worth noting that a running example is provided throughout the paper to demonstrate our theoretical results more intuitively.

{\bf Organization.} The remainder of the paper is structured as follows.
In Section~\ref{sec:2}, we provide preliminary discussions on notations, models, and the underlying problems to be solved in this work.
In Section~\ref{sec:design_sva}, we first present ($\epsilon$,$\delta$)-approximate probabilistic relations between original models and their finite abstractions.
Then, we provide the design of Safe-visor architecture, including the design of the safety advisor and the supervisor. 
Finally, we apply our proposed results to two case studies in Section~\ref{Case_study} and conclude our work in Section~\ref{sec:discussion}.

\section{Problem Formulation}\label{sec:2}
\subsection{Preliminaries}
A probability space in this work is presented by $(\hat \Omega,\mathcal F_{\hat \Omega},\mathbb{P}_{\hat \Omega})$, where $\hat \Omega$ is the sample space,
$\mathcal F_{\hat \Omega}$ is a sigma-algebra on $\hat \Omega$ which comprises subsets of $\hat\Omega$ as events, and $\mathbb{P}_{\hat \Omega}$ is a probability measure that assigns probabilities to events.
Random variables introduced here are measurable functions $X:(\hat \Omega,\mathcal F_{\hat \Omega})\rightarrow(S_X,\mathcal F_X)$ such that any random variable $X$ induces a probability measure on its space $(S_X,\mathcal F_X)$ as $Prob\{A\} = \mathbb{P}_{\hat \Omega}\{X^{-1}(A)\}$ for any $A\in \mathcal F_X$. 
We directly present the probability measure on $(S_X,\mathcal F_X)$ without explicitly mentioning the underlying probability space and the function $X$ itself. 

A topological space $S$ is called a Borel space if it is homeomorphic to a Borel subset of a Polish space (\emph{i.e.,} a separable and completely metrizable space). 
One of the examples of Borel space is Euclidean spaces $\mathbb{R}^n$.
Any Borel space $S$ is assumed to be endowed with a Borel $\sigma$-algebra denoted by $\mathcal{B}(S)$.  
We denote by $\mathbf{P}(S,\mathcal{B}(S))$ the set of probability measures on the measurable space ($S$,$\mathcal{B}(S)$).
A map $f:X\rightarrow Y$ is measurable whenever it is Borel measurable.  
A map $f:X\rightarrow Y$ is universally measurable if the inverse image of every Borel set under $f$ is measurable with respect to every complete probability measure on $X$ that measures all Borel subsets of $X$.

\subsection{Notations}
We use $\mathbb{R}$ and $\mathbb{N}$ to denote sets of real and natural numbers, respectively. 
These symbols are annotated with subscripts to restrict the sets in a usual way, \emph{e.g.,} $\mathbb{R}_{\geq0}$  and $\mathbb{N}_{\geq 1}$ denote sets of non-negative real numbers and positive integers, respectively.
In addition, $\mathbb{R}^{n\times m}$ with $n,m\in \mathbb{N}_{\geq 1}$ denotes the vector space of real matrices with $n$ rows and $m$ columns.
For $a,b\in\mathbb{R}$ (resp. $a,b\in\mathbb{N}$) with $a\leq b$, the closed, open and half-open intervals in $\mathbb{R}$ (resp. $\mathbb{N}$) are denoted by $[a,b]$, $(a,b)$ ,$[a,b)$ and $(a,b]$, respectively. 
Given $N$ vectors $x_i \in \mathbb R^{n_i}$, $n_i\in \mathbb N_{\ge 1}$, and $i\in\{1,\ldots,N\}$, we use $x = [x_1;\ldots;x_N]$ to denote the corresponding column vector of dimension $\sum_i n_i$.
We denote respectively by $\mathbf{0}_n$ and $\mathbf{1}_n$ column vectors in $\R^n$ with all elements equal to 0 and 1. 
Symbols $I_n$ and $0_n$ denote identity and zero matrices in $\R^{n\times n}$, respectively. 
Moreover, $\lVert x\rVert$ denotes Euclidean norm of $x$. 
Given sets $X$ and $Y$, a relation $\mathscr{R} \in X\times Y$ is a subset of the Cartesian product $X\times Y$ that relates $x\in X$ with $y\in Y$ if $(x,y)\in\mathscr{R}$, which is equivalently denoted by $x\mathscr{R}y$.
Given a set $\mathsf{M} =X_1\times X_2\times\ldots\times X_n$ and vector $\mathsf{m}=(x_1,x_2,\ldots,x_n)\in\mathsf{M}$ with $x_i\in X_i$, we define $\mathsf{m}_{X_i}=x_i$.
Moreover, given a set $X$, $X^{\mathbb{N}}$ denotes the Cartesian product among the countable infinite number of set $X$. 
Given functions $f:X\rightarrow Y$ and $g:Y\rightarrow Z$, we denote by $g\circ f : X\rightarrow Z$ the composite function of $f$ and $g$.
Additionally, $\mathcal{N}(\cdot|\mu,\Sigma)$ denotes the normal distribution with mean $\mu$ and covariance matrix $\Sigma$, and $\delta_d(\cdot|c)$ indicates Dirac delta distribution centered at $c$.

\subsection{General Markov Decision Processes} \label{sec:sys_model}
In this work, the stochastic systems we are interested in can be formulated as a class of general Markov decision processes (gMDPs) that evolve over continuous or uncountable state sets.  This class of models generalizes the usual notion of MDPs~\cite{Baier2008Principles} by adding an output set over which properties of interest are defined. 

\begin{definition}\label{def:gMDP}
	\emph{(gMDP)}
	A general Markov decision process is a tuple
	\begin{equation}
	\label{eq:dt-SCS}
	\mathfrak{D} =(X,U,x_0,T,Y,h),
	\end{equation}
	where,
	\begin{itemize}
		\item $X\subseteq \mathbb R^{s}$ is a Borel set as the state set of the system. 
		We denote by $(X, \mathcal B (X))$ the measurable space with $\mathcal B (X)$ being the Borel $\sigma$-algebra on the state set;
		\item $U\subseteq \mathbb R^m$ is a Borel set as the input set of the system;
		\item $x_0\in X$ is the initial state;
		\item $T:\mathcal B(X)\times X\times U\rightarrow[0,1]$ 
		is a conditional stochastic kernel that assigns to any $x \in X$ and $u\in U$, a probability measure $T(\cdot | x,u)$
		on the measurable space $(X,\mathcal B(X))$. 
		This stochastic kernel specifies probabilities over executions $\{x(k),k\in\mathbb N\}$ of the gMDP such that for any set $\mathcal{X} \in \mathcal B(X)$ and any $k\in\mathbb N$, $\mathbb P \big\{x(k+1)\in \mathcal{X}\,|\, x(k),u(k)\big\}=\int_\mathcal{X} T (\mathsf{d}x(k+1)|x(k),u(k))$;
		\item $Y\subseteq \mathbb R^q$ is a Borel set as the output set of the system; 
		\item $h:X\rightarrow Y$ is a measurable function that maps a state $x\in X$ to its output $y = h(x)$.
	\end{itemize}
\end{definition}
Alternatively, a gMDP $\mathfrak{D}$ as in~\eqref{eq:dt-SCS} can be described by difference equations
\begin{equation}\label{eq:gMDP_f}
\mathfrak{D}\!:
\left\{\hspace{-1.5mm}\begin{array}{l}
x(k+1)=f(x(k),u(k),\varsigma(k)),\\
y(k)=h(x(k)),\end{array}\right.\quad k\in\mathbb N,
\end{equation}
where $x(k)\in X$, $u(k)\in U$, $y(k)\in Y$, and $\varsigma:=\{\varsigma(k): \hat\Omega\rightarrow V_{\varsigma}, k\in \N\}$ is a sequence of independent and identically distributed (i.i.d.) random variables from the sample space $\hat\Omega$ to a set $V_{\varsigma}$. 
Then, the evolution of a gMDP can be described by its paths and output sequences as defined below. 

\begin{definition}\label{def:path}
	\emph{(Path)}
	A path of a gMDP $\mathfrak{D}=(X,U,x_0,T,$ $Y,h)$ is 
	\begin{align*}
	\omega\,=\,(x(0),u(0),\ldots,x(k-1),u(k-1),x(k),\ldots),
	\end{align*}
	where $x(k)\in X$ and $u(k)\in U$ with $k\in\N$. 
	We denote by $\omega_{x}=(x(0),x(1),\ldots,$ $x(k),\ldots)$ and $\omega_{u}=(u(0),u(1),\ldots,u(k),\ldots)$ the subsequences of states and inputs in $\omega$, respectively.
	
	\noindent
	\emph{(Output Sequence)} The corresponding output sequence is denoted by 
	\begin{equation*}
	y_{\omega} = (y(0),y(1),\ldots,y(k),\ldots),
	\end{equation*}
	with $y(k)=h(x(k))$.
	Moreover, we denote by $\omega_k$ the path up to the time instant $k$ and, respectively, by $\omega_{xk}$, $\omega_{uk}$, and $y_{\omega k}$ its corresponding state, input, and output (sub)sequences.
\end{definition}
\begin{example}
	\emph{(Running Example) }
	In this paper, a system with two cars, which is adapted from~\cite{Haesaert2020Robust}, is served as a running example.
	\begin{figure}
		\centering
		\includegraphics[width=6cm]{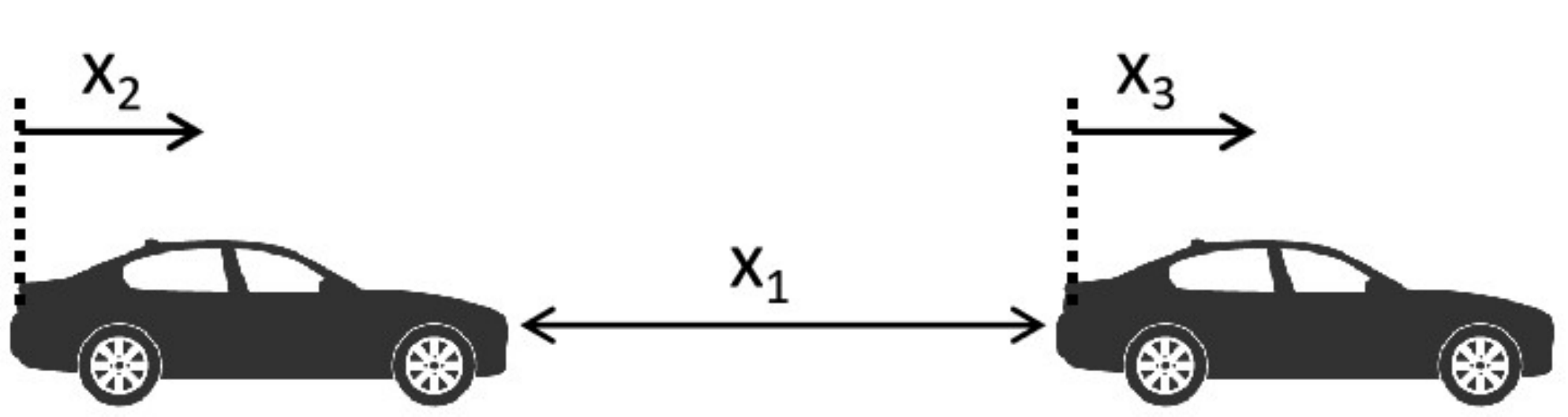}
		\caption{Running example: A system with two cars.} \label{fig1:tag}
	\end{figure}
	This example is depicted in Figure~\ref{fig1:tag} and can be modeled by the following set of stochastic difference equations:
	\begin{small}
		\begin{equation*}
		\mathfrak{D}\!:
		\!\left\{\hspace{-1.5mm}\begin{array}{l}
		\begin{aligned}
		x(k+1) = Ax(k)&+Bu(k)+R\varsigma(k),
		\end{aligned}\\
		y(k)=Cx(k),\end{array}\right.\quad k\in\mathbb N,
		\end{equation*}
	\end{small}
	with
	\begin{align*}
	&A \!=\!  \begin{bmatrix}
	\begin{smallmatrix}1\ &-0.15\ &0.15\\0\ &0.6\ &0\\0\ &0\ &0.6\end{smallmatrix}
	\end{bmatrix}\!,
	~B \!=\! [-0.03\,;1\,;0],
	~R \!=\! [
	0.006\,;0\,;0.1
	],
	~C\!=\![
	1\,;0\,;0
	]^T.
	\end{align*}
	Here, $x(k) = [x_1(k);x_2(k);x_3(k)]$ is the state of the system, in which $x_1(k)$, $x_2(k)$, and $x_3(k)$ denote the distance between cars, and velocities of follower and leader cars, respectively. 
	Input $u(k)\in[-8,8]$ is the external actuation of the follower car. 
	Besides, $\varsigma(k)$ is a sequence of standard Gaussian random variable that models the unpredictable changes in the leader car's velocity and in the distance between two cars, and $y(k)$ represents the output of the system.
\end{example}

The space of all infinite paths $\Omega = (X\times U)^{\N}$ along with their product $\sigma$-algebra $(\mathcal B(X)\times \mathcal B(U))^{\N}$ is called a canonical sample space for the gMDP. 
In this paper, we are interested in Markov policies for controlling gMDPs, which determine the input at the time instant $k$ only based on the state at the same time instant, \emph{i.e.,} $x(k)$.
This is formalized in the next definition.

\begin{definition}\label{def_mp}
	\emph{(Markov Policy~\cite{Shreve1978Stochastic})}
	Consider a gMDP $\mathfrak{D} =(X,U,x_0, T,Y,h)$.
	A Markov policy $\rho$ is a sequence $\rho\,=\,(\rho_0,\,\rho_1,\,\ldots)$ of universally measurable maps $\rho_k\!\!:X\,\rightarrow\,U$ with
	\begin{equation*}
	\rho_k(U \,\big |\, \omega_k)=\rho_k(U \,\big |\, \omega_{xk}(k)) = 1,
	\end{equation*}
	for all $\omega_k\in\Omega$ with $k\in\N$. 
	We use $\mathcal P$ to denote the set of all Markov policies.
	Accordingly, we denote by $\mathcal P^H$ the set of all Markov policies within time horizon $[0,H-1]$.
\end{definition}

Next, we propose a more general set of control strategies, with which inputs are provided based on paths of the gMDP via a memory state.
The definition here is adapted from~\cite{Haesaert2020Robust} by allowing the memory and output update map to be time dependent.
\begin{definition}\label{def_p}
	\emph{(Control Strategy)} A control strategy for a gMDP $\mathfrak{D} =(X,U,x_0,$ $T,Y,h)$ is a tuple
	\begin{align}
	\mathbf{C} = (\mathsf{M},\mathsf{U},\mathsf{Y},\mathsf{H},\mathsf{m}_{0},\pi_{\mathsf{M}},\pi_{\mathsf{Y}}),
	\end{align}
	where, 
	\begin{itemize}
		\item $\mathsf{M}$ is a Borel set as the \emph{memory state set};
		\item $\mathsf{U}\subseteq\mathbb{R}^s$ is a Borel set as the \emph{observation set}; 
		\item $\mathsf{Y}\subseteq U$ is a Borel set as the \emph{output set}; 
		\item $\mathsf{H}\subseteq \N$ is the \emph{time domain};
		\item $\mathsf{m}_{0}\in\mathsf{M}$ is the \emph{initial memory state}; 
		\item $\pi_{\mathsf{M}}\!: \!\mathsf{M}\times\mathsf{U}\times\mathsf{H}\!\rightarrow\! \mathbf{P}(\mathsf{M},\mathcal{B}(\mathsf{M}))$ is a \emph{memory update function};
		\item $\pi_{\mathsf{Y}}:\mathsf{M}\times\mathsf{H}\rightarrow\mathbf{P}(\mathsf{Y},\mathcal{B}(\mathsf{Y}))$ is an \emph{output update function}.
	\end{itemize}
\end{definition}
Given a gMDP $\mathfrak{D}$ and a control strategy $\mathbf{C}$, the controlled gMDP is denoted by $\mathbf{C}\times \mathfrak{D}$.
Additionally, we denote by $\mathbb{P}_{\mathfrak{D}}$ (resp. $\mathbb{P}_{\mathbf{C}\times \mathfrak{D}}$) the probability measure over the space of output sequences of $\mathfrak{D}$ (resp. $\mathbf{C}\times \mathfrak{D}$).
In the next subsection, we proceed with presenting the safety properties of interest in this paper.

\subsection{Deterministic Finite Automata}\label{sec:DFA}
As mentioned in the introduction, we focus on a larger class of specifications in this work (compared to~\cite{Zhong2019Sandboxing}) to describe the safety properties. 
Such specifications can be characterized by the accepting languages of DFAs, as formalized in the following definition.
\begin{definition}
	\emph{(DFA)}
	A DFA is a tuple $\mathcal{A} = (Q, q_0, \Pi,$ $ \tau, F)$, where $Q$ is a finite set of states, $q_0\in Q$ is the initial state, $\Pi$ is a finite set of alphabet, $\tau : Q \times \Pi \rightarrow Q$ is a transition function, and $F\subseteq Q$ is a set of accepting states.
\end{definition}

A finite word $\sigma = (\sigma_0, \sigma_1,\ldots,\sigma_{k-1})\in \Pi^k$ is accepted by $\mathcal{A}$ if there exists a finite state run $q = (q_0,q_1,\ldots,q_k)\in Q^{k+1}$ such that $q_{i+1} = \tau(q_i,\sigma_i)$, $\sigma_i \in \Pi$ for all $0\leq i<k$ and $q_k\in F$.
The set of words accepted by $\mathcal{A}$ is called the \emph{language of $\mathcal{A}$} and denoted by $\mathcal{L}(\mathcal{A})$.
Next, we show that how a gMDP $\mathfrak{D}$ as in \eqref{eq:dt-SCS} can be connected to a DFA $\mathcal{A}$ using a measurable labelling function.

\begin{definition}\label{def:sactisfaction_DFA}
	\emph{(Labelling Function)}
	Given a gMDP $\mathfrak{D} =(X,U,x_0,T,Y,h)$ and a DFA $\mathcal{A} = (Q, q_0, \Pi,$ $\tau, F)$, we define a measurable labelling function $L: Y\rightarrow \Pi$ and a function $L_H:Y^H\rightarrow \Pi^H$ as follows. 
	Consider a finite output sequence $y_{\omega (H-1)}=\big(y(0),y(1),\ldots,y(H-1)\big)\in Y^H$ of $\mathfrak{D}$ with some $H\in \mathbb{N}_{\geq 1}$.
	The trace of $y_{\omega (H-1)}$ over $\Pi$ is $\sigma = L_H(y_{\omega (H-1)}) = (\sigma_0,\sigma_1,\ldots,\sigma_{H-1})$, where $\sigma_k=L(y(k))$ for all $k\in[0,H-1]$.
	Moreover, $y_{\omega (H-1)}$ is accepted by $\mathcal{A}$, denoted by $y_{\omega (H-1)}\models \mathcal{A}$, if $L_H(y_{\omega (H-1)})\in \mathcal{L}(\mathcal{A})$. 
\end{definition}

Since we model the desired safety property with a DFA $\mathcal{A}$, we evaluate the performance of the gMDP $\mathfrak{D}$ concerning this property in terms of $\mathbb{P}_{\mathfrak{D}}\{y_{\omega (H-1)}\models \mathcal{A} \}$ within a bounded-time horizon. 
To this end, we construct a product gMDP based on $\mathfrak{D}$ and $\mathcal{A}$, which is defined as follows.
\begin{definition}\label{def:product_gmdp_dfa}
	\emph{(Product gMDP~\cite{Haesaert2020Robust})}Given a gMDP $\mathfrak{D} =(X,U,x_0,T,Y,h)$, a DFA $\mathcal{A} = (Q, q_0, \Pi,$ $\tau, F)$, and a labelling function $L: Y\rightarrow \Pi$, the product of $\mathfrak{D}$ and $\mathcal{A}$ is a gMDP and defined as
	\begin{equation*}
	\mathfrak{D}\otimes\mathcal{A} = \{\bar{X},\,\bar{U},\,\bar{x}_0,\,\bar{T},\,\bar{Y},\,\bar{h}\},
	\end{equation*}
	where
	\begin{itemize}
		\item $\bar{X}:=X\times Q$ is the state set;
		\item $\bar{U}:= U$ is the input set; 
		\item $\bar{x}_0:=(x_0,\bar{q}_0)$ is the initial state with
		\begin{equation}
		\bar{q}_0 = \tau\big(q_0,L\circ h(x_0)\big);\label{eq:compute_initial_q0}
		\end{equation}
		\item $\bar{T}(dx'\times\{q'\}\,|\,x,q,u)$ is the stochastic kernel that assigns for any $(x,q)\in\bar{X}$ and $u\in\bar{U}$, the probability $\bar{T}(dx'\times\{q'\}|x,q,u):=\textbf{1}_{\{q^+\}}(q')T(dx'|x,u)$,	
		with $q^+ =\tau(q,L\circ h(x'))$, where $\textbf{1}_{Q'}(\cdot)$ is the indicator function for a set $Q'\subseteq Q$, i.e., if $q'\in Q'$, then $\textbf{1}_{Q'}(q')=1$, otherwise $\textbf{1}_{Q'}(q')=0$;
		\item $\bar{Y}:=Y$ is the output set;
		\item $\bar{h}(x,q):=h(x)$ is the output map.
	\end{itemize}
\end{definition}

\subsection{Problem Formulation}\label{sec:problem_for}
In this paper, we focus on two different problems according to different types of safety specifications of interest.
In the following, we denote by $\eta$ the \emph{maximal tolerable probability of violating the safety specification} which is acceptable.
For some safety specifications, \emph{e.g.,} those expressed as co-safe-LTL$_F$ properties~\cite{Faruq2018Simultaneous}, all infinite output sequences that satisfy this type of specifications have a finite good prefix. 
In this case, we build a DFA $\mathcal{A}$ which accepts all good prefixes to model the safety specifications of this kind.
Accordingly, we focus on a problem of \emph{robust satisfaction} as follows.
\begin{problem}\label{prob:robust_satisfaction}
	\emph{(Robust Satisfaction)} Consider a gMDP $\mathfrak{D}$ as in \eqref{eq:dt-SCS}. 
	The~\emph{problem of robust satisfaction} with respect to the parameter $\eta$ is to design a Safe-visor architecture as in Figure~\ref{fig1:Svisor_arc} for $\mathfrak{D}$ such that
	\begin{align}\label{problem1}
	\mathbb{P}_{\mathfrak{D}}\Big\{y_{\omega H}\models \mathcal{A}\Big\} \geq 1-\eta,
	\end{align}
	where $\mathcal{A}$ is a DFA that accepts all good prefixes of the desired safety specification.
\end{problem}
Meanwhile, for some other safety specifications, \emph{e.g.,} those expressed as safe-LTL$_F$ properties~\cite{Saha2014Automated}, all infinite output sequences that violate these specifications have a finite bad prefix. 
In this case, we model these specifications with a DFA $\mathcal{A}$ that accepts all bad prefixes.
Correspondingly, we are interested in a problem of \emph{worst-case violation} as defined below.
\begin{problem}\label{prob:worst_case_violation}
	\emph{(Worst-case Violation)} 
	Consider a gMDP $\mathfrak{D}$ as in \eqref{eq:dt-SCS}. 
	The~\emph{Problem of worst-case violation} with respect to the parameter $\eta$ is to design a Safe-visor architecture as in Figure~\ref{fig1:Svisor_arc} for $\mathfrak{D}$ such that
	\begin{align}\label{problem2}
	\mathbb{P}_{\mathfrak{D}}\Big\{y_{\omega H}\models \mathcal{A}\Big\} \leq \eta,
	\end{align}
	where $\mathcal{A}$ is a DFA that accepts all bad prefixes of the desired safety specification.
\end{problem}

\addtocounter{example}{-1}
\begin{example}[continued]
	\begin{figure}
		\centering
		\includegraphics[width=7cm]{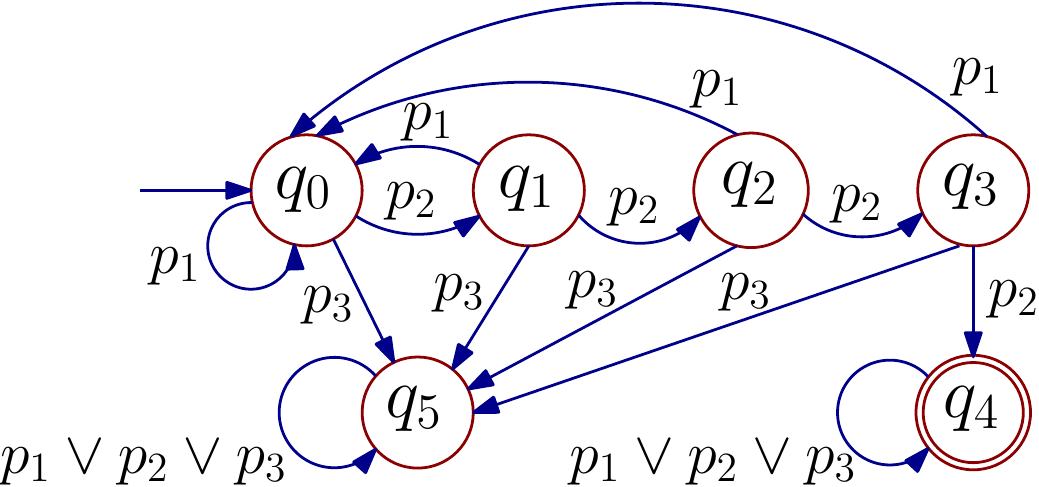}
		\caption{DFA modeling $\psi$, with accepting state $q_4$, alphabet $\Pi=\{p_1,p_2,p_3\}$ and labelling function $L:Y \rightarrow \Pi$, where $L(y)=p_1$ when $y \in (3, 10]$, $L(y)=p_2$ when $y \in [0, 3]$ and $L(y)=p_3$ when $y \in (-\inf, 0)\cup (10,+\inf)$.} \label{fig:case}
	\end{figure}
	We are interested in the following safety specification $\psi$: (i) Within $8$ time instances, the system output should reach $[0,3]$ and then stay within $[0,3]$ for 3 time instances after it reaches $[0,3]$; (ii) the output should not exceed $[0,10]$.
	The DFA modeling $\psi$ is shown in Figure~\ref{fig:case}.
	Accordingly, we focus on the problem of robust satisfaction corresponding to this DFA.	
\end{example}

\section{Design of Safe-visor Architecture} \label{sec:design_sva}
In this section, we discuss the design of Safe-visor architecture in great detail. 
The proposed framework consists of designing a safety advisor based on a given safety specification, and a supervisor that detects the unverified controller's harmful behaviors and decides the input fed to the system (either the one from the unverified controller or the safety advisor).
Prior to designing the architecture, we first present an approximate probabilistic relation that plays a crucial role in providing a formal safety guarantee in our setting.

\subsection{Approximate Probabilistic Relations between gMDPs}\label{sec:APR}
First, we define the notation of $\delta$-lifted relation over general state spaces, which lays the foundation of defining the approximate probabilistic relation between gMDPs.
\begin{definition}\label{lifting}
	\emph{($\delta$-Lifted Relation~\cite{Haesaert2017Verification})}
	Let $X, \hat X$ be two sets with associated measurable spaces $(X, \mathcal B(X))$ and $(\hat X, \mathcal B(\hat X))$, 
	and $\mathscr{\bar R}_{\delta}\subseteq\mathbf{P}(X, \mathcal B(X))\times\mathbf{P}(\hat X,$ $\mathcal B(\hat X))$ be a $\delta$-lifted relation between two probability distributions over $\mathcal B(X)$ and $\mathcal B(\hat X)$.
	Given a relation $\mathscr{R}\subseteq X\times\hat X$ that is measurable, i.e., $\mathscr{R} \in \mathcal B(X \times \hat X)$, and probability distribution $\Phi$ and $\Theta$ defined over $(X, \mathcal B(X))$ and  $(\hat X, \mathcal B(\hat X))$, respectively,  
	we have $(\Phi,\Theta)\in\mathscr{\bar R}_{\delta}$, denoted by $\Phi\mathscr{\bar R}_{\delta}\Theta$, if there exists a lifting $\mathscr{L}$ with a probability space $(X \times \hat X, \mathcal B(X \times \hat X), \mathscr{L})$ such that
	\begin{itemize}
		\item $\forall \mathcal{X} \in \mathcal B(X), ~\mathscr{L}(\mathcal{X} \times \hat X) = \Phi (\mathcal{X})$;
		\item $\forall \mathcal{\hat X} \in \mathcal B(\hat X), ~\mathscr{L}(X \times \mathcal{\hat X}) = \Theta (\mathcal{\hat X})$;
		\item $\mathscr{L}(\mathscr{R})\geq 1-\delta$, i.e., for the probability space $(X \times \hat X, \mathcal B(X \times \hat X), \mathscr{L})$, it holds that $x\mathscr{R} \hat x$ with the probability of at least $1-\delta$.
	\end{itemize}
\end{definition}
Next, we provide conditions under which we establish an ($\epsilon, \delta$)-approximate probabilistic relation between two gMDPs based on the $\delta$-lifted relation between their probability measures. 

\begin{figure}[ht]
	\begin{center}
		\includegraphics[width=7.5cm]{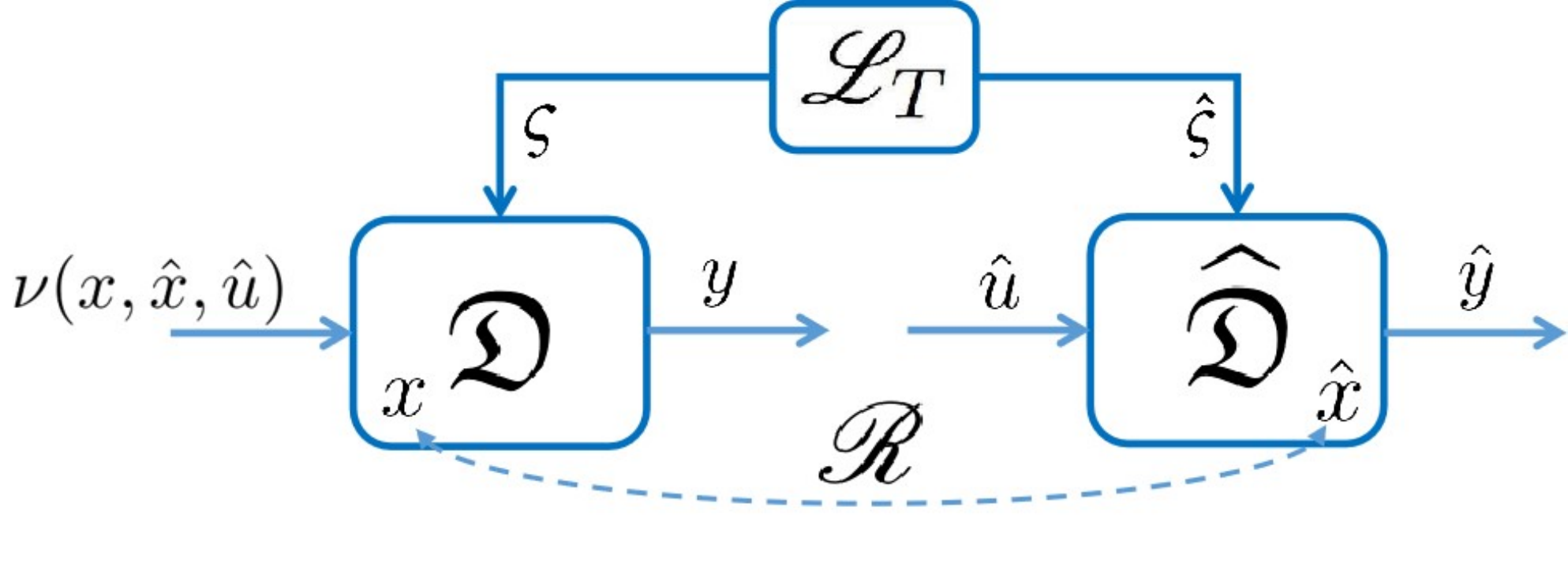}
		\caption{
			Relation $\mathscr{R}$ is defined over the state sets of $\mathfrak{D}$ and $\widehat{\mathfrak{D}}$. Stochastic kernel $\mathscr{L}_{T}$ specifies the relation between $\varsigma$ and $\hat{\varsigma}$ which affects $\mathfrak{D}$ and $\widehat{\mathfrak{D}}$, respectively. Interface function $\nu(x,\hat x,\hat u)$ is employed for the controller refinement.}
		\label{fig:controller_equivalent}
	\end{center}
\end{figure}

\begin{definition}\label{Def: apr}
	\emph{(($\epsilon, \delta$)-Approximate Probabilistic Relation~\cite{Haesaert2020Robust})}
	Consider two gMDPs $\mathfrak{D} =(X,U,x_0,T,Y,h)$ and $\widehat{\mathfrak{D}} =(\hat X,\hat U,\hat{x}_0,$ $\hat T,Y,\hat h)$ with the same output set. 
	System $\widehat{\mathfrak{D}}$ is ($\epsilon, \delta $)-stochastically simulated by $\mathfrak{D}$, denoted by $ \widehat{\mathfrak{D}}\preceq_{\epsilon}^{\delta}\mathfrak{D} $, if there exist a measurable relation $\mathscr{R}\subseteq X \times \hat X$ and a Borel measurable stochastic kernel $\mathscr{L}_{T}(\cdot~|~ x, \hat x, \hat u)$ on $X \times \hat X$ such that: 
	\begin{itemize}
		\item $\forall (x,\hat x) \in \mathscr{R}$, $\Vert y- \hat y \Vert \leq \epsilon$, with $y = h(x)$ and $\hat y = \hat h (\hat x)$;
		\item $\forall (x,\hat x)\!\in\! \mathscr{R}$ and $\forall \hat u \!\in\! \hat U$, there exists $u\!\in\!U$ such that $T(\cdot~|~ x, u)~\mathscr{\bar R}_{\delta} ~ \hat T(\cdot~|~ \hat x, \hat u)$ with lifting $\mathscr{L}_{T}(\cdot~|~ x, \hat x,\hat u)$;
		\item $ x_0 \mathscr{R}\hat{x}_0 $.
	\end{itemize}
\end{definition}
Consider a gMDP $\mathfrak{D}$ that models the original stochastic systems.
When building the Safe-visor architecture, we would first construct a finite abstraction of $\mathfrak{D}$, denoted by $\widehat{\mathfrak{D}}$, which is also a gMDP but with finite state and input sets.
Accordingly, the ($\epsilon, \delta$)-approximate probabilistic relation in Definition~\ref{Def: apr} is used to quantify the similarity between $\mathfrak{D}$ and $\widehat{\mathfrak{D}}$, which is the key insight for providing a formal safety guarantee.
The second condition of Definition~\ref{Def: apr} implicitly implies that there exists an \emph{interface function}~\cite{Girard2009Hierarchical} 
\begin{equation}\label{eq:interface}
\nu: X\times\hat{X}\times\hat{U}\rightarrow U,
\end{equation}
such that the state probability measures are in the $\delta$-lifted relation after one transition, when the control input $\hat u$ for $\widehat{\mathfrak{D}}$ is refined to $u$ for $\mathfrak{D}$ using this function.
Intuitively, $\delta$-lifted relation between the state probability measures of $\mathfrak{D}$ and $\widehat{\mathfrak{D}}$ indicates that the Euclidean distance between their outputs, \ie, $y$ and $\hat{y}$, is not larger than $\epsilon$, with a probability of at least $1-\delta$ at each time step.
This distance-based $\delta$-lifted relation is the key for providing the safety guarantee.
With this relation, one can control $y$ indirectly by controlling $\hat{y}$ while keeping $y$ sufficiently closed to $\hat{y}$ with some probability.
It is also worth noting that we need to ensure $u$ belongs to $U$ for $\mathfrak{D}$ so that the second condition in Definition~\ref{Def: apr} is fulfilled.
Therefore, we need to use the set
\begin{equation}\label{eq:U'}
\hat{U}'= \{\hat{u}\in\hat{U}\,|\,\nu(x,\hat{x},\hat{u})\in U\ \text{for all}\ (x,\hat{x})\in\mathscr{R}\},
\end{equation}
as the input set of the finite abstraction when synthesizing controllers for the abstraction.
To ensure a non-empty $\hat{U}'$, one can adopt the methods in~\cite[Section 4.3]{Zhong2021Automata}, in which elements in $\hat{U}'$ are first decided and then accommodated in the search for the ($\epsilon, \delta$)-approximate probabilistic relation.
We illustrate the ingredients of Definition~\ref{Def: apr} in Figure~\ref{fig:controller_equivalent} to provide more intuition for the ($\epsilon, \delta$)-approximate probabilistic relation between two gMDPs. 

So far, we provided the required conditions for establishing ($\epsilon, \delta$)-approximate probabilistic relations. 
The remaining problem is how to build a finite abstraction for the original model so that such a relation between two systems exists. 
To do so, one can employ a similar approach as in~\cite{Tkachev2013Quantitative}, in which continuous state and input states of gMDPs are partitioned via finite numbers of bounded cells. 
Representative points are then selected for these partitions, based on which the finite abstraction is constructed.
We refer the interested readers to~\cite{Tkachev2013Quantitative} for more technical details. 
For more details on establishing an ($\epsilon, \delta$)-approximate probabilistic relation between the original model and its constructed finite abstraction, we refer the interested readers to~\cite[Section 4]{Haesaert2020Robust} for linear stochastic systems; or to~\cite[Section 4]{Zhong2021Automata} for nonlinear stochastic systems with slope restrictions on nonlinearity~\cite{Lavaei2019Compositionalc}. 
We do not provide technical details of those methods in this work due to space limitations. 
In particular, our proposed approach is independent of the methods for finding such a relation. 
Instead, we focus on synthesizing the architecture based on the finite abstraction given the ($\epsilon, \delta$)-approximate probabilistic relation.

Note that $\delta$ in the ($\epsilon, \delta$)-approximate probabilistic relation can be employed to quantify the probabilistic deviation, which can be leveraged to reduce state and input sets dimension in the abstraction procedure.
This allows us to build a reduced-order model of the original system, based on which we construct a finite abstraction for synthesizing the safety advisor. 
Since the space complexity of building the finite abstraction grows exponentially w.r.t. the dimension of the system's state set, constructing a reduced-order model alleviates the encountered complexity.
As a result, one can leverage the proposed approach in this work and deal with systems with high-dimensional state sets.
This issue is shown via the running example.
\begin{remark}
	The degree of model-order reduction depends on the specific methods for constructing the reduced-order model, as well as the topology of the original system.
	We refer interested readers to~\cite{Touze2021Model,Baur2014Model,Girard2009Hierarchical} and references therein for more details.
	Additionally, compositional techniques (see~\cite{lavaei2021automated} and references therein) can also be leveraged to further alleviate the complexity issue.
	As a future direction, we plan to design the Safe-visor architecture based on finite abstractions that are constructed in a compositional manner.
\end{remark}
\addtocounter{example}{-1}
\begin{example}[continued]
	For constructing the finite abstraction, we follow the methods in~\cite[Section 4.1]{Zhong2021Automata}.
	We first construct a reduced-order version for the original model with scalar state and input sets via balance truncation as implemented in MATLAB.
	This results in a reduced-order model 
	\begin{align*}
	\widehat{\mathfrak{D}}_{\textsf r}\!:\left\{\hspace{-1.5mm}\begin{array}{l}\hat{x}_{\textsf r}(k+1) = \hat{A}_{\textsf r}\hat{x}_{\textsf r}(k)+\hat{B}_{\textsf r}\hat{u}_{\textsf r}(k)+\hat{R}_{\textsf r}\varsigma(k),\\
	\hat{y}_{\textsf r}(k)  = \hat{C}_{\textsf r}\hat{x}_{\textsf r}(k),\end{array}\right.\quad k\in\mathbb N,
	\end{align*}
	where $\hat{A}_{\textsf r} = 1$, $\hat{B}_{\textsf r} = 0.3469$, and $\hat{C}_{\textsf r} = 1$ (the index ${\textsf r}$ signifies the reduced-order version of the original model), with a reduction matrix $P = [1\,;0.76\,;0]$.
	Additionally, $\hat{R}_{\textsf r}$ would be selected later when establishing the approximate probabilistic relation.
	Then, we construct a finite abstraction of the reduced-order model by considering $\hat{X}_{\textsf r} = [0,10]$ as state set of $\widehat{\mathfrak{D}}_{\textsf r}$, and uniformly dividing this set into 100 partitions. 
	Moreover, we uniformly partition the input set into cells whose lengths are $0.03$.	
	For establishing ($\epsilon$,$\delta$)-approximate probabilistic relation, we apply the method in~\cite[Section 4.3]{Zhong2021Automata}.
	We set $\hat{U}'=\{\hat{u}\in\hat{U}|-0.3\leq \hat{u}\leq 0.3\}$, $\delta = 0.01$, and the expected range of $\epsilon$ as $[0.05,1]$.
	Then, the finite abstraction is $(\epsilon,\delta)$-stochastically simulated by original model with the relation $\mathscr{R} = \{(x,\hat{x})|(x-P\hat{x})^TM(x-P\hat{x})\leq \epsilon^2\}$	where
	\begin{align*}
	M=\begin{bmatrix}\begin{smallmatrix}2.4021\ &-0.2239\ &0.2239\\-0.2239\ &0.03576\ &-0.03576\\0.2239\ &-0.03576\ &0.03576\end{smallmatrix} \end{bmatrix},
	\end{align*}
	$\epsilon = 0.7984$, and the interface function is $\nu(x,\hat{x},\hat{u}):= K(x-P\hat{x}) + D\hat{x} + \tilde{R}\hat{u}$ with $K=[7.5764\,;-1.2399\,;1.2399]^T$, $D=0.1852$, and $\tilde{R} = 0.2530$.
	Moreover, we select $\hat{R}_{\textsf r}\!=\!0.0159$ as in~\cite[(4.39)]{Zhong2021Automata}, and construct the lifting stochastic kernel for this relation such that the noise term in the original and finite systems are the same.
	
	With the running example, we also show how one can improve the scalability issue by constructing a reduced-order model.
	We now construct the finite abstraction of the original model without any model order reduction by considering $X = [0,10]^3$.
	Then, we uniformly partition this region with girds whose sizes are $(0.1,0.1,0.1)$ and divide the original input set into 20 partitions for a fair comparison with the reduced-order model setting.
	The number of states and the required memory~\footnote{We allocate $8$ bytes for each entry of the stochastic kernel to be stored as a double-precision floating-point.}  for saving the stochastic kernels of the finite abstractions in both cases are shown in Table~\ref{tab_scalbility}. 
	When a reduced-order model is built, we are able to construct the Safe-visor architecture with respect to the desired safety specification (see the discussion in Section~\ref{sec:running}).
	Meanwhile, we only need around $1.55$ MB to save the stochastic kernel thanks to the reduced-order model and, hence, the reduction in the number of states of the finite abstraction.
	On the contrary, without constructing a reduced-order model, we need $10^8$ times more memory to save the stochastic kernel, which is not tractable.
	\begin{table}
		\caption{Required memory for the construction of finite abstraction from reduced-order and original models. \vspace{0.2cm}}\label{tab_scalbility}
		\begin{tabular}{|p{3.5cm}|p{3cm}|p{4cm}|}
			\hline
			System model&  Number of states &  Required memory (GB)\\
			\hline
			Reduced-order model& 101 & $1.55\times 10^{-3}$ \\
			\hline
			Original model & $10^6$ & $1.49\times 10^5$ \\
			\hline
		\end{tabular}
	\end{table}
\end{example}
\begin{remark}
	In general, by reducing the quantization parameters (i.e. the size of the cells for constructing the finite abstraction) in partitioning the continuous state sets, a better safety guarantee can be provided by the safety advisor, but at the cost of increasing the execution time of the supervisor.
	This is shown with an example in Section~\ref{sec:dcmotor}.
\end{remark}

\subsection{Safety Advisor}\label{sec:safetyadv}
In this work, we synthesize a robust controller with a method adapted from~\cite[Section 5]{Zhong2021Automata} as the safety advisor in the Safe-visor architecture. 
Note that~\cite{Zhong2021Automata} focuses on synthesizing robust controllers for two and a half player games~\cite{Zhu2015Game}(\emph{e.g.,} stochastic systems with adversary input selected by rational agent).
However, gMDP in this paper can be treated as a game without any adversary inputs (\emph{i.e.,} one and a half player games) so that the method in~\cite{Zhong2021Automata} is still applicable to the problem of interest in this paper.
\begin{figure}[htbp]
	\centering
	\subfigure{
		\includegraphics[width=0.12\textwidth]{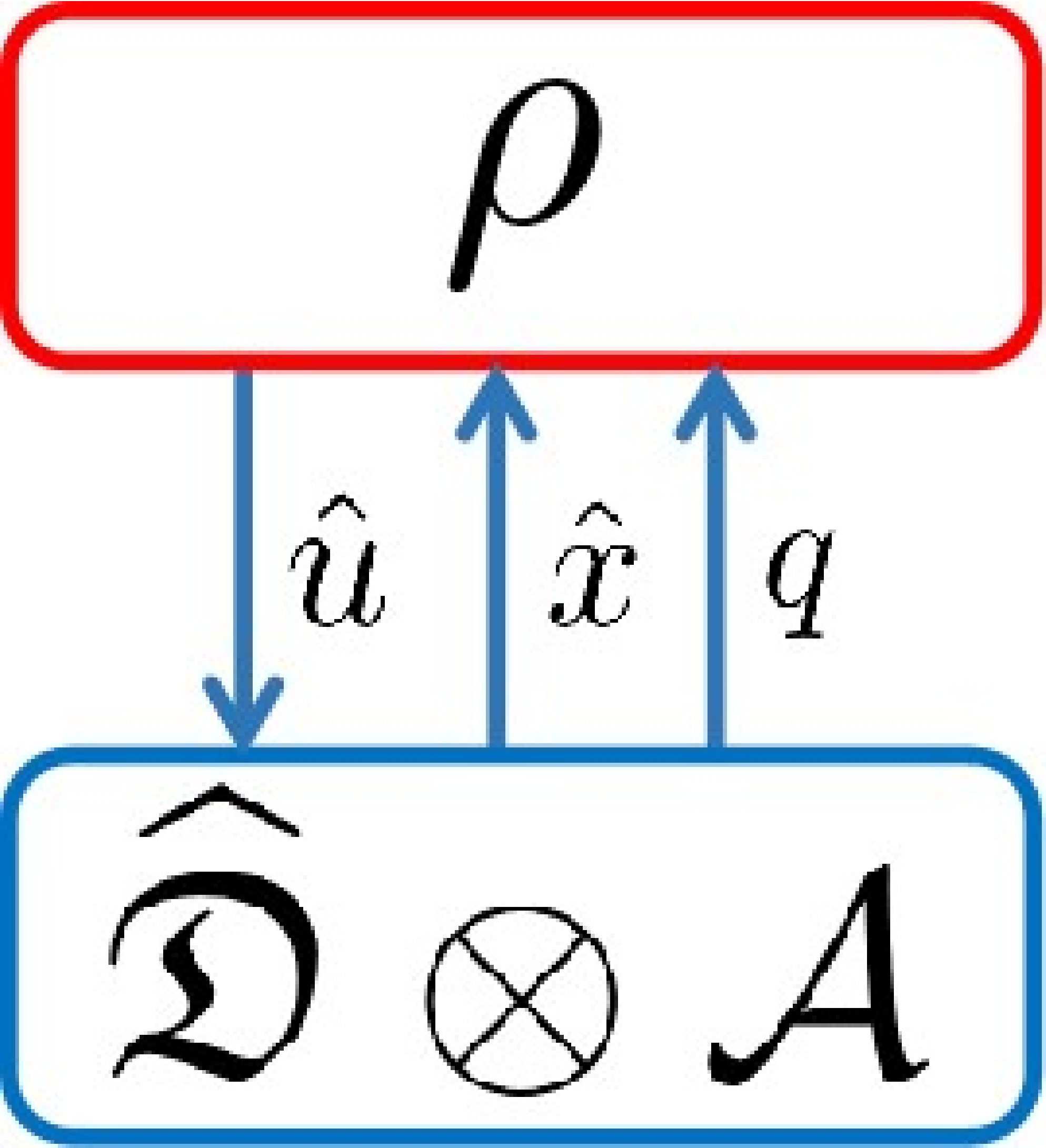}
	}
	\quad
	\subfigure{
		\includegraphics[width=0.35\textwidth]{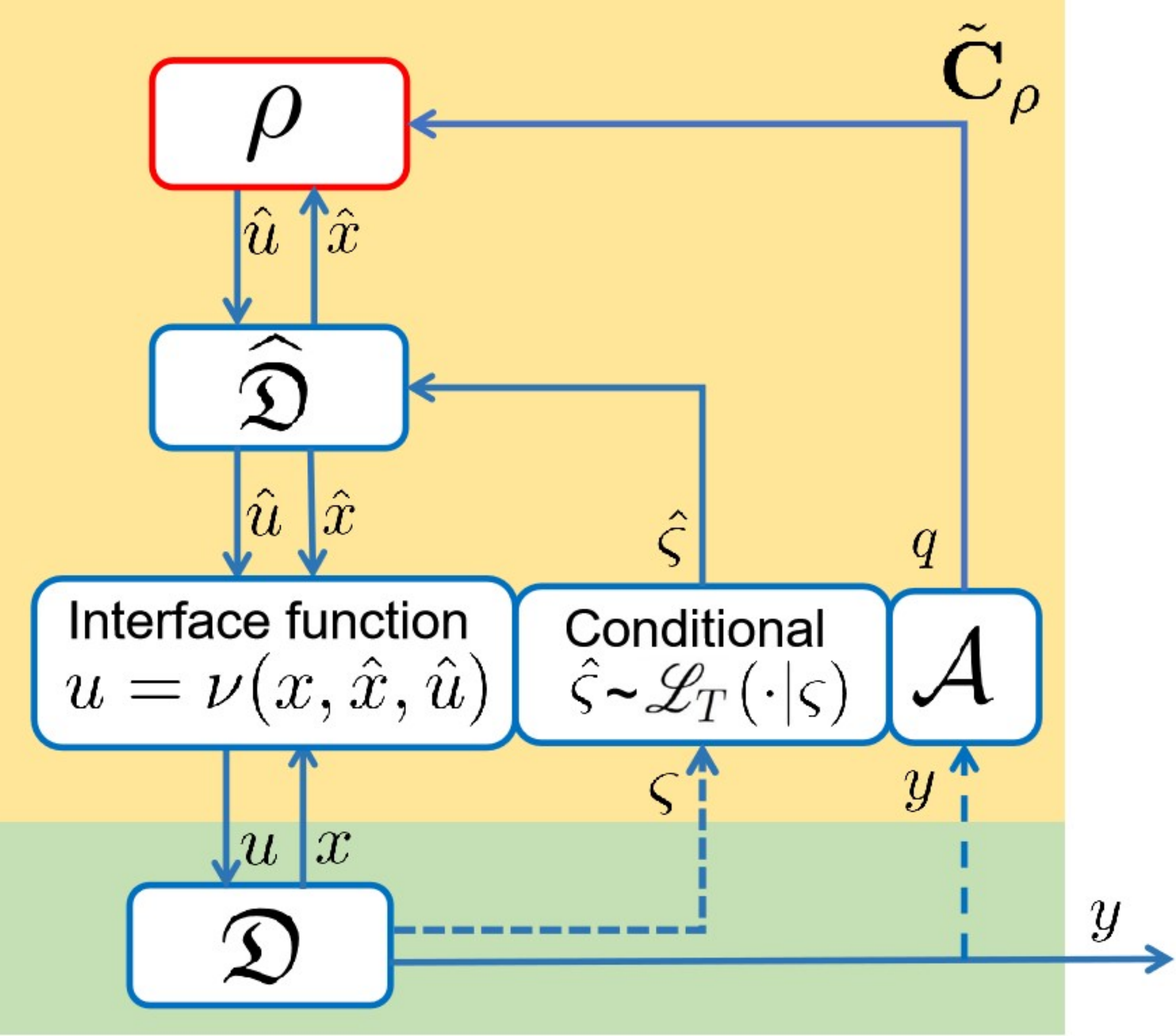}
	}
	\caption{\textbf{Left:} Synthesizing Markov policy for $\widehat{\mathfrak{D}}\otimes\mathcal{A}$.~\textbf{Right:} Construction of $\tilde{\mathbf{C}}_{\rho}$ (the yelow region).}
	\label{fig:idea}
\end{figure}
As shown in Figure~\ref{fig:idea} (left), we first synthesize a Markov policy $\rho$ for the gMDP $\widehat{\mathfrak{D}}\otimes\mathcal{A}$.
We then construct a control strategy $\tilde{\mathbf{C}}_{\rho}$ based on $\rho$ as depicted in Figure~\ref{fig:idea} (right).
In runtime, when a state $x$ of $\mathfrak{D}$ is fed to $\tilde{\mathbf{C}}_{\rho}$:
\begin{enumerate}[(i)]
	\item State $\hat{x}$ of $\widehat{\mathfrak{D}}$ is first updated according to $x$ and the conditional kernel $\mathscr{L}_T$.
	Then, the state $q$ of $\mathcal{A}$ is updated according to the output function $h(x)$ of $\mathfrak{D}$ and transition function $\tau$ of $\mathcal{A}$;
	\item $\hat{u}$ is provided by $\rho$ based on $\hat{x}$ and $q$, and is refined to $\mathfrak{D}$ via the interface function $\nu$ as in~\eqref{eq:interface}.
\end{enumerate}

Here, we formally define the construction of $\tilde{\mathbf{C}}_{\rho}$ as follows.
\begin{definition}\label{def:C_rho}
	\emph{(Construction of $\tilde{\mathbf{C}}_{\rho}$)} Given a Markov policy $\rho=(\rho_{0},\rho_{1},\ldots,$ $\rho_{H-1})$ for $\widehat{\mathfrak{D}}\otimes\mathcal{A}$, we construct $\tilde{\mathbf{C}}_{\rho}= (\tilde{\mathsf{M}},\tilde{\mathsf{U}},\tilde{\mathsf{Y}},\tilde{\mathsf{H}},\tilde{\mathsf{m}}_{0},\tilde{\pi}_{\mathsf{M}},\tilde{\pi}_{\mathsf{Y}})$ for $\mathfrak{D}$, with   
	\begin{itemize}
		\item $\tilde{\mathsf{M}}=X\times\hat{X}\times Q$;
		\item $\tilde{\mathsf{U}}=X$; 
		\item $\tilde{\mathsf{Y}}=U$;
		\item $\tilde{\mathsf{H}}=[0,H-1]$;
		\item $\tilde{\mathsf{m}}_0\!\!=\!\!\big(\tilde{\mathsf{m}}_{X}(0),\tilde{\mathsf{m}}_{\hat{X}}(0),\tilde{\mathsf{m}}_{Q}(0)\big)\!\!\in\!\tilde{\mathsf{M}}$ with $\tilde{\mathsf{m}}_{X}(0)=x_0$, $\tilde{\mathsf{m}}_{\hat{X}}(0)=\hat{x}_0$ and $\tilde{\mathsf{m}}_{Q}(0)=\tau\big(q_0,L\circ h(\tilde{\mathsf{m}}_{X}(0))\big)$;
		\item $\tilde{\pi}_{\mathsf{M}}$ updates $\big(\tilde{\mathsf{m}}_{X}(k),\tilde{\mathsf{m}}_{\hat{X}}(k),\tilde{\mathsf{m}}_{Q}(k)\big)\in\tilde{\mathsf{M}}$ at all time instant $k\in\mathsf{H}\backslash\{0\}$ with the following steps:
		\begin{enumerate}[(i)]
			\item update $\tilde{\mathsf{m}}_{\hat{X}}(k)$ according to the conditional stochastic kernel
			\begin{align}
			\mathscr{L}_{T}\big(d\hat{x}\,\big|\,\tilde{\mathsf{m}}_{\hat X}(k-1),\tilde{\mathsf{m}}_{X}(k-1),x(k),\hat{u}(k-1)\big),\label{eq:condition_kernel}
			\end{align}
			with $x(k)$ which is the state of $\mathfrak{D}$ at time instant $k$ and $\hat{u}(k-1)=\tilde{\rho}_k(\tilde{\mathsf{m}}_X(k-1),\tilde{\mathsf{m}}_{\hat X}(k-1),\tilde{\mathsf{m}}_{Q}(k-1))$;
			\item update $\tilde{\mathsf{m}}_{X}(k)$ as $\tilde{\mathsf{m}}_{X}(k)=x(k)$;
			\item update $\tilde{\mathsf{m}}_{Q}(k)$ as $\tilde{\mathsf{m}}_{Q}(k)=\tau\big(\tilde{\mathsf{m}}_{Q}(k-1),L\circ h(\tilde{\mathsf{m}}_{X}(k))\big)$;
		\end{enumerate}
		\item $\tilde{\pi}_{\mathsf{Y}}$ updates $\mathsf{y}(k)\in\mathsf{Y}$ at time instant $k\in\mathsf{H}$ as $\mathsf{y}(k)=\nu\big(\tilde{\mathsf{m}}_X(k),\tilde{\mathsf{m}}_{\hat X}(k),$ $\rho_k(\tilde{\mathsf{m}}_{\hat X}(k),\tilde{\mathsf{m}}_{Q}(k))\big)$, where $\nu$ is the interface function associated with the ($\epsilon$-$\delta$)-approximate probabilistic relation.	
	\end{itemize}
\end{definition}
\begin{remark}
	The conditional stochastic kernel~\eqref{eq:condition_kernel} can be obtained by decomposing the lifting kernel $\mathscr{L}_{T}(dx'\times d\hat{x}'~|~ x,\hat{x},\hat{u})$ in Definition~\ref{Def: apr} as
	\begin{align}
	\mathscr{L}_{T}\big(dx'|x,\hat{x},\hat{x}',\nu(x,\hat{x},\hat{u})\big)\hat{T}(d\hat{x}'|\hat{x},\hat{u}).
	\end{align}
	This decomposition is feasible according to~\cite[Corollary 3.1.2]{Borkar2012Probability}.
\end{remark}

Before synthesizing $\rho$ for $\widehat{\mathfrak{D}}\otimes\mathcal{A}$, we raise the following assumption that is required for the ($\epsilon$,$\delta$)-approximate probabilistic relation. 

\begin{assumption}\label{asp1}
	Consider gMDPs $\mathfrak{D}= (X,U,x_0,T,Y,h)$ and $\widehat{\mathfrak{D}}= (\hat X,\hat U, \hat{x}_0,$ $\hat T, Y,\hat h)$ with $\widehat{\mathfrak{D}}\preceq^{\delta}_{\epsilon}\mathfrak{D}$ as in Definition~\ref{Def: apr}.
	We assume that 
	\begin{equation*}
	\int_{\bar{\mathscr{R}}_{\hat{x}'}}\mathscr{L}_{T}\big(dx'|x,\hat{x},\hat{x}',\nu(x,\hat{x},\hat{u})\big)\geq 1-\delta
	\end{equation*}
	holds for all $\hat{x},\hat{x}'\in\hat{X}$, with $\bar{\mathscr{R}}_{\hat{x}'}=\{x' \in X | (x',\hat{x}')\in \mathscr{R}\}$ and $\mathscr{L}_{T}(dx'|x,\hat{x},\hat{x}',$ $\nu(x,\hat{x},\hat{u}))$ being the conditional probability of $x'\in X$ given $x$, $\hat{x}$, $\hat{x}'$ and the interface function $\nu(x,\hat{x},\hat{u})$.
\end{assumption}

Assumption~\ref{asp1} is similar to~\cite[Assumption 5.2]{Zhong2021Automata}.
This assumption assumes that all states in the abstraction are coupled into the $\delta$-lifted relation, and at every time instant $k$, $\mathbb{P}\{(x',\hat{x}')\in\mathscr{R}~|~(x,\hat{x})\in \mathscr{R}\}\geq 1-\delta$ holds for all $\hat{x}\in\hat{X}$ with the interface function, where $(x,\hat{x})$, $(x',\hat{x}')$ are the state pairs at time instants $k$ and $k+1$, respectively. 
Considering all available results for constructing ($\epsilon$,$\delta$)-approximate probabilistic relation~\cite{Haesaert2020Robust,Lavaei2019Compositionala,Huijgevoort2020Similarity,Zhong2021Automata}, 
Assumption~\ref{asp1} does not introduce extra subtlety in practice.
In all those results, although they do not explicitly need this assumption, the existence of an ($\epsilon$,$\delta$)-approximate probabilistic relation is in fact guaranteed by forcing Assumption~\ref{asp1} (c.f.~\cite[Section 4, Condition A3]{Haesaert2020Robust},~\cite[Theorem 5.3, equation(5.5g)]{Lavaei2019Compositionala},~\cite[Theorem 3]{Huijgevoort2020Similarity} and~\cite[equation(A.1)]{Zhong2021Automata}).
Here, we adopt the results in~\cite{Zhong2021Automata} for synthesizing the safety advisor since they provide a less conservative guarantee than the ones proposed in~\cite {Haesaert2020Robust} (see~\cite[Lemma 5.6, Corollary 5.7]{Zhong2021Automata}).
Next, we proceed with discussing how to synthesize a Markov policy $\rho$ for the gMDP $\widehat{\mathfrak{D}}\otimes \mathcal{A}$ given an ($\epsilon, \delta$)-approximate probabilistic relation.

{\bf Problem of Robust Satisfaction.}
For the problem of robust satisfaction as in Problem~\ref{prob:robust_satisfaction}, consider a gMDP $\mathfrak{D}= (X,U,x_0,T,Y,h)$ and its finite abstraction $\widehat{\mathfrak{D}}= (\hat X,\hat U,\hat{x}_0, \hat T,Y,\hat h)$, a DFA $\mathcal{A}$ that characterizes the desired safety specification and the product gMDP $\widehat{\mathfrak{D}}\otimes\mathcal{A} = \{\bar{X},\,\bar{U},\,\bar{x}_0,\,\bar{T},\,\bar{Y},\,\bar{h}\}$ as in Definition~\ref{def:product_gmdp_dfa}.
Given a Markov policy $\rho=(\rho_0,\rho_1,\ldots,\rho_{H-1})$ defined over the time horizon $[0,H]$, we define a cost-to-go function $\bar{V}^{\rho}_n:\hat{X}\times Q \rightarrow [0,1]$ which assigns a real number to states of $\widehat{\mathfrak{D}}\otimes\mathcal{A}$ at time instant $H-n$.
We initialize this function with $\bar{V}_0^{\rho}(\hat{x},q)=1$ when $q\in F$, and $\bar{V}_0^{\rho}(\hat{x},q)=0$ when $q\notin F$, and recursively compute $\bar{V}_{n+1}^{\rho}(\hat{x},q)$ as 
\begin{align}
\bar{V}_{n+1}^{\rho}(\hat{x},q):=\left\{ 
\begin{aligned}
(1-\delta)\sum_{\hat{x}'\in \hat{X}}\bar{V}^{\rho}_{n}\big(\hat{x}',\underline{q}(\hat{x}',q)\big)&\hat{T}\big(\hat{x}'\,\big|\,\hat{x},\rho_{H-n-1}(\hat{x},q)\big), \text{ if } q\notin F;\\
&1, \quad \quad \quad\quad \quad\quad\quad\quad \quad\  \text{   if } q\in F,
\end{aligned}\right.\label{eq:P_general}
\end{align}
with
\begin{equation}\label{eq:underline_p}
\underline{q}(\hat{x}',q) := \mathop{\arg\min}_{q'\in Q'_{\epsilon}(\hat{x}')}\bar{V}^{\rho}_{n}(\hat{x}',q'),
\end{equation}
where 
\begin{equation}
Q'_{\epsilon}(\hat{x}') \!:=\! \big\{ q'\in Q\,\big|\,~\exists x\in X\,s.t.\,q'=\tau(q,L\circ h(x))\ \text{with}\ h(x)\in {N}_{\epsilon}(\hat{h}(\hat{x}')) \big\},\label{Q_eps}
\end{equation}
and 
\begin{equation}\label{eq:N_eps}
N_{\epsilon}(\hat{y}):=\{y\in Y~|~\lVert y-\hat{y}\rVert\leq \epsilon \}.
\end{equation}
With this notion, we propose the following theorem that provides the safety guarantee associated with $\rho$ for the problem of robust satisfaction.
\begin{theorem}\label{thm:gua_prmax}
	Consider gMDPs $\mathfrak{D}= (X,U,x_0,T,Y,h)$ and $\widehat{\mathfrak{D}}= (\hat X,\hat U, \hat{x}_0, \hat T, $ $Y,\hat h)$ with $\widehat{\mathfrak{D}}\preceq_{\epsilon}^{\delta}\mathfrak{D}$, and a DFA $\mathcal{A}=(Q, q_0, \Pi,\tau, F)$ characterizing the desired safety property.
	For any Markov policy $\rho$ designed for $\widehat{\mathfrak{D}}\times\mathcal{A}$ and a control strategy $\tilde{\mathbf{C}}_{\rho}$ of $\mathfrak{D}$ constructed based on $\rho$ as in Definition~\ref{def:C_rho}, we obtain
	\begin{equation}\label{eq:thm_sad_pmax}
	\mathbb{P}_{\tilde{\mathbf{C}}_{\rho}\times \mathfrak{D}}\{\exists k\leq H, y_{\omega k}\models \mathcal{A}\}\geq \bar{V}_H^{\rho}(\hat{x}_0,\bar{q}_0),
	\end{equation}
	with $y_{\omega H}$ being output sequences of $\mathfrak{D}$ up to the time instant $H$, $\bar{V}_H^{\rho}(\hat{x}_0,\bar{q}_0)$ computed as in~\eqref{eq:P_general}, and $\bar{q}_0 = \tau(q_0,L\circ h(x_0))$.
\end{theorem}
Theorem~\ref{thm:gua_prmax} is adapted from~\cite[Theorem 5.3]{Zhong2021Automata} with some modification and can therefore be proved similarly.
Since the safety advisor is responsible for maximizing the safety probability, we are interested in a $\rho\in\mathcal P^H$ that maximizes $\bar{V}_H^{\rho}(\hat{x}_0,\bar{q}_0)$ as in \eqref{eq:thm_sad_pmax}.
The following proposition shows how such a Markov policy can be synthesized.
\begin{proposition}
	Consider gMDPs $\mathfrak{D}= (X,U,x_0,T,Y,h)$ and $\widehat{\mathfrak{D}}= (\hat X,\hat U, \hat{x}_0,$ $\hat T, $ $Y,\hat h)$ with $\widehat{\mathfrak{D}}\preceq_{\epsilon}^{\delta}\mathfrak{D}$, and a DFA $\mathcal{A}=(Q, q_0, \Pi,\tau, F)$.
	The Markov policy $\rho^*=(\rho_0^*,\rho_1^*,\ldots,\rho_{H-1}^*)$ maximizes $\bar{V}_H^{\rho}(\hat{x}_0,\bar{q}_0)$ as in \eqref{eq:thm_sad_pmax}, with 
	\begin{align}
	\rho^*_{H-n-1}\in\mathop{\arg\max}_{\rho_{H-n-1}\in \mathcal P}(1-\delta)\!\sum_{\hat{x}'\in \hat{X}}\!\bar{V}^*_{n}\big(\hat{x}',\underline{q}^*(\hat{x}',q)\big)\hat{T}\big(\hat{x}'\,\big|\,\hat{x},\rho_{H-n-1}(\hat{x},q)\big),\label{eq:policy_opt_sac_ab}
	\end{align}
	for all $n\in[0,H-1]$. 
	Here, we denote by $\bar{V}_{n}^*(\hat{x},q)$ the cost-to-go function associated with $\rho^*$, and this function can be recursively computed as
	\begin{align}
	\bar{V}_{n+1}^*(\hat{x},q)\!:=\!\left\{ 
	\begin{aligned}
	\max_{\rho_{H-n-1}\in \mathcal P}(1-\delta)&\sum_{\hat{x}'\in \hat{X}}\bar{V}^*_{n}\big(\hat{x}'\!,\underline{q}^*(\hat{x}'\!,q)\big)\hat{T}\big(\hat{x}'\big|\hat{x},\rho_{H-n-1}(\hat{x},q)\big), \text{ if } q\notin F;\\
	&1, \quad \quad \quad\quad \quad\quad\quad\quad\quad\quad \text{   if } q\in F,
	\end{aligned}\right.\label{eq:P_opt_sac}
	\end{align}
	initialized by $\bar{V}_0^*(\hat{x},q)=1$, when $q\in F$, and $\bar{V}_0^*(\hat{x}_0,q)=0$ otherwise, where 
	\begin{equation}\label{eq:underline_p_star}
	\underline{q}^*(\hat{x}',q) := \mathop{\arg\min}_{q'\in Q'_{\epsilon}(\hat{x}')}\bar{V}^{*}_{n}(\hat{x}',q'),
	\end{equation}
	and $Q'_{\epsilon}(\hat{x}')$ is the set as in~\eqref{Q_eps}.
\end{proposition}

{\bf Problem of Worst-case Violation}.
For the problem of worst-case violation as in Problem~\ref{prob:worst_case_violation}, consider a Markov policy $\rho=(\rho_0,\rho_1,\ldots,$ $\rho_{H-1})$ defined over the time horizon $[0,H]$.
We define a cost-to go function $\ul{V}^{\rho}_n:\hat{X}\times Q\rightarrow [0,1]$ which maps each state of $\widehat{\mathfrak{D}}\otimes\mathcal{A}$ at the time instant $H-n$ to a real number.
Here, $\ul{V}^{\rho}_{n+1}(x,q)$ is recursively computed as
\begin{align}
\ul{V}_{n+1}^{\rho}(\hat{x},q)\! :=\! \left\{
\begin{aligned}
(1-\delta)\!\!\sum_{\hat{x}'\in \hat{X}}\ul{V}^{\rho}_{n}\big(\hat{x}',\bar{q}(\hat{x}',q)\big)&\hat{T}\big(\hat{x}'\,\big|\,\hat{x},\rho_{H-n-1}(\hat{x},q)\big)+\delta,\text{ if } q\notin F;\\
&1, \quad \quad \quad\quad \quad\quad\quad\quad\quad\quad \quad \text{   if } q\in F,
\end{aligned}\right.\label{eq:T_general}
\end{align}
initialized by $\ul{V}_0^{\rho}(\hat{x},q)=1$, when $q\in F$, and $\ul{V}_0^{\rho}(\hat{x},q)=0$ otherwise, where
\begin{equation}\label{eq:overline_p}
\bar{q}(\hat{x}',q) := \mathop{\arg\max}_{q'\in Q'_{\epsilon}(\hat{x}')}\ul{V}^{\rho}_{n}(\hat{x}',q'),
\end{equation}
and $Q'_{\epsilon}(\hat{x}')$ is the set as in~\eqref{Q_eps}.
Similar to Theorem~\ref{thm:gua_prmax}, next theorem provides the safety guarantee associated with $\rho$ for the problem of worst-case violation.
\begin{theorem}\label{thm:gua_prmin}
	Consider gMDPs $\mathfrak{D}= (X,U,x_0,T,Y,h)$ and $\widehat{\mathfrak{D}}= (\hat X,\hat U, \hat{x}_0, \hat T, $ $Y,\hat h)$ with $\widehat{\mathfrak{D}}\preceq_{\epsilon}^{\delta}\mathfrak{D}$, and a DFA $\mathcal{A}=(Q, q_0, \Pi,\tau, F)$ characterizing the desired safety property.
	For any Markov policy $\rho$ designed for $\widehat{\mathfrak{D}}\times\mathcal{A}$ and a control strategy $\tilde{\mathbf{C}}_{\rho}$ of $\mathfrak{D}$ constructed based on $\rho$ as in Definition~\ref{def:C_rho}, we obtain
	\begin{equation}\label{eq:thm_sad_pmin}
	\mathbb{P}_{\tilde{\mathbf{C}}_{\rho}\times \mathfrak{D}}\{\exists k\leq H, y_{\omega k}\models \mathcal{A}\}\leq \ul{V}_H^{\rho}(\hat{x}_0,\bar{q}_0),
	\end{equation}
	where $y_{\omega H}$ is the output sequences of $\mathfrak{D}$ up to the time instant $H$, $\ul{V}_H^{\rho}(\hat{x}_0,\bar{q}_0)$ is computed as in~\eqref{eq:T_general}, and $\bar{q}_0 = \tau(q_0,L\circ h(x_0))$.
\end{theorem}
Theorem~\ref{thm:gua_prmin} is adapted from~\cite[Theorem 5.8]{Zhong2021Automata} with some modification and its proof is omitted here due to the lack of space.
As for the problem of worst-case violation, the safety advisor is responsible for minimizing the probability of violating the desired safety specifications.
Thus, we are interested in a $\rho\in \mathcal P^H$ that minimizes $\ul{V}_H^{\rho}(\hat{x}_0,\bar{q}_0)$ in \eqref{eq:thm_sad_pmin}.
We discuss in the following proposition how to synthesize such a Markov policy.

\begin{proposition}
	Consider gMDPs $\mathfrak{D}= (X,U,x_0,T,Y,h)$ and $\widehat{\mathfrak{D}}= (\hat X,\hat U, \hat{x}_0,$ $\hat T, $ $Y,\hat h)$ with $\widehat{\mathfrak{D}}\preceq_{\epsilon}^{\delta}\mathfrak{D}$, and a DFA $\mathcal{A}=(Q, q_0, \Pi,\tau, F)$.
	The Markoc policy $\rho_{*}=(\rho_{*_0},\rho_{*_1},\ldots,\rho_{*_{H-1}})$ minimizes $\ul{V}_H^{\rho}(\hat{x}_0,\bar{q}_0)$ in \eqref{eq:thm_sad_pmin}, with 
	\begin{align}\label{eq:policy_opt_vio_ab}
	\rho_{*_{H-n-1}}\in\mathop{\arg\min}_{\rho_{H-n-1}\in \mathcal P}\Big(&(1-\delta)\!\!\sum_{\hat{x}'\in \hat{X}}\!\!\ul{V}_{*,n}\big(\hat{x}',\bar{q}_*(\hat{x}',q)\big)\hat{T}\big(\hat{x}'\,\big|\,\hat{x},\rho_{H-n-1}(\hat{x},q)\big)+\delta\Big) ,
	\end{align}
	for all $n\in[0,H-1]$.
	Here, we denote by $\ul{V}_{*,n}(\hat{x},q)$ the cost-to-go function associated with $\rho_*$.
	We initialize it with $\ul{V}_{*,0}(\hat{x},q)=1$, when $q\in F$, $\ul{V}_{*,0}(\hat{x},q)=0$, when $q\notin F$, and recursively compute it as
	\begin{align}
	\ul{V}_{*,n+1}(\hat{x},q) :=\left\{
	\begin{aligned}
	\min_{\rho_{H-n-1}\in \mathcal P}\Big(&(1-\delta)\!\!\sum_{\hat{x}'\in \hat{X}}\!\!\ul{V}_{*,n}\big(\hat{x}'\!,\bar{q}_*(\hat{x}'\!,q)\big)\hat{T}\big(\hat{x}'\,\big|\,\hat{x},\rho_{H-n-1}(\hat{x},q)\big)+\delta\Big),\text{ if } q\notin F;\\
	&\quad \  1, \quad \quad \quad\quad \quad\quad\quad\quad\quad\quad \quad\, \text{   if } q\in F,
	\end{aligned}\right.\label{eq:P_opt_vio}
	\end{align}
	where
	\begin{equation}\label{eq:overline_p_star}
	\bar{q}_*(\hat{x}',q) := \mathop{\arg\max}_{q'\in Q'_{\epsilon}(\hat{x}')}\ul{V}_{*,n}(\hat{x}',q'),
	\end{equation}
	and $ Q'_{\epsilon}(\hat{x}')$ is a set as in~\eqref{Q_eps}.
\end{proposition}

Finally, we summarize the construction of safety advisor as follows:
\begin{itemize}
	\item For the problem of robust satisfaction, we synthesize a Markov policy $\rho^*$ as in~\eqref{eq:policy_opt_sac_ab}.
	Then, we construct a control strategy $\tilde{\mathbf{C}}_{\rho^*}$ as in Definition~\ref{def:C_rho} based on $\rho^*$ and apply $\tilde{\mathbf{C}}_{\rho^*}$ as the safety advisor.
	\item As for the problem of worst-case violation, we construct a control strategy $\tilde{\mathbf{C}}_{\rho_*}$ as in Definition~\ref{def:C_rho} based on a Markov policy $\rho_*$ synthesized as in~\eqref{eq:policy_opt_vio_ab}, and employ $\tilde{\mathbf{C}}_{\rho_*}$ as the safety advisor.
\end{itemize}
Since both $\rho^*$ and $\rho_*$ are look-up tables that are computed offline, the safety advisor can be applied efficiently at runtime.

\subsection{Design of Supervisor}\label{sec:supervisor}
In general, the proposed supervisor takes two steps to decide whether to accept the input from the unverified controller, denoted by $u_{uc}(k)$, at every time instant $k$:
\begin{itemize}
	\item Step 1: Check whether the ($\epsilon$,$\delta$)-approximate probabilistic relation will still hold between the finite abstraction and the original system assuming that $u_{uc}(k)$ is accepted.
	If the relation will not hold, reject $u_{uc}(k)$ without going through Step 2 and feed input provided by the safety advisor, denoted by $u_{\text{safe}}$, to the system $\mathfrak{D}$.
	\item Step 2: Estimate the probability of violating the desired safety specification, denoted by $\mathcal{E}_{pv}(k)$, assuming that $u_{uc}(k)$ is accepted.
	Accept $u_{uc}(k)$ only if $\mathcal{E}_{pv}(k)\leq \eta$; otherwise, feed $u_{\text{safe}}$ to the system $\mathfrak{D}$.
\end{itemize}
\begin{figure}[htbp]
	\centering
	\subfigure{
		\includegraphics[width=0.55\textwidth]{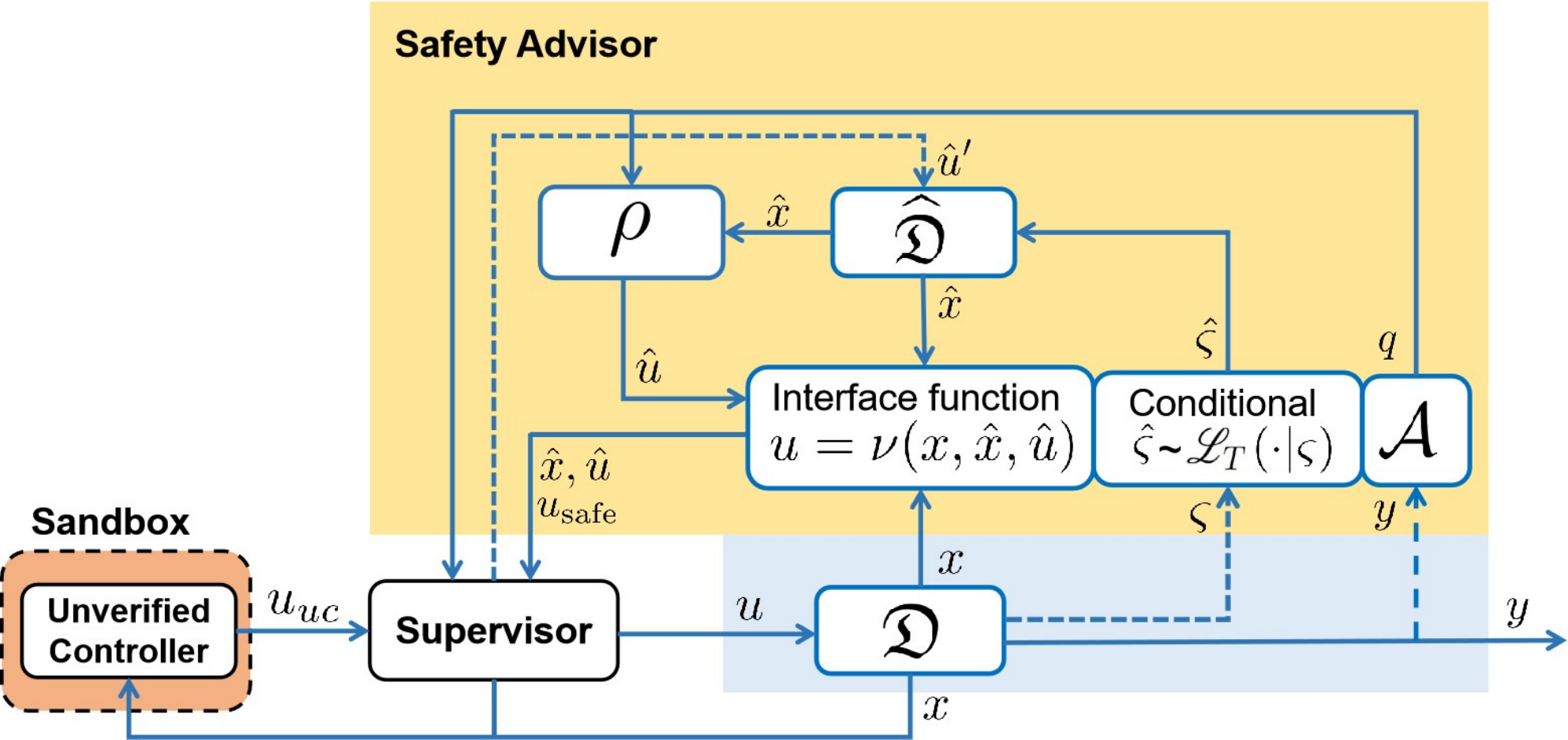}
	}
	\quad
	\subfigure{
		\includegraphics[width=0.25\textwidth]{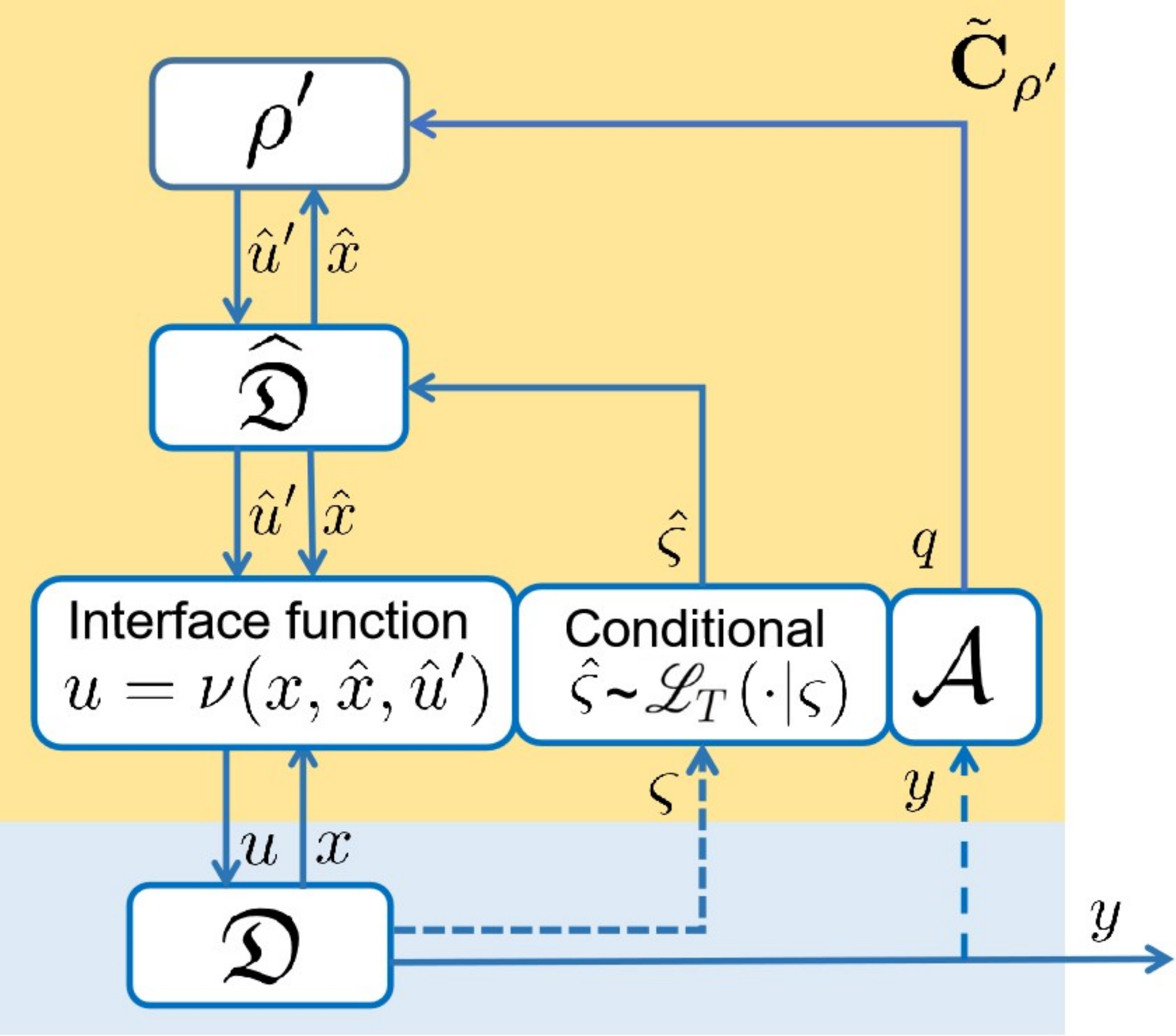}
	}
	\caption{\textbf{Left:} Safe-visor architecture designed based on an ($\epsilon$,$\delta$)-approximate probabilistic relation.\textbf{Right:} Controlled gMDP $\tilde{C}_{\rho'}\times\mathfrak{D}$ that is equivalent to $\mathfrak{D}$ operating in the Safe-visor \\ architecture.}
	\label{fig1:svalifting}
\end{figure}
We illustrate in Figure~\ref{fig1:svalifting} (left) the Safe-visor architecture with the safety-advisor proposed in Section~\ref{sec:safetyadv} and the Supervisor in this subsection. 
Different from Figure~\ref{fig:idea} (right), $\hat{u}'$ that is fed to the abstraction $\widehat{\mathfrak{D}}$ is decided by the Supervisor, instead of $\rho$.
Concretely, if $u_{uc}(k)$ is rejected, we have $\hat{u}'=\hat{u}$, with $\hat{u}$ being the input provided by $\rho$; otherwise, we need to select $\hat{u}'\in \hat{U}'$ for $\widehat{\mathfrak{D}}$, with $\hat{U}'$ as in~\eqref{eq:U'} (we would elaborate on how to select such $\hat{u}'$ later).
Step 1 ensures that $u_{uc}(k)$ would only be accepted if there exists $\hat{u}'$ such that the safety guarantee based on the ($\epsilon, \delta$)-approximate probabilistic relation still holds.
Checking Step 1 can be performed by virtue of the following proposition.
\begin{proposition}\label{PR_for_UC}
	Consider a gMDP $\mathfrak{D}=(X,U,x_0,T,Y,h)$ and its finite abstraction $\widehat{\mathfrak{D}}= (\hat X,\hat U,\hat{x}_0, \hat T, Y,\hat h)$ with $\widehat{\mathfrak{D}}\preceq^{\delta}_{\epsilon}\mathfrak{D}$.
	If $u_{uc}(k)$ is applied to $\mathfrak{D}$ at the time instant $k$, the ($\epsilon$,$\delta$)-approximate probabilistic relation still holds between $\mathfrak{D}$ and $\widehat{\mathfrak{D}}$ at the time instant $k+1$ if the set
	\begin{align}\label{eq:Uf}
	U_f\!=\!\Big\{\hat{u}\in \hat{U}'&\,\big|\, \mathbb{P}\big\{(x',\hat{x}')\in \mathscr{R}\big\}\geq 1-\delta, \text{ with } x'= f(x(k),u_{uc}(k),\varsigma(k)),\hat{x}'=\hat{f}(\hat{x}(k),\hat{u},\hat{\varsigma}(k))\Big\}
	\end{align}
	is not empty, with $\hat{U}'$ as in~\eqref{eq:U'}.
	Here, $f$ and $\hat{f}$ are transition maps of $\mathfrak{D}$ and $\widehat{\mathfrak{D}}$ as in \eqref{eq:gMDP_f}, respectively. 
	Moreover, $x(k)\in X$, $\hat{x}(k)\in \hat{X}$, $\varsigma(k)$ and $\hat{\varsigma}(k)$  are current states of $\mathfrak{D}$ and $\widehat{\mathfrak{D}}$, noise affecting $\mathfrak{D}$ and  $\widehat{\mathfrak{D}}$, respectively.
\end{proposition}
Intuitively, the non-emptiness of $U_f$ ensures that there exists at least one $\hat{u}\in \hat{U}'$ corresponding to $u_{uc}(k)$ such that $(x(k+1),\hat{x}(k+1))\in \mathscr{R}$ holds with the probability of at least $1-\delta$. 
Accordingly, if $U_f$ is not empty, and $u_{uc}(k)$ is accepted after going through Step 2, we need to use $\hat{u}\in U_f$ for $\widehat{\mathfrak{D}}$.
The selection of $\hat{u}$ is related to the estimation $\mathcal{E}_{pv}(k)$ in Step 2 which is discussed later.
Prior to discussing how to obtain $\mathcal{E}_{pv}(k)$, we show how to check whether $U_f$ is empty with the help of Proposition~\ref{PR_for_UC} through the running example.
\addtocounter{example}{-1}
\begin{example}[continued]
	Consider the current state $x(k)$ of the original model, $\hat{x}(k)$ of the finite abstraction, and input $u_{uc}(k)$ provided by the unverified controller.
	To show whether $U_f$ for the running example is empty, we need to show whether there exists $\hat{u}\in\hat{U}'$ such that 
	\begin{equation}\label{eq:check1}
	\lVert Ax(k)+Bu_{uc}(k)+R\varsigma(k)\!-\!P\big(\hat{A}_{\textsf r}\hat{x}(k)+\hat{B}_{\textsf r}\hat{u}+\hat{R}_{\textsf r}\hat{\varsigma}(k)\big) \rVert_M\leq \epsilon
	\end{equation}
	holds for all $\lVert\varsigma\rVert\leq\chi^{-1}_1(1-\delta)$ with $\delta = 0.01$ (cf. Section~\ref{sec:APR}, Example 1), $\hat{\varsigma}(k) = \varsigma(k)$, $\lVert\bar{x}\rVert_M=\sqrt{\bar{x}^TM\bar{x}}$, and $\chi^{-1}_1(\cdot)$ being the chi-square inverse cumulative distribution function with one degree of freedom.
	Note that~\eqref{eq:check1} holds when
	\begin{equation}\label{eq:check2}
	\lVert \varphi - P\hat{B}_{\textsf r}\hat{u}\rVert_M\leq \epsilon - \gamma,
	\end{equation}
	with $\varphi = Ax(k)+Bu_{uc}(k)-P\hat{A}_{\textsf r}\hat{x}_{\textsf r}(k)$ and 
	\begin{equation*}
	\gamma = \max_{\lVert\varsigma\rVert\leq\chi^{-1}_1(0.99),\beta\in\Delta}\lVert(R-P\hat{R}_{\textsf r})\varsigma + P\beta\rVert_M.
	\end{equation*}
	Here, $\Delta$ denotes the set of all possible quantization errors introduced by discretization of the original state set~\cite[Section 4.1, (4.14)]{Zhong2021Automata}.
	Note that $\hat{U}'$ contains a finite number of $\hat{u}$ (usually not too large in practice), $\varphi$ can be readily computed at runtime, and $\gamma$ can be computed offline after constructing the finite abstraction.
	Therefore, we can find out whether there exists $\hat{u}\in\hat{U}'$ such that~\eqref{eq:check2} holds at runtime efficiently.
\end{example}

Next, we proceed with discussing how to obtain $\mathcal{E}_{pv}(k)$ in Step 2.
First, the Safe-visor architecture as in Figure~\ref{fig1:svalifting} (left) can be equivalently described by a controlled gMDP $\tilde{\mathbf{C}}_{\rho'}\times \mathfrak{D}$ as in Figure~\ref{fig1:svalifting} (right) according to its running mechanism.
Here, $\tilde{\mathbf{C}}_{\rho'}$ (the yellow region) is a control strategy constructed as in Definition~\ref{def:C_rho} based on a Markov policy $\rho'$, which differs from  $\rho^*$ as in~\eqref{eq:policy_opt_sac_ab} (for problem of robust satisfaction) or $\rho_*$ as in~\eqref{eq:policy_opt_vio_ab} (for problem of worst-case violation) due to the acceptance of unverified controllers.
Consider the maximal tolerable probability of violating the desired safety specification (\emph{i.e.,} $\eta$) within a time horizon $[0,H]$.
For the problem of robust satisfaction, if the supervisor is designed such that we have
\begin{equation}\label{req_for_sp0}
1-\bar{V}_H^{\rho'}(\hat{x}_0,\bar{q}_0)\leq \eta,
\end{equation}
we are then able to ensure~\eqref{problem1} by combining~\eqref{req_for_sp0} and Theorem~\ref{thm:gua_prmax}.
Similarly, if we design the supervisor for the problem of worst-case violation such that
\begin{equation}\label{req_for_sp}
\ul{V}_H^{\rho'}(\hat{x}_0,\bar{q}_0)\leq \eta,
\end{equation}
we can guarantee~\eqref{problem2} by considering~\eqref{req_for_sp} and Theorem~\ref{thm:gua_prmin}.\footnote{In both~\eqref{req_for_sp0} and~\eqref{req_for_sp}, we only focus on the case in which $\bar{q}_0\notin F$, since $\bar{q}_0\in F$ indicates that the accepting states are reached at the initial state, which is not of interest.}
Thus, a direct idea is to set $\mathcal{E}_{pv}(k) = 1-\bar{V}_H^{\rho'}(\hat{x}_0,\bar{q}_0)$ for the problem of robust satisfaction and $\mathcal{E}_{pv}(k) = \ul{V}_H^{\rho'}(\hat{x}_0,\bar{q}_0)$ for the problem of worst-case violation.
Note that when the initial state $(\hat{x}_0,\bar{q}_0)$ and the time horizon $[0,H]$ are fixed, $\bar{V}^{\rho}_H(\hat{x}_0.\bar{q}_0)$ and $\ul{V}^{\rho}_H(\hat{x}_0.\bar{q}_0)$ can be computed given $\rho'$. 
The remaining problem is how to determine $\rho'$ at runtime.

In general, determining $\rho'$ at runtime is very challenging.
At each time instant $k\in [0, H-2]$ in each individual execution, $\rho_z'$ are unknown for all time instant $z \in (k, H-1]$ since the supervisor does not know which inputs the unverified controller will provide in the future.
Moreover, the supervisor does not have complete information about $\rho'_z$ for all $z\in [0,k]$, either.
In each individual execution, the system only reaches one state at each time step.
Therefore, the supervisor does not know what inputs the unverified controller would offer if the system reached $\hat{X}\times Q\backslash \{(\hat{x}(z),q(z))\}$ at time instants $z\in [0,k]$, let alone whether they would be accepted and what $\hat{u}'$ would accordingly be fed to $\widehat{\mathfrak{D}}$.
As a result, $\rho'_z(\tilde{x},\tilde{q})$ is unknown for all $(\tilde{x},\tilde{q}) \in \hat{X}\times Q\backslash \{(\hat{x}(z),q(z))\}$, with $\hat{x}(z)$ and $q(z)$ being the state of $\widehat{\mathfrak{D}}$ and $\mathcal{A}$ at time instant $z\in [0,k]$, respectively.

In this paper, instead of determining $\rho'$ and computing $\bar{V}^{\rho'}_H(\hat{x}_0,\bar{q}_0)$ or $\ul{V}^{\rho'}_H(\hat{x}_0,$ $\bar{q}_0)$ at runtime, we propose a \emph{history-based supervisor} that  can provide an estimation of $\mathcal{E}_{pv}(k)$ without requiring to know $\rho'$ at runtime.
This estimation is given with the help of history paths $\bar{\omega}_k$ of the Safe-visor architecture as in Figure~\ref{fig1:svalifting} (left), which is defined below.
\begin{definition}\label{def:path_new}
	\emph{(History Path of Safe-visor Architecture)}
	Consider a finite abstraction $\widehat{\mathfrak{D}}= (\hat X,\hat U,\hat{x}_0, \hat T, Y,\hat h)$, a DFA $\mathcal{A}=(Q, q_0, \Pi,\tau, F)$ characterizing the safety property of interest and the corresponding  Safe-visor architecture as in Figure~\ref{fig1:svalifting} (left).
	The history path of the Safe-visor architecture up to the time instant $k$ is denoted by 
	\begin{align*}
	\bar{\omega}_k\,=\,\Big(\hat{x}(0),q(0),\hat{u}(0),\,\hat{x}(1),q(1),\hat{u}(1),\ldots,\hat{x}(k-1),q(k-1),\hat{u}(k-1),\hat{x}(k),q(k)\Big),
	\end{align*}
	where $\hat{x}(k)\in\hat{X}$, $q(k)\in Q$, and $\hat{u}(k)\in \hat{U}$ are states of $\widehat{\mathfrak{D}}$, $\mathcal{A}$, and the input fed to $\widehat{\mathfrak{D}}$ at the time instant $k$, respectively. 
	Moreover, we denote by $\bar{\omega}_{\hat{x}k}$, $\bar{\omega}_{qk}$, and $\bar{\omega}_{uk}$ the subpath of $\hat{x}$, $q$, and $\hat{u}$ corresponding to $\bar{\omega}_k$, respectively.
\end{definition}

In general, \emph{history-based supervisor} provides $\mathcal{E}_{pv}(k)$ such that~\eqref{req_for_sp0} or~\eqref{req_for_sp} can be respected.
Consider a gMDP $\mathfrak{D}=(X,U,x_0,T,Y,h)$ and its finite abstraction $\widehat{\mathfrak{D}}= (\hat X,\hat U,\hat{x}_0, \hat T, Y,\hat h)$ with $\widehat{\mathfrak{D}}\preceq^{\delta}_{\epsilon}\mathfrak{D}$, a DFA $\mathcal{A}=(Q, q_0, \Pi,\tau, F)$, a labelling function $L: Y\rightarrow \Pi$ associated with $\mathcal{A}$ as in Definition~\ref{def:sactisfaction_DFA}, and $\eta$ to be the maximal tolerable probability of violating the desired safety specification. 
We discuss how such a supervisor is designed for both problems of interest in the rest of this section.

{\bf Supervisor for the Problem of Robust Satisfaction.}
Design of supervisor for the problem of robust satisfaction is formally proposed as follows.
\begin{definition}\label{def:History-based Supervisor_rs}
	\emph{(History-based Supervisor for the Problem of Robust Satisfaction)}
	At each time instant $k\in [0,H-1]$, given the history path $\bar{\omega}_k$ as in Definition~\ref{def:path_new}, the feasibility of an input $u_{uc}(k)$ from the unverified controller is checked by the following two steps:
	\begin{enumerate}[(i)]
		\item Check set $U_f$ as in~\eqref{eq:Uf} and reject $u_{uc}(k)$ if $U_f$ is empty;
		\item If $U_f$ is not empty, estimate $\mathcal{E}_{pv}(k)$ as
		\begin{align}
		\mathcal{E}_{pv}(k) &=\,\bar{\mathcal{E}}_{pv}(k) =\prod_{z=1}^{k}\Big((1-\delta)\!\!\!\!\!\!\!\!\!\!\sum_{\hat{x}\in \hat{X}'_{\epsilon}(q(z-1))}\!\!\!\!\!\!\!\!\!\!\hat{T}(\hat{x}\,\big|\,\hat{x}(z-1), \hat{u}(z-1))\Big)\nonumber\\
		&\times(1-\delta)\Big(1-\sum_{\hat{x}\in \hat{X}}\bar{V}^*_{H-k-1}\big(\hat{x},\underline{q}^*(\hat{x}(k),q(k))\big)\hat{T}(\hat{x}\,\big|\,\hat{x}(k),\hat{u}^*)\Big)\nonumber\\
		&+\delta+\sum_{j=1}^{k}\Big(\delta\prod_{z=1}^{j}\big((1-\delta)\!\!\!\!\!\!\!\!\!\sum_{\hat{x}\in \hat{X}'_{\epsilon}(q(z-1))}\!\!\!\!\!\!\!\!\!\hat{T}(\hat{x}\,\big|\,\hat{x}(z-1), \hat{u}(z-1))\big)\Big),\label{eq:supv_cosafe}
		\end{align}
		with $\bar{V}^*_{H-k-1}$ as in \eqref{eq:P_opt_sac} associated with the safety advisor, $\underline{q}^*$ as in \eqref{eq:underline_p_star},
		\begin{equation}\label{eq:u*}
		\hat{u}^* = \argmax_{\hat{u}\in U_f}\bar{\mathcal{E}}_{pv}(k),
		\end{equation}
		and
		\begin{align}\label{eq:xeps}
		\hat{X}'_{\epsilon}(q(z-1)) \!:=\!\Big\{\hat{x}\!\in\!\hat{X}\big|\exists x\!\in\! X, \tau(q(z\!-\!1),L\circ h(x))\!\notin\! F, h(x)\!\in\!{N}_{\epsilon}(\hat{h}(\hat{x}))\Big\},
		\end{align}
		where ${N}_{\epsilon}(\hat{h}(\hat{x}))$ is as in~\eqref{eq:N_eps}.
		If $\mathcal{E}_{pv}(k)\leq \eta$, the supervisor accepts $u_{uc}(k)$ and feeds $\hat{u}' = \hat{u}^*$ as in~\eqref{eq:u*} to $\widehat{\mathfrak{D}}$; otherwise, it rejects $u_{uc}(k)$ and feeds $\hat{u}' = \hat{u}$ to $\widehat{\mathfrak{D}}$, with $\hat{u}$ provided by $\rho^*_k$ as in~\eqref{eq:policy_opt_sac_ab} associated with the safety advisor.
	\end{enumerate}
\end{definition}
By applying the history-based supervisor as in Definition~\ref{def:History-based Supervisor_rs}, we are able to guarantee~\eqref{req_for_sp0} so that the problem of robust satisfaction can be solved.
We formally propose this result in the next theorem.
\begin{theorem}\label{theorem:guarantee for rs}
	Consider a gMDP $\mathfrak{D}=(X,U,x_0,T,Y,h)$ and a DFA $\mathcal{A}=(Q, q_0, \Pi,\tau, F)$ that characterizes the desired safety specification. 
	Employing the supervisor as in Definition~\ref{def:History-based Supervisor_rs} at all time instants $k\in[0,H-1]$ in the Safe-visor architecture, one has
	\begin{equation}\label{eq:gua_cosafe}
	\mathbb{P}_{\mathfrak{D}}\Big\{y_{\omega H}\models \mathcal{A}\Big\}\geq 1-\eta,
	\end{equation}
	for the problem of robust satisfaction as in Problem~\ref{prob:robust_satisfaction}, with $y_{\omega H}$ being output sequences of $\mathfrak{D}$ up to the time instant $H$.
\end{theorem}
The proof of Theorem~\ref{theorem:guarantee for rs} is provided in~\ref{proof}. 
Here, we provide some intuition for different parts in $\mathcal{E}_{pv}(k)$ as in~\eqref{eq:supv_cosafe} and discuss the complexity for computing them at runtime.
Generally, the estimation of $\mathcal{E}_{pv}(k)$ in~\eqref{eq:supv_cosafe} can be divided into three parts:
\begin{itemize}
	\item Part 1: $\prod_{z=1}^{k}\big((1-\delta)\sum_{\hat{x}\in \hat{X}'_{\epsilon}(q(z-1))}\hat{T}(\hat{x}\,\big|\,\hat{x}(z-1), \hat{u}(z-1))\big)$: \\
	Given the history path $\bar{\omega}_k$, Part 1 denotes the maximal probability of $\mathfrak{D}$ not being accepted by $\mathcal{A}$ within the time horizon $[0,k]$, while $x(z)\mathscr{R}\hat{x}(z)$ holds for all $z\in[0,k]$.
	As defined in~\eqref{eq:xeps}, for all $\hat{x}\in\hat{X}'_{\epsilon}(q(z-1))$ with $q(z-1)\in Q$, there exits at least one $x$ with $x\mathscr{R}\hat{x}$, such that $F$ is not reachable with $x(z)=x$.
	In another word, if $\hat{x}(z)\in\hat{X}'_{\epsilon}(q(z-1))$, $\mathfrak{D}$ may not be accepted by $\mathcal{A}$ even when we have $x(z)\mathscr{R}\hat{x}(z)$.
	Therefore, given $\hat{x}(z-1)$ and $\hat{u}(z-1)$, $(1-\delta)\sum_{\hat{x}\in \hat{X}'_{\epsilon}(q(z-1))}\hat{T}(\hat{x}\,\big|\,\hat{x}(z-1), \hat{u}(z-1))$ denotes the maximal probability of $\mathfrak{D}$ not being accepted by $\mathcal{A}$ at the time instant $z$ while $x(z)\mathscr{R}\hat{x}(z)$ still holds.
	As for the complexity of computing Part $1$ at runtime, we have 
	\begin{small}
		\begin{align*}
		&\prod_{z=1}^{k}\Big((1-\delta)\sum_{\hat{x}\in \hat{X}'_{\epsilon}(q(z-1))}\hat{T}(\hat{x}\,\big|\,\hat{x}(z-1), \hat{u}(z-1))\Big)\\
		&\begin{matrix}
		=\underbrace{\prod_{z=1}^{k-1}\Big((1-\delta)\!\!\!\!\!\!\!\!\!\!\!\!\sum_{\hat{x}\in \hat{X}'_{\epsilon}(q(z-1))}\!\!\!\!\!\!\!\!\!\!\!\!\hat{T}(\hat{x}\big|\hat{x}(z-1), \hat{u}(z-1))\Big)}\\ \text{Term 1}
		\end{matrix}
		\begin{matrix}
		\times\underbrace{(1-\delta)\!\!\!\!\!\!\!\!\!\!\!\!\sum_{\hat{x}\in \hat{X}'_{\epsilon}(q(k-1))}\!\!\!\!\!\!\!\!\!\!\!\!\hat{T}(\hat{x}\big|\hat{x}(k-1), \hat{u}(k-1))}.\\\text{Term 2}
		\end{matrix}
		\end{align*}
	\end{small}
	On one hand, $\forall q\in Q$, set $\hat{X}'_{\epsilon}(q)$ can be computed offline, and $\hat{T}$ is readily computed when synthesizing the safety advisor. 
	On the other hand, at the time instant $k-1$, Term 1 has already been computed at time instant $k-2$.
	Therefore, the number of operations required for computing Part 1 at time instant $k-1$ is proportional to the number of states in the set $\hat{X}'_{\epsilon}(q(k-1))$.
	\item Part 2: $(1-\delta)\big(1-\sum_{\hat{x}\in \hat{X}}\bar{V}^*_{H-k-1}\big(\hat{x},\underline{q}^*(\hat{x}(k),q(k))\big)\hat{T}(\hat{x}\,\big|\,\hat{x}(k),\hat{u}^*)\big)$: \\
	Part 2 quantifies the probability of $\mathfrak{D}$ not being accepted by $\mathcal{A}$ within time horizon $[k+1,H]$, while (i) $x(k+1)\mathscr{R}\hat{x}(k+1)$ holds, given $\hat{x}(k)$, $q(k)$, and $\hat{u}(k)=\hat{u}^*$; (ii) $\mathfrak{D}$ is controlled by the safety advisor within $[k+1,H]$.
	Since $\bar{V}^*_{H-k-1}$ and $\hat{T}$ are readily computed when synthesizing the safety advisor, the number of operations required for computing Part 2 is proportional to the number of elements in sets $\hat{X}$ and $U_f$.
	\item Part 3: $\delta+\sum_{j=1}^{k}\big(\delta\prod_{z=1}^{j}\big((1-\delta)\sum_{\hat{x}\in \hat{X}'_{\epsilon}(q(z-1))}\hat{T}(\hat{x}\,\big|\,\hat{x}(z-1), \hat{u}(z-1))\big)\big)$:\\
	Given the history path $\bar{\omega}_k$, Part 3 quantifies the maximal probability of $\mathfrak{D}$ not being accepted by $\mathcal{A}$ within the time horizon $[0,k]$ while $\exists z\in [1,k+1]$ such that $x(z)\mathscr{R}\hat{x}(z)$ does not hold.
	Concretely, the first $\delta$ in Part 3 quantifies $\mathbb{P}\{(x(1),\hat{x}(1))\notin \mathscr{R}~|~(x(0),\hat{x}(0))\in \mathscr{R}\}$.	
	Moreover, the term $\delta\prod_{z=1}^{j}\big((1-\delta)\sum_{\hat{x}\in \hat{X}'_{\epsilon}(q(z-1))}\hat{T}(\hat{x}\,\big|\,\hat{x}(z-1), \hat{u}(z-1))\big)$ represents the maximal probability of $\mathfrak{D}$ not being accepted by $\mathcal{A}$ within the time horizon $[0,j+1]$, while (i) $x(z)\mathscr{R}\hat{x}(z)$ holds for all $z\in[0,j]$; (ii) $x(j+1)\mathscr{R}\hat{x}(j+1)$ does not hold.
	One may notice that $(1-\delta)\sum_{\hat{x}\in \hat{X}'_{\epsilon}(q(z-1))}\hat{T}(\hat{x}\,\big|\,\hat{x}(z-1), \hat{u}(z-1))$ is the same as Term 2 in Part 1.
	Therefore, the number of operations needed for computing Part 3 at the time instant $k-1$ is also proportional to the number of states in the set $\hat{X}'_{\epsilon}(q(k-1))$.
\end{itemize}
In conclusion, the number of operations required for computing $\mathcal{E}_{pv}(k)$ as in~\eqref{eq:supv_cosafe} is proportional to the number of elements in $\hat{X}$ and $U_f$.
Consequently, $\mathcal{E}_{pv}(k)$ can be computed efficiently at runtime.
We show the real-time applicability of the supervisor in the experiments in Section~\ref{Case_study}.

{\bf Supervisor for the Problem of Worst-case Violation.}
Next, we proceed with proposing the supervisor for the problem of worst-case violation.
\begin{definition}\label{def:History-based Supervisor_wcv}
	\emph{(History-based Supervisor for the Problem of Worst-case Violation)}
	At each time instant $k\in [0,H-1]$, given the history path $\bar{\omega}_k$ as in Definition~\ref{def:path_new}, the validity of an input $u_{uc}(k)$ from the unverified controller is checked by the following two steps:
	\begin{enumerate}[(i)]
		\item Check the set $U_f$ as in~\eqref{eq:Uf} and reject $u_{uc}(k)$ if $U_f$ is empty;
		\item If $U_f$ is not empty, estimate $\mathcal{E}_{pv}(k)$ as
		\begin{align}
		\mathcal{E}_{pv}&(k) =\,\ul{\mathcal{E}}_{pv}(k) =1 - \prod_{z=1}^{k}\Big((1-\delta)\!\!\!\!\!\!\!\!\!\!\sum_{\hat{x}\in \hat{X}'_{-\epsilon}(q(z-1))}\!\!\!\!\!\!\!\!\!\!\hat{T}(\hat{x}\,\big|\,\hat{x}(z-1), \hat{u}(z-1))\Big)\nonumber\\
		&\times(1-\delta)\Big(1-\!\!\sum_{\hat{x}\in \hat{X}}\ul{V}_{*,H-k-1}(\hat{x},\bar{q}_*(\hat{x}(k),q(k)))\hat{T}(\hat{x}\,\big|\,\hat{x}(k),\hat{u}^*)\Big),\label{eq:supv_safe}
		\end{align}
		where $\ul{V}_{*,H-k-1}$ is as in \eqref{eq:P_opt_vio} associated with the safety advisor, $\bar{q}_*$ is as in \eqref{eq:overline_p_star},
		\begin{equation}\label{eq:u*2}
		\hat{u}^* = \argmax_{\hat{u}\in U_f}\ul{\mathcal{E}}_{pv}(k),
		\end{equation}
		and 
		\begin{align}\label{eq:xeps-}
		\hat{X}'_{-\epsilon}(q(z-1))\!\!:=\!\Big\{\hat{x}\!\in\!\hat{X}\big|\forall x\in X,\tau(q(z-1),L\!\circ\! h(x))\!\notin\!F, h(x)\!\in\!{N}_{\epsilon}(\hat{h}(\hat{x}))\!\Big\},
		\end{align}
		with ${N}_{\epsilon}(\hat{h}(\hat{x}))$ as in~\eqref{eq:N_eps}.
		If $\mathcal{E}_{pv}(k)\!\leq\! \eta$, the supervisor accepts $u_{uc}(k)$ and feeds $\hat{u}' = \hat{u}^*$ as in~\eqref{eq:u*2} to $\widehat{\mathfrak{D}}$; otherwise, it rejects $u_{uc}(k)$ and feeds $\hat{u}' = \hat{u}$ to $\widehat{\mathfrak{D}}$, with $\hat{u}$ provided by $\rho_{*_k}$ as in~\eqref{eq:policy_opt_vio_ab} associated with the safety advisor.
	\end{enumerate}
\end{definition}
By employing the history-based supervisor as in Definition~\ref{def:History-based Supervisor_wcv}, we are able to guarantee~\eqref{req_for_sp}, which solves the problem of worst-case violation.
This is formalized in the next theorem.
\begin{theorem}\label{theorem:guarantee for wcv}
	Consider a gMDP $\mathfrak{D}=(X,U,x_0,T,Y,h)$ and a DFA $\mathcal{A}=(Q, q_0, \Pi,\tau, F)$ characterizing the desired safety specification. 
	Utilizing the supervisor as in Definition~\ref{def:History-based Supervisor_wcv} at all time instants $k\in[0,H-1]$ in the Safe-visor architecture, one has
	\begin{equation}\label{eq:gua_safe}
	\mathbb{P}_{\mathfrak{D}}\Big\{y_{\omega H}\models \mathcal{A}\Big\}\leq \eta,
	\end{equation}
	for the problem of worst-case violation as in Problem~\ref{prob:worst_case_violation}, where $y_{\omega H}$ are output sequences of $\mathfrak{D}$ up to the time instant $H$.
\end{theorem}
The proof of Theorem~\ref{theorem:guarantee for wcv} is provided in~\ref{proof}. 
Note that $\mathcal{E}_{pv}(k)$ in~\eqref{eq:supv_safe} denotes the maximal probability of $\mathfrak{D}$ not being accepted by $\mathcal{A}$ within the time horizon $[0,H]$, given the history path $\bar{\omega}_k$. 
The term after the first minus sign in~\eqref{eq:supv_safe} can be divided into two components as follows:
\begin{itemize}
	\item Component 1: $\prod_{z=1}^{k}\big((1-\delta)\sum_{\hat{x}\in \hat{X}'_{-\epsilon}(q(z-1))}\hat{T}(\hat{x}\,\big|\,\hat{x}(z-1), \hat{u}(z-1))\big)$:\\
	Given the history path $\bar{\omega}_k$, Component 1 denotes the minimal probability of $\mathfrak{D}$ not being accepted by $\mathcal{A}$ within the time horizon $[0,k]$, while $x(z)\mathscr{R}\hat{x}(z)$ holds for all time instant $z\in[0,k]$.
	According to~\eqref{eq:xeps-}, for all $\hat{x}\in\hat{X}'_{-\epsilon}(q(z-1))$ with $q(z-1)\in Q$, there is no $x$ with $x\mathscr{R}\hat{x}$ such that $F$ is reached at time step $z$.
	In other words, if $\hat{x}(z)\in\hat{X}'_{-\epsilon}(q(z-1))$, we can ensure that $\mathfrak{D}$ will not be accepted by $\mathcal{A}$ at the time instant $z$ if one can ensure $x(z)\mathscr{R}\hat{x}(z)$.
	Therefore, given $\hat{x}(z-1)$ and $\hat{u}(z-1)$, $(1-\delta)\sum_{\hat{x}\in \hat{X}'_{-\epsilon}(q(z-1))}\hat{T}(\hat{x}\,\big|\,\hat{x}(z-1), \hat{u}(z-1))$ denotes the minimal probability of $\mathfrak{D}$ not being accepted by $\mathcal{A}$ at $z$ while $x(z)\mathscr{R}\hat{x}(z)$ still holds.
	\item Component 2: $\!(1-\delta)\big(\!1-\sum_{\hat{x}\in \hat{X}}\ul{V}_{*,H-k-1}(\hat{x},\bar{q}_*(\hat{x}(k),q(k)))\hat{T}(\hat{x}\,\big|\,\hat{x}(k),\hat{u}^*)\big)$:\\
	Component 2 denotes the probability of $\mathfrak{D}$ not being accepted by $\mathcal{A}$ within the time horizon $[k+1,H]$, while (i) $x(k+1)\mathscr{R}\hat{x}(k+1)$ holds, given $\hat{x}(k)$, $q(k)$, and $\hat{u}(k)=\hat{u}^*$; (ii) $\mathfrak{D}$ is controlled by the safety advisor within $[k+1,H]$.
\end{itemize}
\begin{algorithm}
	\caption{Running mechanism of Safe-visor architecture.}
	\label{tbl:safe-visor_mech}
	\KwIn{gMDP $\mathfrak{D}=(X,U,x_0,T,Y,h)$, DFA $\mathcal{A}=(Q, q_0, \Pi,\tau, F)$, safety advisor $\tilde{\mathbf{C}}_{\rho^*}$ for the problem of robust satisfaction (resp. $\tilde{\mathbf{C}}_{\rho_*}$ for the problem of worst-case violation) as in Section~\ref{sec:safetyadv}, and the supervisor as in Definition~\ref{def:History-based Supervisor_rs} (resp. as in Definition~\ref{def:History-based Supervisor_wcv}).}
	$k=0$, $x(0) = x_0$\\
	\While{$k<H$}
	{
		\uIf{$k=0$}{\label{sucess_start}
			Initialize $\hat{x}(0)=\hat{x}_0$ such that $ x_0 \mathscr{R}\hat{x}_0 $\\
			Update $q(0)$ as $\bar{q}_0=\tau (q_0, L\circ h(x_0))$
		}
		\Else{
			Update the state $x(k)$ of $\mathfrak{D}$ from the measurement\\
			Update $q(k)$ with $q(k)=\tau\big(q(k-1),L\circ h(x(k))\big)$\\
			Update $\hat{x}(k)$ with $\mathscr{L}_{T}(d\hat{x}|x(k), \hat{x}(k-1),\hat{u}(k-1))$
		}
		Update $u_{uc}(k)$ from the unverified controller\\
		Update $u_{\text{safe}}(k)$ from $\tilde{\mathbf{C}}_{\rho^*}$ (resp. $\tilde{\mathbf{C}}_{\rho_*}$) according to $\hat{x}(k)$, $x(k)$, and $q(k)$\\
		Feed $\hat{x}(k)$, $x(k)$, $q(k)$, $u_{uc}(k)$, and $u_{\text{safe}}(k)$ to the supervisor\\
		Update $u(k)$ and $\hat{u}(k)$ according to the decision of the supervisor\\
		$k=k+1$
	}
	\KwOut{$u(k)$ for controlling the system $\mathfrak{D}$ at each time instant $k$.}
\end{algorithm}
Similar to the computation of~\eqref{eq:supv_cosafe}, $\mathcal{E}_{pv}(k)$ as in~\eqref{eq:supv_safe} can be efficiently computed at runtime.
In brief, the required number of operations for computing $\mathcal{E}_{pv}(k)$ in~\eqref{eq:supv_safe} is also proportional to the number of elements in $\hat{X}$ and $U_f$.
The real-time applicability of the supervisor is shown in Section~\ref{Case_study} through an example. 
Finally, we summarize in Algorithm~\ref{tbl:safe-visor_mech} the running mechanism of the Safe-visor architecture equipped with the safety advisor proposed in Section~\ref{sec:safetyadv} and supervisor proposed in Section~\ref{sec:supervisor}.

\section{Case Studies}\label{Case_study}
In this section, we apply our proposed approaches to two case studies, including the running example and a control problem regarding a DC motor. 
We simulate each case study with $1.0 \times 10^5$ empirical Monte Carlo runs and analyze accordingly the percentage of output sequences satisfying the desired safety specifications in the given time horizon to illustrate our theoretical bound.
Additionally, we also compute the \emph{acceptance rate} of the unverified controller in each case, \ie, the average percentage of inputs from the unverified controller being accepted among these runs.
For comparison, we simulate them (i) using only the unverified controller and (ii) using only the safety advisor in the architecture. 
We show that the desired safety probability can be guaranteed by sandboxing the unverified controller with our proposed architecture. 
Furthermore, we compute the average execution time for our proposed architecture in both cases to show the feasibility of running this architecture in real-time. 
We also investigate the effect of the abstraction resolution on the safety guarantees provided by the safety advisor, as well as the execution speed of the supervisor.
The experiments are performed via MATLAB 2019b, on a machine with Windows 10 operating system (Intel(R) Xeon(R) E-2186G CPU (3.8 GHz) and 32 GB of RAM).

\subsection{Running Example}\label{sec:running}
\begin{figure}[htbp]
	\centering
	\includegraphics[width=8cm]{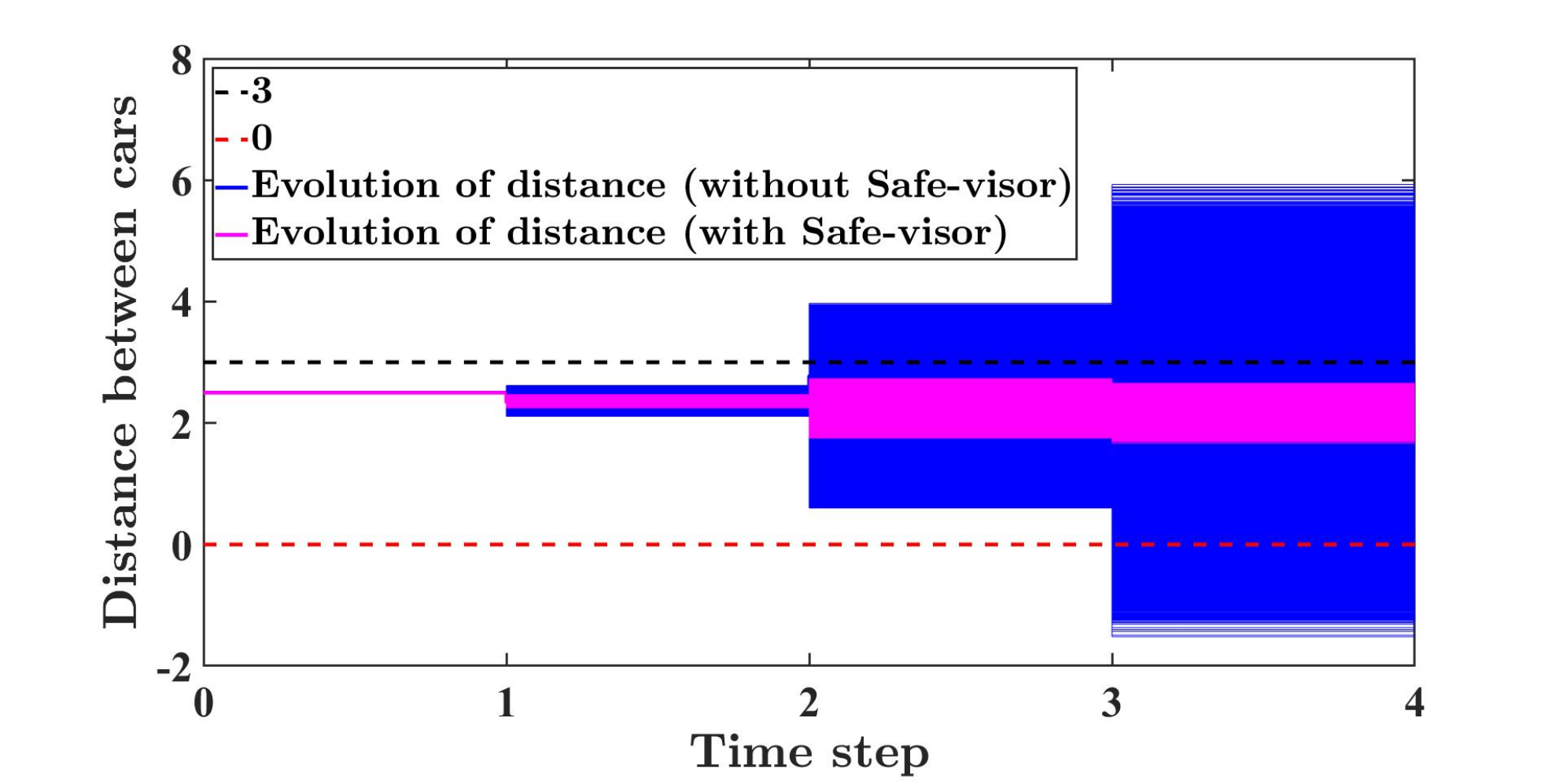}
	\caption{Simulation results for the running example.}\label{fig:temp_simulation}
\end{figure}
For simulating Safe-visor architecture regarding the safety specification $\psi$ as described in Section~\ref{sec:2}, we initialize the system at $x_0\!\!=\![2.5\,; 2.4\,; 1.5]$ and select $\eta =0.1$. 
To ensure the last condition in Definition~\ref{Def: apr}, we initialize the finite abstraction with $\hat{x}_0 = 2.55$.
Moreover, to model the unverified controller that would endanger the system, we use a controller that randomly selects input at each time instant following a uniform distribution within the input range. 
Simulation results are shown in Figure~\ref{fig:temp_simulation}, and summarized in Table~\ref{tab1}. 
One can readily verify that the desired safety probability specified by $\eta$ is respected.

\subsection{DC Motor}\label{sec:dcmotor}
In the second case study, we focus on a DC motor as in Figure~\ref{fig1:case1} (a), which can be described by the following difference equations:
\begin{figure}
	\centering
	\subfigure[DC Motor.]{
		\includegraphics[width=5.3cm]{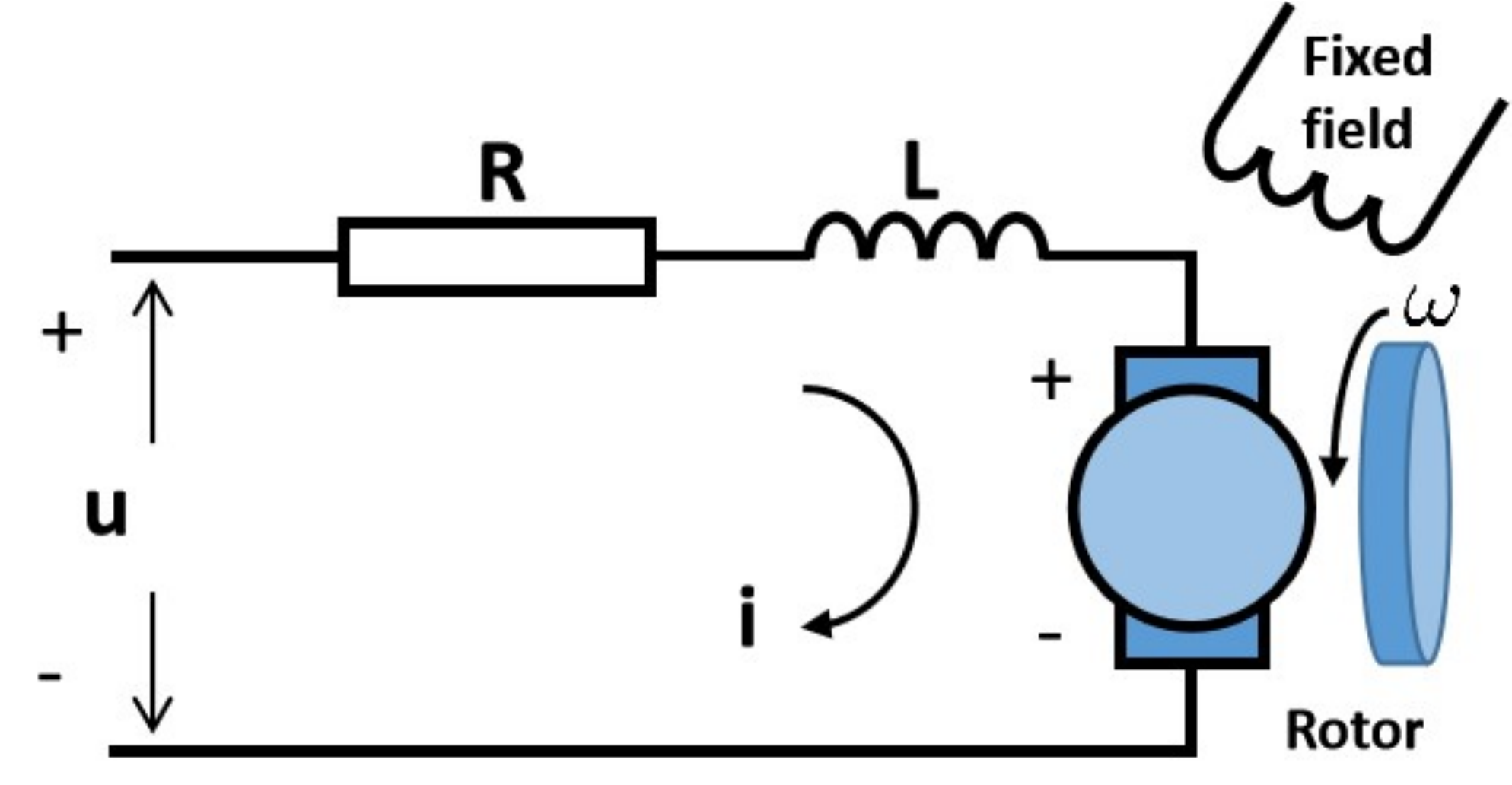}
	}
	\subfigure[DFA modelling $\psi_{dc}$.]{
		\includegraphics[width=5.3cm]{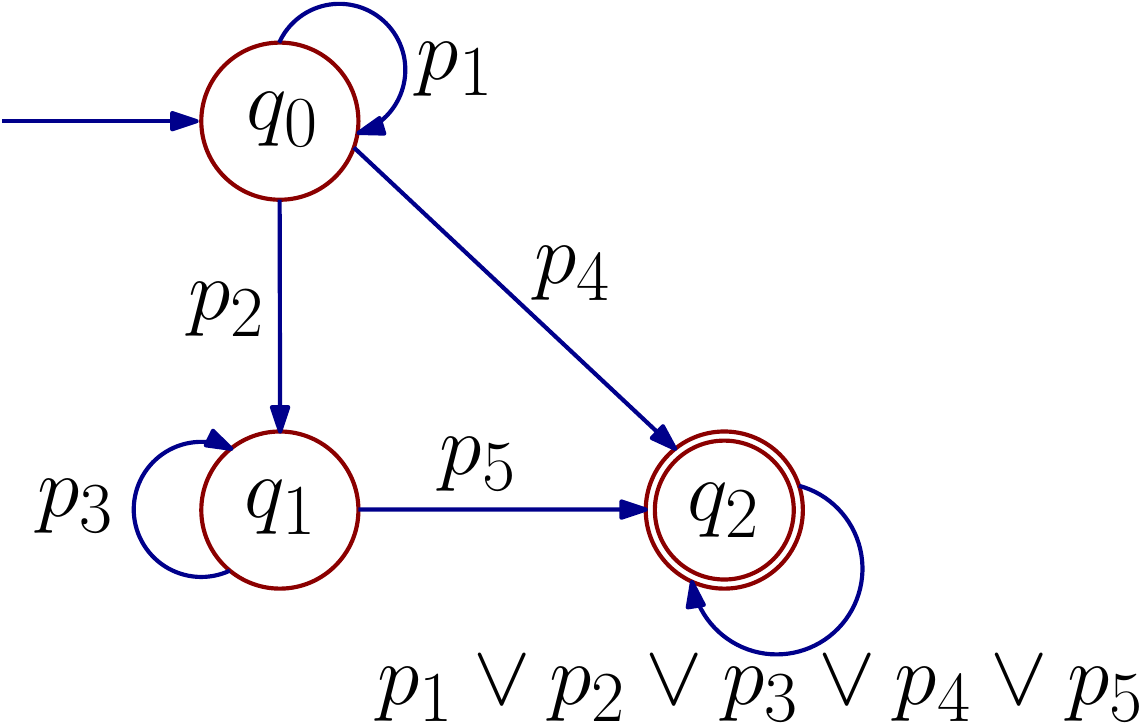}
	}
	\caption{DC Motor and DFA describing $\psi_{dc}$ for the DC motor, with $q_2$ being the accepting state, alphabet $\Pi=\{p_1,p_2,p_3,p_4,p_5\}$, and labelling function $L:Y \rightarrow \Pi$ with $L(y)=p_1$, when $y \in \big([1.5\pi, 1.875\pi)\cup(2.125\pi, 2.5\pi]\big)\times [0,2.4]$; 
		$L(y)=p_2$, when $y \in [1.875\pi, 2.125\pi]\times [0,2.4]$; 
		$L(y)=p_3$, when $y \in [1.75\pi, 2.25\pi]\times [0,2.4]$; 
		$L(y)=p_4$, when $y \in \R^2\backslash([1.5\pi,2.5\pi]\times[0,2.4])$; 
		$L(y)=p_5$, when $y \in \R^2\backslash([1.75\pi,2.25\pi]\times[0,2.4])$.}\label{fig1:case1}
\end{figure}
\begin{align}\label{eq:DCmotor}
\mathfrak{D}\!:\left\{\hspace{-1.5mm}\begin{array}{l}x(k+1) = Ax(k)+Bu(k) + Ee^{Fx(k)}+ R\varsigma(k),\\
y(k) = x(k),\end{array}\right.\quad k\in\mathbb N,
\end{align}
with 
\begin{align*}
A =& \begin{bmatrix}\begin{smallmatrix}0.6387\ &\ 0.0080\\-0.1606\ &\ -0.0020\end{smallmatrix}\end{bmatrix}\!,
~B = [0.3996\,;0.4011],
~E = [-0.2\,;0],\\\notag
~F =& [-0.0796\,; 0]^T, 
~R = \begin{bmatrix}\begin{smallmatrix}0.01\ &\ 0\\0\ &\ 0.01\end{smallmatrix}\end{bmatrix}\!.
\end{align*} 
Here, $x(k) = [x_1(k)\,;x_2(k)]$ is the state of the DC motor, in which $x_1(k)$ and $x_2(k)$ are the angular velocity and the armature current of the motor, respectively.
Input $u(k)\in [0,9]$ is the voltage source applied to the motor's armature. 
Additionally, we have $\varsigma(k) = [\varsigma_{1}(k)\,; \varsigma_{2}(k)]$, where $\varsigma_{1}(k)$ and $\varsigma_{2}(k)$ are standard Gaussian random variables that affect $x_1(k)$ and $x_2(k)$, respectively. 
This model is adapted from a continuous-time model of a DC-motor as in~\cite{Zaccarian2012DC} by discretizing it with a sampling time $\tau=0.02$s and including stochasticity in the model as an additive noise. 
In this case study, the DC motor is required to satisfy a safety specification $\psi_{dc}$ within 7 minutes: 
(i) the armature current should be within $[0,2.4]$; 
(ii) the angular velocity should stay within $[1.5\pi, 2.5\pi]$; 
(iii) additionally, if angular velocity reaches $[1.875\pi, 2.215\pi]$, it should stay within $[1.75\pi, 2.25\pi]$ afterwards, instead of $[1.5\pi, 2.5\pi]$. 
Accordingly, we build a DFA which accepts all bad prefixes of output sequences that violate $\psi_{dc}$, as shown in Figure~\ref{fig1:case1} (b), and we accordingly focus on the problem of worst-case violation.

Now, we design the safety advisor for the DC motor following the proposed approach in Section~\ref{sec:safetyadv}.
Here, we construct a finite abstraction for the model of DC motor using the results in~\cite[Section 4.1]{Zhong2021Automata}. 
First, we uniformly partition the input set $[0,9]$ into $40$ partitions and select $X = [1.5\pi, 2.5\pi]\times [0,2.4]$. 
Then, we uniformly partition this region into $40$ cells on both dimensions, which results in a finite gMDP with $1601$ discrete states ($1600$ states correspond to the representative points of partitions, and 1 sink state) and $40$ inputs. 

As for establishing an ($\epsilon$,$\delta$)-approximate probabilistic relation, we apply the method in~\cite[Section 4.3]{Zhong2021Automata}.
We set $\hat{U}'=\hat{U}$, $\delta = 0$, and the expected range of $\epsilon$ as $[0.05,1]$.
Then, this finite abstraction is $(\epsilon,\delta)$-stochastically simulated by the original model as in \eqref{eq:DCmotor} w.r.t. the relation
$\mathscr{R} = \{(x,\hat{x})\,|\,(x-\hat{x})^T(x-\hat{x})\leq \epsilon^2\}$ with $\delta = 0$ and $\epsilon = 0.1138$, when the interface function
\begin{align*}
\nu(x,\hat{x},\hat{u}):= (K+bL)(x-\hat{x}) + \hat{u}+ Ge^{F\hat{x}},
\end{align*}
is employed for the controller refinement.
$\mathscr{R} = \{(x,\hat{x})\,|\,(x-\hat{x})^TM(x-\hat{x})\leq \epsilon^2\}$
Here, $\hat{x}$ is the state of the finite abstraction, $K = [-0.5948\,;-0.0110]^T$, $L = [-0.0452\,; 0.0039]^T$, $G = 3.0954$, and $b = \frac{e^{Fx}-e^{F\hat{x}}}{F(x-\hat{x})}$.
The lifting stochastic kernel for this relation is constructed such that the noise terms in the original and the finite systems are the same.
Now, we are ready to synthesize the safety advisor as in Section~\ref{sec:safetyadv}. 
\begin{figure}[htbp]
	\centering
	\subfigure[Evolution of the angular velocity.]{
		\includegraphics[width=8cm]{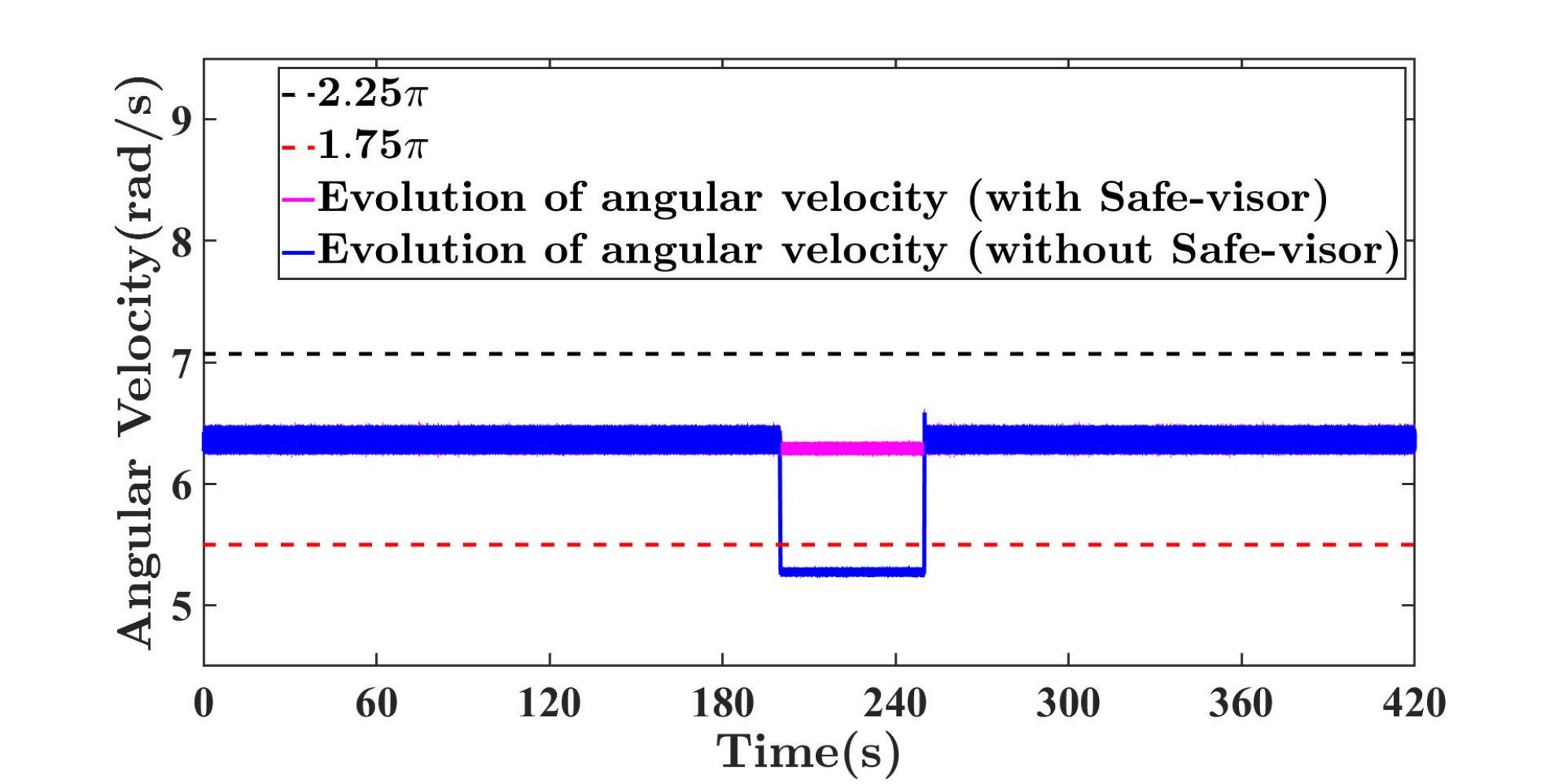}\hspace{-0.7cm}
		%\caption{fig1}
	}
	\quad
	\subfigure[Evolution of the armature current.]{
		\includegraphics[width=8cm]{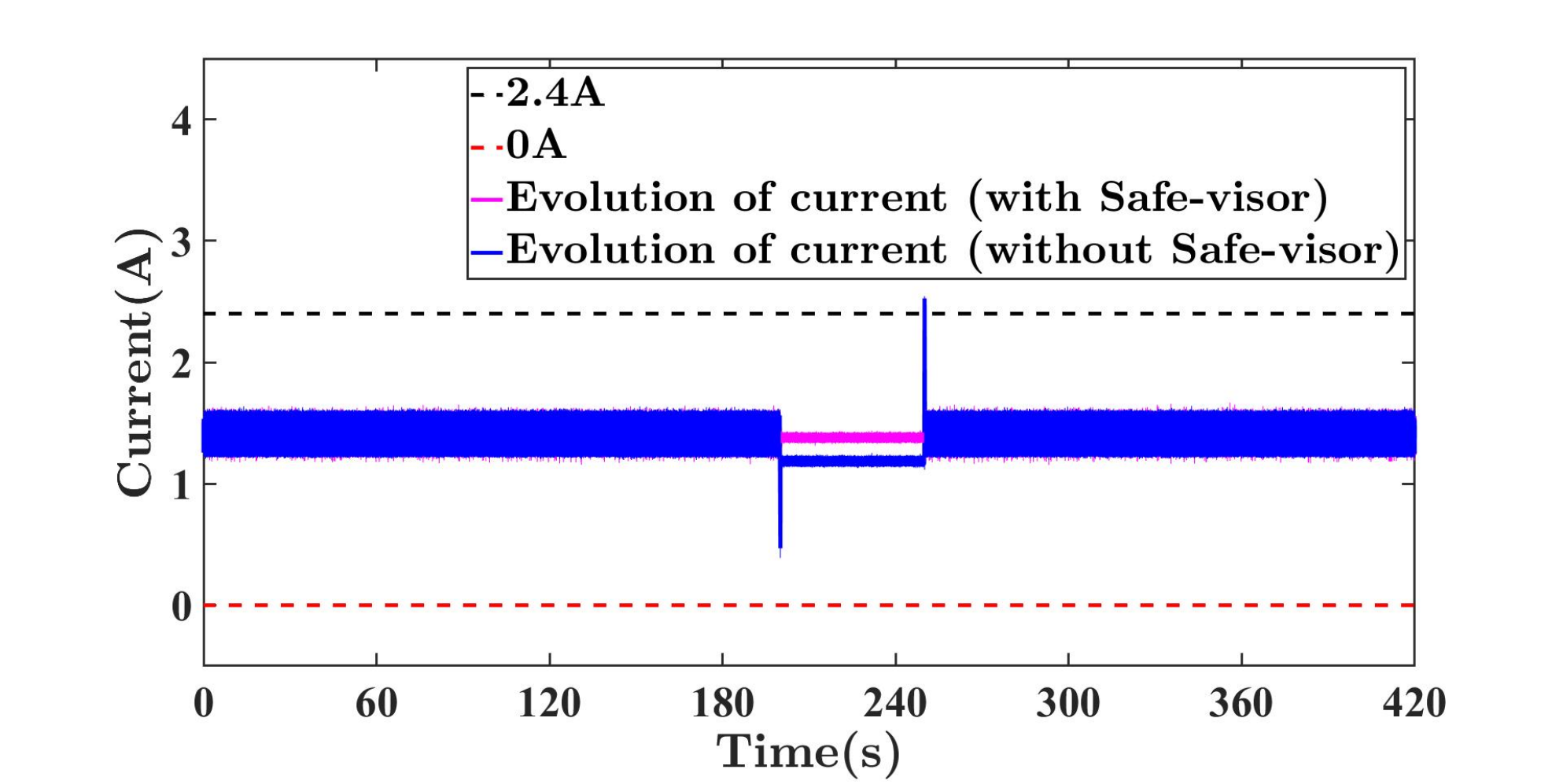}
	}
	\caption{Simulation results for DC motor for safety specification $\psi_{dc}$.}\label{fig:dc_simulation}
\end{figure}

As for the simulation, we initialize the system at $x_0=[2\pi \,; 1.256]$ and set $\eta =0.001$.
To ensure the last condition in Definition~\ref{Def: apr}, we initialize the finite abstraction with $\hat{x}_0 = [6.3225\,; 1.23]$.
As for the unverified controller, we train a controller using deep reinforcement learning with deep deterministic policy gradient (DDPG) algorithms~\cite{Lillicrap2016Continuous} for tracking the desired angular velocity, denoted by $x_{1_d}$. 
We are not providing details for the training procedure since designing and improving the performance of an (AI-based) unverified controller are out of the scope of this paper.
The AI-controller is applied here for demonstration purposes.
In the simulation, we set $x_{1_d}=2.1\pi\,rad/s$ when $k\in[0,10000]\cup[12500,21000]$ and $x_{1_d}=1.78\pi\,rad/s$ when $k\in[10000,12500]$. 
The results are shown in Figure~\ref{fig:dc_simulation} and summarized in Table~\ref{tab1}.
Without sandboxing the unverified controller, all output sequences violate $\psi_{dc}$.
Meanwhile, by sandboxing the unverified controller, $100\%$ of output sequences satisfy $\psi_{dc}$ while $84.73\%$ of inputs from the unverified controller can be accepted.
One can readily see that the safety probability specified by $\eta$ is respected. 
Simultaneously, the unverified controller can still be applied most of the time when it does not endanger the system's safety.

\begin{table}
	\caption{Simulation results for both case studies.}\label{tab1}\vspace{0.2cm}
	\begin{tabular}{|p{8.9cm}|p{1cm}|p{1cm}|}
		\hline
		&  $\psi$ & $\psi_{dc}$ \\
		\hline
		Percentage of satisfaction (with Safe-visor architecture)& 100\% & 100\% \\
		\hline
		Average acceptance rate of the unverified controller & 27.27\% & 84.73\%\\
		\hline
		Percentage of satisfaction (without Safe-visor architecture)&  55.87\% & 0\%\\
		\hline
		Percentage of satisfaction \par(when system is fully controlled by the safety advisor) & 100\% & 100\%\\
		\hline
		Average execution time of the supervisor (ms) & 0.1005 & 0.3782\\
		\hline
	\end{tabular}
\end{table}
\begin{figure}[htbp]
	\centering
	\subfigure[Number of Partitions: $10\times 10$]{
		\includegraphics[width=0.49\textwidth]{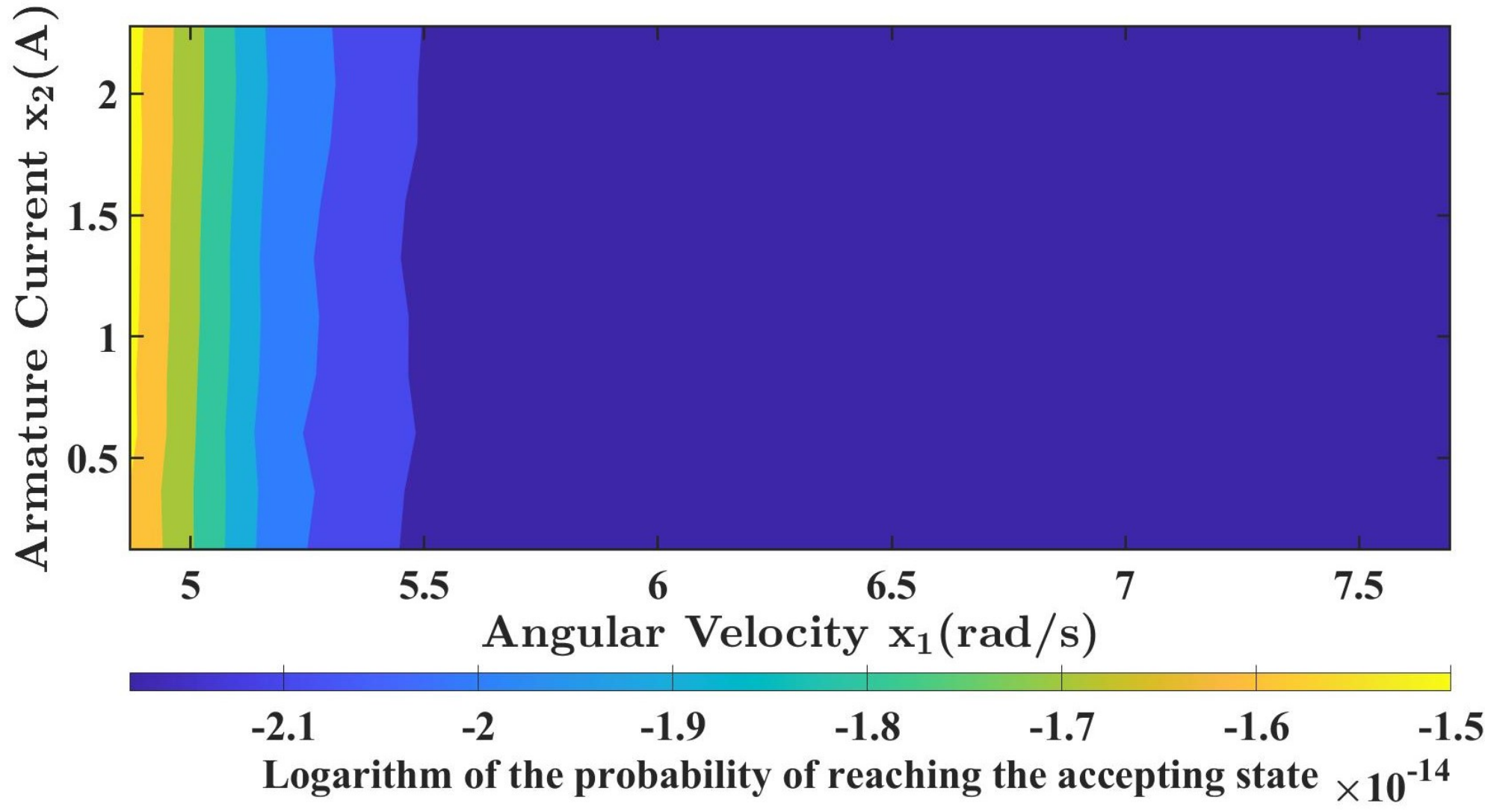}\hspace{-0.7cm}
	}
	\quad
	\subfigure[Number of Partitions: $20\times 20$]{
		\includegraphics[width=0.49\textwidth]{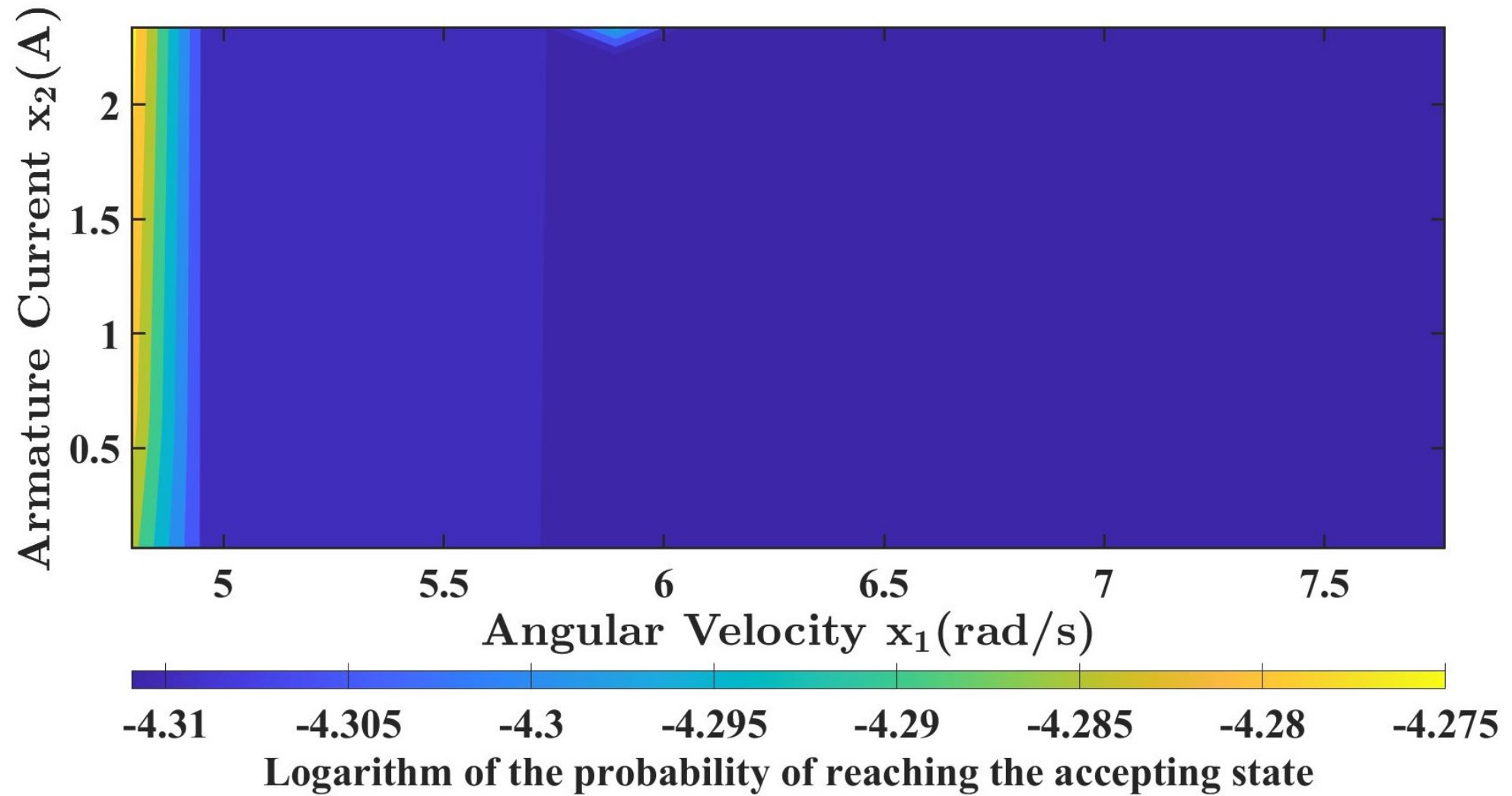}
	}
	\quad
	\subfigure[Number of Partitions: $30\times 30$]{
		\includegraphics[width=0.48\textwidth]{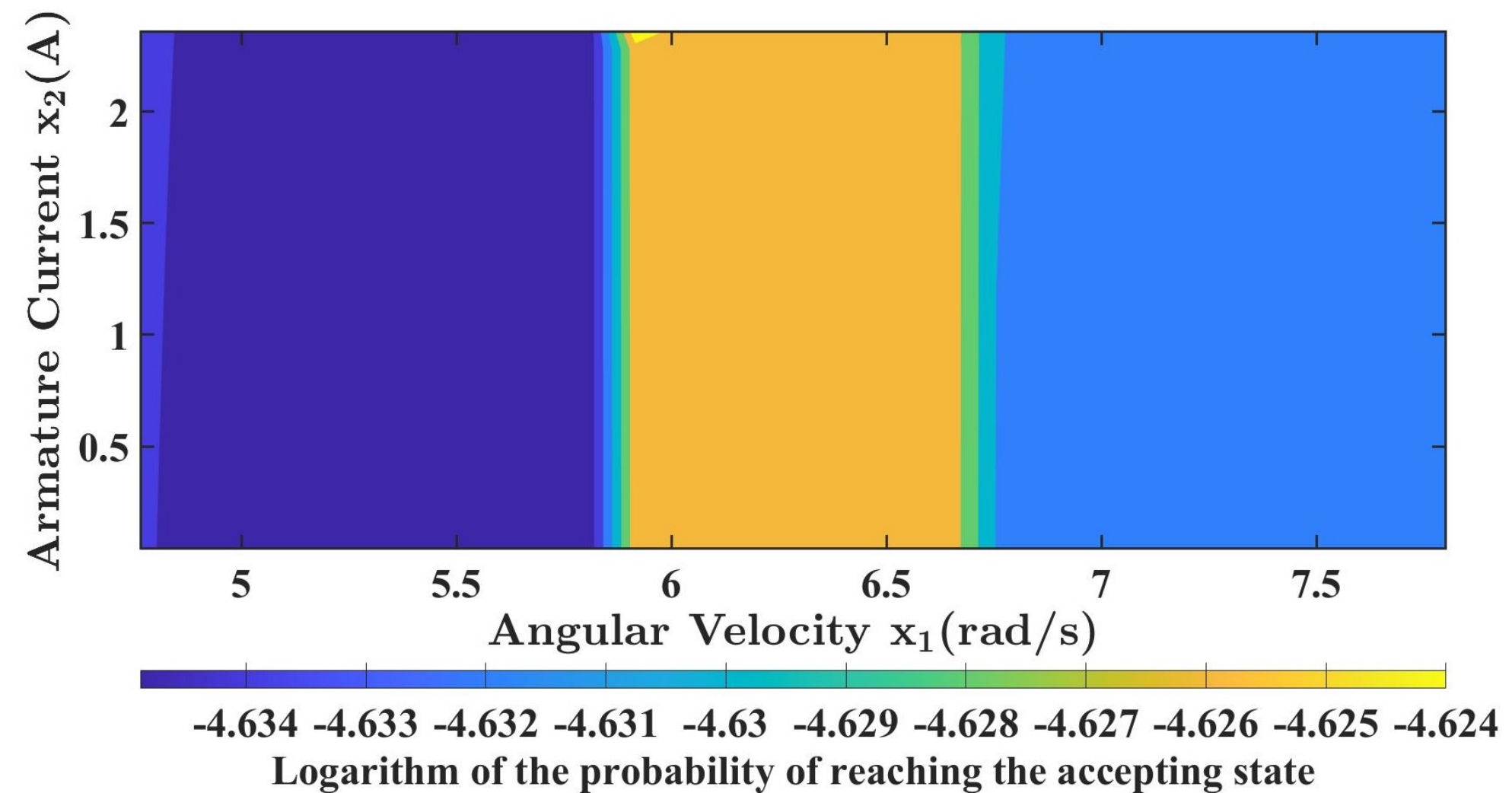}\hspace{-0.7cm}
	}
	\quad
	\subfigure[Number of Partitions: $40\times 40$]{
		\includegraphics[width=0.48\textwidth]{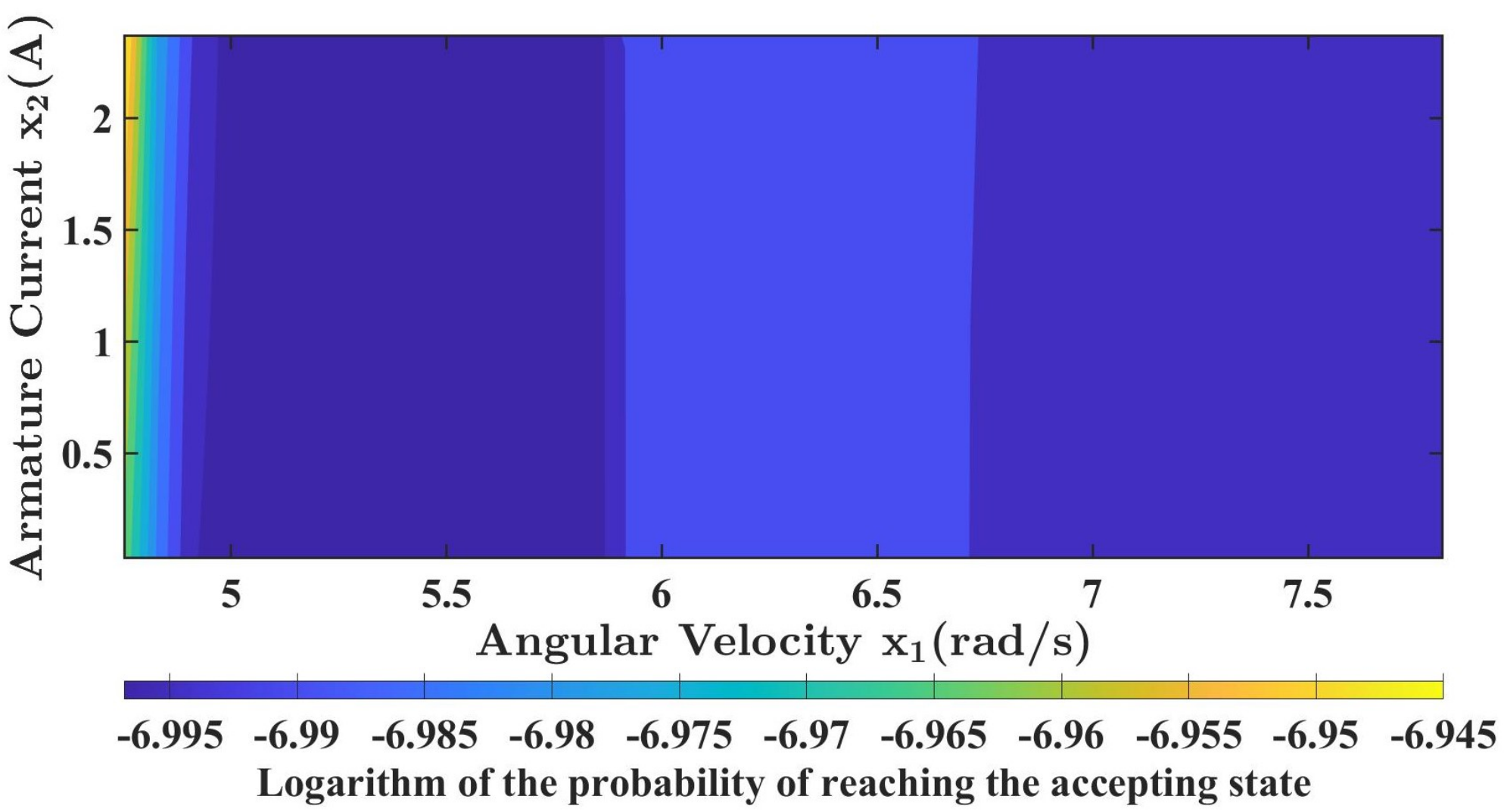}
	}
	\quad
	\subfigure[Number of Partitions: $50\times 50$]{
		\includegraphics[width=0.48\textwidth]{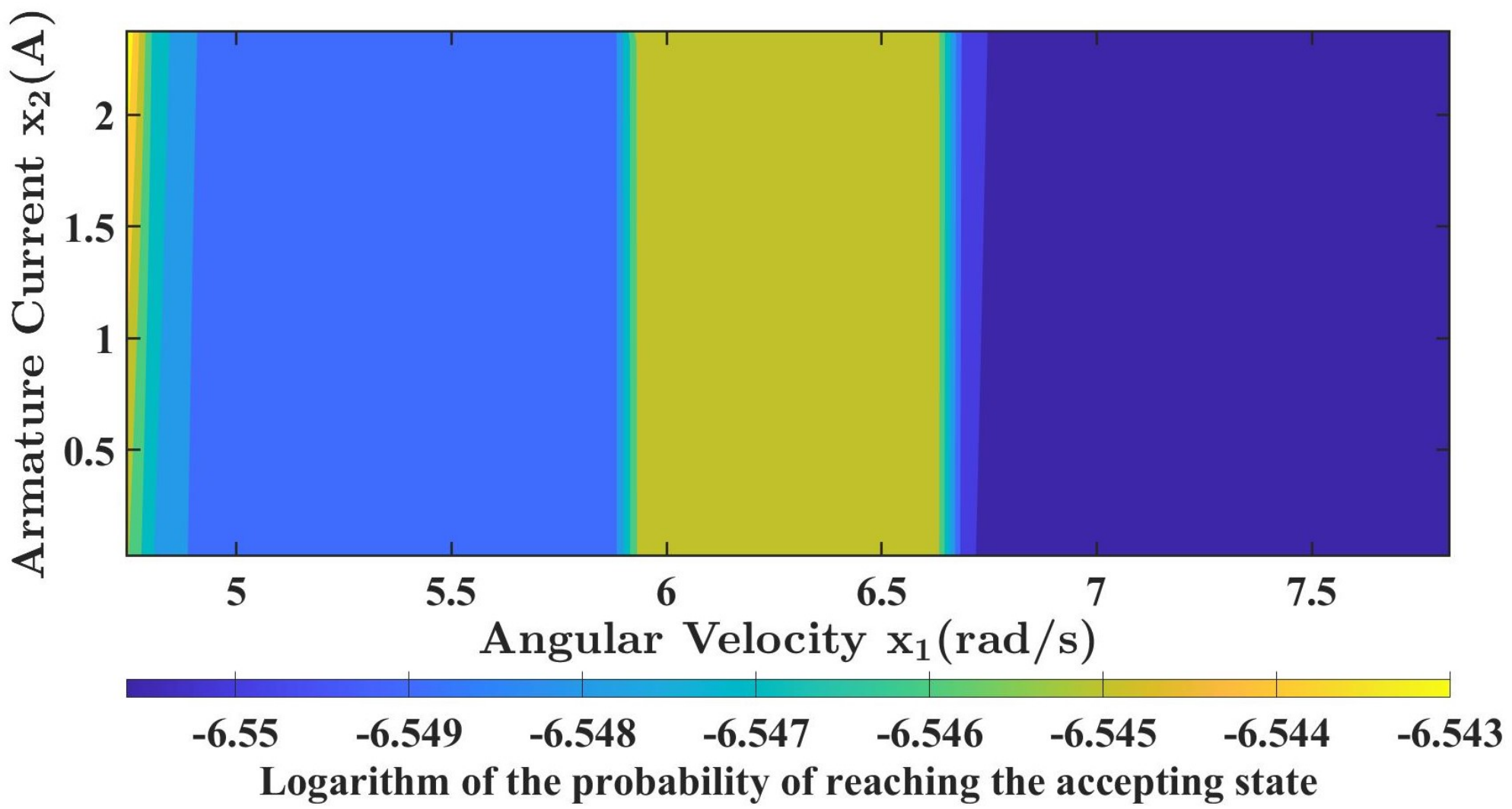}\hspace{-0.7cm}
	}
	\quad
	\subfigure[Number of Partitions: $60\times 60$]{
		\includegraphics[width=0.48\textwidth]{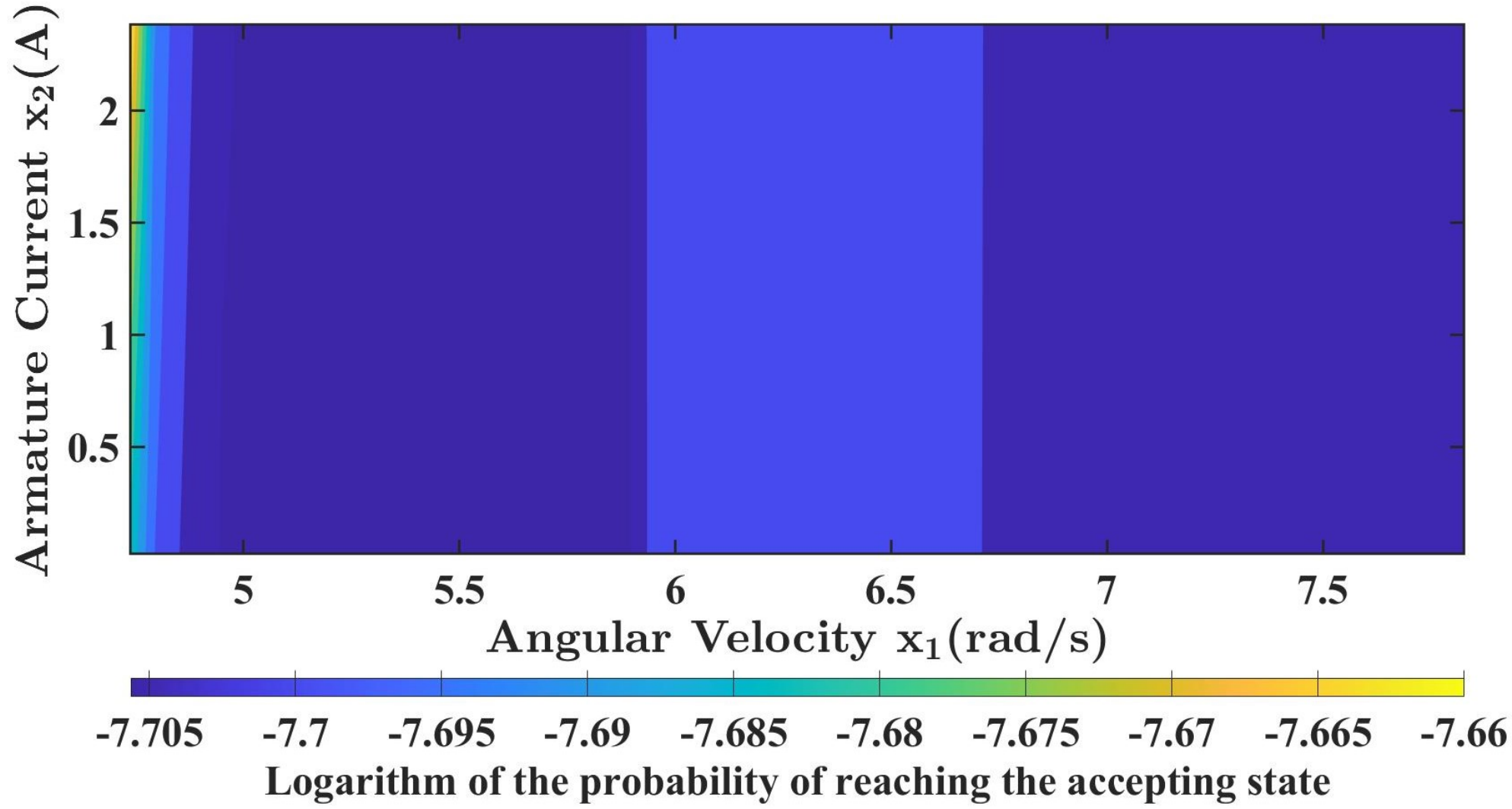}
	}
	\caption{Contours of the safety guarantee provided by the safety advisor.}\label{fig:safety_guarantee}
\end{figure}
	In this case study, we also show the effect of the abstraction resolution on the safety guarantee provided by the safety advisor and the execution speed of the supervisor.
	To this end, we select 6 different abstraction resolutions for the continuous state sets (see Table~\ref{tab2}, $10\times 10$ means uniformly partitioning $X= [1.5\pi, 2.5\pi]\times [0,2.4]$ into $10$ cells on both dimensions, and so on).
	We initialize the system at $x_0=[2\pi \,; 1.256]$ and simulate each case with $1.0 \times 10^4$ empirical Monte Carlo runs.
	The average execution time of the supervisor and the probabilistic guarantee provided by the safety advisor in each case are shown in Table~\ref{tab2} and Figure~\ref{fig:safety_guarantee}, respectively.
	According to Table~\ref{tab2}, the average execution time of the supervisor is increasing by reducing the quantization parameters in constructing finite abstractions.
	However, as shown in Figure~\ref{fig:safety_guarantee}, the upper bound of the probability of reaching the accepting state, which indicates a violation of the desired specification as in Figure~\ref{fig1:case1} (b), reduces by reducing the quantization parameters.
\begin{table}
	\caption{Average execution time of the supervisor with different size of grids for partitioning the continuous state set.}\label{tab2}\vspace{0.2cm}
	\begin{tabular}{|p{4.5cm}|p{1cm}|p{1cm}|p{1cm}|p{1cm}|p{1cm}|p{1cm}|}
			\hline
			Number of partitions & $10\!\times\! 10$& $20\!\times\! 20$& $30\!\times\! 30$&$40\!\times\! 40$&$50\!\times\! 50$ & $60\!\times\! 60$\\
			\hline
			Number of states & 101 & 401 &  901 & 1601& 2501 & 3601 \\
			\hline
			$\epsilon$ of the relation & 0.5021 & 0.2297 & 0.1518 & 0.1138 & 0.0911 & 0.0759 \\
			\hline
			Average execution time (ms)& 0.2388 & 0.2579 & 0.3174 & 0.3763 & 0.4431 & 0.5041\\
			\hline
	\end{tabular}
\end{table}

\section{Conclusion}\label{sec:discussion}
In this paper, we proposed a Safe-visor architecture for sandboxing (AI-based) unverified controllers in stochastic CPSs while enforcing safety properties characterized by accepting languages of deterministic finite automata. 
To construct the safety advisor that provides a fallback in the proposed architecture, we employed a robust controller regarding the desired safety specification. 
For constructing such a controller, we first synthesized a controller based on a finite abstraction that is $(\epsilon,\delta)$-stochastically simulated by the original model. 
The synthesized controller is then refined over the original dynamics based on an ($\epsilon$,$\delta$)-approximate probabilistic relation, which is the main key for providing the safety guarantees. 
To reach a compromise between safety and functionality, we utilized a history-based supervisor which checks the input provided by the unverified controller based on the history paths at runtime. 
The main idea for this checking was to estimate the probability of violating the desired safety specification assuming the input from the unverified controller is accepted, and compare this estimation with the maximal tolerable probability of violation. 
Finally, we applied our proposed architecture to two case studies. 
Extending the proposed methods to those systems modeled as partially observable Markov decision processes (POMDPs) is under investigation as a future work.

\section{Acknowledgment}
This work was supported in part by the H2020 ERC Starting Grant AutoCPS (grant agreement No 804639) and by an Alexander von Humboldt Professorship endowed by the German Federal Ministry of Education and Research.

%\noindent

%\bibliographystyle{plain}        % Include this if you use bibtex 
\bibliographystyle{my-elsarticle-num}
%\biboptions{sort&compress}
%\bibliography{autosam}           % and a bib file to produce the     
\bibliography{mybibfile}  

\begin{thebibliography}{10}
\expandafter\ifx\csname url\endcsname\relax
  \def\url#1{\texttt{#1}}\fi
\expandafter\ifx\csname urlprefix\endcsname\relax\def\urlprefix{URL }\fi
\expandafter\ifx\csname href\endcsname\relax
  \def\href#1#2{#2} \def\path#1{#1}\fi

\bibitem{Zhong2019Sandboxing}
B.~Zhong, M.~Zamani, M.~Caccamo, Sandboxing controllers for stochastic
  cyber-physical systems, in: International Conference on Formal Modeling and
  Analysis of Timed Systems, Springer, 2019, pp. 247--264.

\bibitem{Reis2009Browser}
C.~Reis, A.~Barth, C.~Pizano, Browser security: lessons from google chrome,
  Communications of the ACM 52~(8) (2009) 45--49.

\bibitem{Haesaert2017Verification}
S.~Haesaert, S.~E. Zadeh~Soudjani, A.~Abate, Verification of general {Markov}
  decision processes by approximate similarity relations and policy refinement,
  SIAM Journal on Control and Optimization 55~(4) (2017) 2333--2367.

\bibitem{Baier2008Principles}
C.~Baier, J.-P. Katoen, Principles of model checking, MIT press, 2008.

\bibitem{DeGiacomo2013Linear}
G.~De~Giacomo, M.~Y. Vardi, Linear temporal logic and linear dynamic logic on
  finite traces, in: Proceedings of the Twenty-Third International Joint
  Conference on Artificial Intelligence, 2013, p. 854–860.

\bibitem{Bojarski2016End}
M.~Bojarski, D.~Del~Testa, D.~Dworakowski, B.~Firner, B.~Flepp, P.~Goyal, L.~D.
  Jackel, M.~Monfort, U.~Muller, J.~Zhang, et~al., End to end learning for
  self-driving cars, arXiv:1604.07316(2016) .

\bibitem{Julian2019Deep}
K.~D. Julian, M.~J. Kochenderfer, M.~P. Owen, Deep neural network compression
  for aircraft collision avoidance systems, Journal of Guidance, Control, and
  Dynamics 42~(3) (2019) 598--608.

\bibitem{Papernot2016limitations}
N.~Papernot, P.~McDaniel, S.~Jha, M.~Fredrikson, Z.~B. Celik, A.~Swami, The
  limitations of deep learning in adversarial settings, in: IEEE European
  symposium on security and privacy, IEEE, 2016, pp. 372--387.

\bibitem{NTSB2018Preliminary}
{National Transportation Safety Board},
  \href{https://www.urbanismnext.org/resources/preliminary-report-highway-hwy18mh010}{Preliminary
  report highway-wy18mh010} (2018).
\newline\urlprefix\url{https://www.urbanismnext.org/resources/preliminary-report-highway-hwy18mh010}

\bibitem{Kwiatkowska2019Safety}
M.~Z. Kwiatkowska, {Safety Verification for Deep Neural Networks with Provable
  Guarantees}, in: 30th International Conference on Concurrency Theory (CONCUR
  2019), Vol. 140, Springer, 2019, pp. 1--5.

\bibitem{Claviere2020Safety}
A.~Clavi{\`e}re, E.~Asselin, C.~Garion, C.~Pagetti, Safety verification of
  neural network controlled systems, arXiv:2011.05174(2020) .

\bibitem{Bloem2015Shield}
R.~Bloem, B.~K{\"o}nighofer, R.~K{\"o}nighofer, C.~Wang, Shield synthesis:
  Runtime enforcement for reactive systems, in: Tools and Algorithms for the
  Construction and Analysis of Systems, Springer Berlin Heidelberg, 2015, pp.
  533--548.

\bibitem{Humphrey2016Synthesis}
L.~Humphrey, B.~K{\"o}nighofer, R.~K{\"o}nighofer, U.~Topcu, Synthesis of
  admissible shields, in: Hardware and Software: Verification and Testing. HVC
  2016. Lecture Notes in Computer Science, 2016, pp. 134--151.

\bibitem{Bharadwaj2019Synthesis}
S.~Bharadwaj, R.~Bloem, R.~Dimitrova, B.~K{\"o}nighofer, U.~Topcu, Synthesis of
  minimum-cost shields for multi-agent systems, in: 2019 American Control
  Conference (ACC), IEEE, 2019, pp. 1048--1055.

\bibitem{Alshiekh2018Safe}
M.~Alshiekh, R.~Bloem, R.~Ehlers, B.~K{\"o}nighofer, S.~Niekum, U.~Topcu, Safe
  reinforcement learning via shielding, in: Thirty-Second AAAI Conference on
  Artificial Intelligence, Vol.~32, 2018.

\bibitem{Sha2001Using}
L.~Sha, Using simplicity to control complexity, IEEE Software (2001) 20--28.

\bibitem{Crenshaw2007Simplex}
T.~L. Crenshaw, E.~Gunter, C.~L. Robinson, L.~Sha, P.~R. Kumar, The simplex
  reference model: Limiting fault-propagation due to unreliable components in
  cyber-physical system architectures, in: 28th IEEE International Real-Time
  Systems Symposium (RTSS 2007), IEEE, 2007, pp. 400--412.

\bibitem{Wang2013L1Simplex}
X.~Wang, N.~Hovakimyan, L.~Sha, L1simplex fault-tolerant control of
  cyber-physical systems, in: Proceedings of the ACM/IEEE 4th International
  Conference on Cyber-Physical Systems, ACM, 2013, pp. 41--50.

\bibitem{Wang2018RSimplex}
X.~Wang, N.~Hovakimyan, L.~Sha, Rsimplex: A robust control architecture for
  cyber and physical failures, ACM Transactions on Cyber-Physical Systems 2~(4)
  (2018) 1--26.

\bibitem{Yao2013NetSimplex}
J.~Yao, X.~Liu, G.~Zhu, L.~Sha, Netsimplex: Controller fault tolerance
  architecture in networked control systems, IEEE Transactions on Industrial
  Informatics 9~(1) (2013) 346--356.

\bibitem{Abdi2018Preserving}
F.~Abdi, C.-Y. Chen, M.~Hasan, S.~Liu, S.~Mohan, M.~Caccamo, Preserving
  physical safety under cyber attacks, IEEE Internet of Things Journal (2018)
  6285--6300.

\bibitem{Bak2014Real}
S.~Bak, T.~T. Johnson, M.~Caccamo, L.~Sha, Real-time reachability for verified
  simplex design, in: 2014 IEEE Real-Time Systems Symposium, IEEE, 2014, pp.
  138--148.

\bibitem{Bak2011Sandboxing}
S.~Bak, K.~Manamcheri, S.~Mitra, M.~Caccamo, Sandboxing controllers for
  cyber-physical systems, in: 2011 IEEE/ACM Second International Conference on
  Cyber-Physical Systems, ACM, 2011, pp. 3--12.

\bibitem{Abdi2017Application}
F.~Abdi, R.~Tabish, M.~Rungger, M.~Zamani, M.~Caccamo, Application and
  system-level software fault tolerance through full system restarts, in: 2017
  ACM/IEEE 8th International Conference on Cyber-Physical Systems (ICCPS), ACM,
  2017, pp. 197--206.

\bibitem{Kenton2019Generalizing}
Z.~Kenton, A.~Filos, O.~Evans, Y.~Gal, Generalizing from a few environments in
  safety-critical reinforcement learning, arXiv:1907.01475(2019) .

\bibitem{Ma2018Improved}
X.~Ma, K.~Driggs-Campbell, M.~J. Kochenderfer, Improved robustness and safety
  for autonomous vehicle control with adversarial reinforcement learning, in:
  2018 IEEE Intelligent Vehicles Symposium (IV), IEEE, 2018, pp. 1665--1671.

\bibitem{Thananjeyan2020Safety}
B.~Thananjeyan, A.~Balakrishna, U.~Rosolia, F.~Li, R.~McAllister, J.~E.
  Gonzalez, S.~Levine, F.~Borrelli, K.~Goldberg, Safety augmented value
  estimation from demonstrations {(SAVED)}: Safe deep model-based {RL} for
  sparse cost robotic tasks, IEEE Robotics and Automation Letters 5~(2) (2020)
  3612--3619.

\bibitem{Fisac2019Bridging}
J.~F. Fisac, N.~F. Lugovoy, V.~Rubies-Royo, S.~Ghosh, C.~J. Tomlin, Bridging
  {Hamilton-Jacobi} safety analysis and reinforcement learning, in: 2019
  International Conference on Robotics and Automation (ICRA), IEEE, 2019, pp.
  8550--8556.

\bibitem{Shyamsundar2016Reinforcement}
S.~Shyamsundar, T.~Mannucci, E.-J. van Kampen, Reinforcement learning based
  algorithm with safety handling and risk perception, in: 2016 IEEE Symposium
  Series on Computational Intelligence (SSCI), IEEE, 2016, pp. 1--7.

\bibitem{Cheng2019End}
R.~Cheng, G.~Orosz, R.~M. Murray, J.~W. Burdick, End-to-end safe reinforcement
  learning through barrier functions for safety-critical continuous control
  tasks, in: Proceedings of the AAAI Conference on Artificial Intelligence,
  Vol.~33, 2019, pp. 3387--3395.

\bibitem{Deshmukh2019Learning}
J.~V. Deshmukh, J.~P. Kapinski, T.~Yamaguchi, D.~Prokhorov, Learning deep
  neural network controllers for dynamical systems with safety guarantees, in:
  2019 IEEE/ACM International Conference on Computer-Aided Design (ICCAD),
  IEEE, 2019, pp. 1--7.

\bibitem{Yang2019Safety}
Y.~Yang, Y.~Yin, W.~He, K.~G. Vamvoudakis, H.~Modares, D.~C. Wunsch,
  Safety-aware reinforcement learning framework with an actor-critic-barrier
  structure, in: 2019 American Control Conference (ACC), IEEE, 2019, pp.
  2352--2358.

\bibitem{Huh2020Safe}
S.~Huh, I.~Yang, Safe reinforcement learning for probabilistic reachability and
  safety specifications: A lyapunov-based approach, arXiv:2002.10126(2020) .

\bibitem{Larsen2017Safe}
R.~B. Larsen, A.~Carron, M.~N. Zeilinger, Safe learning for distributed systems
  with bounded uncertainties, IFAC-PapersOnLine 50~(1) (2017) 2536--2542.

\bibitem{Fisac2018general}
J.~F. Fisac, A.~K. Akametalu, M.~N. Zeilinger, S.~Kaynama, J.~Gillula, C.~J.
  Tomlin, A general safety framework for learning-based control in uncertain
  robotic systems, IEEE Transactions on Automatic Control 64~(7) (2018)
  2737--2752.

\bibitem{Wabersich2018Linear}
K.~P. Wabersich, M.~N. Zeilinger, Linear model predictive safety certification
  for learning-based control, in: 2018 IEEE Conference on Decision and Control
  (CDC), IEEE, 2018, pp. 7130--7135.

\bibitem{Wabersich2018Safe}
K.~P. Wabersich, M.~N. Zeilinger, Safe exploration of nonlinear dynamical
  systems: A predictive safety filter for reinforcement learning,
  arXiv:1812.05506(2018) .

\bibitem{Muntwiler2019Distributed}
S.~Muntwiler, K.~P. Wabersich, A.~Carron, M.~N. Zeilinger, Distributed model
  predictive safety certification for learning-based control,
  arXiv:1911.01832(2019) .

\bibitem{Hewing2020Learning}
L.~Hewing, K.~P. Wabersich, M.~Menner, M.~N. Zeilinger, Learning-based model
  predictive control: Toward safe learning in control, Annual Review of
  Control, Robotics, and Autonomous Systems 3 (2020) 269--296.

\bibitem{Li2019Temporal}
X.~Li, C.~Belta, Temporal logic guided safe reinforcement learning using
  control barrier functions, arXiv:1903.09885v1(2019) .

\bibitem{Lavaei2020Formal}
A.~Lavaei, F.~Somenzi, S.~Soudjani, A.~Trivedi, M.~Zamani, Formal controller
  synthesis for continuous-space mdps via model-free reinforcement learning,
  in: 2020 ACM/IEEE 11th International Conference on Cyber-Physical Systems
  (ICCPS), IEEE, 2020, pp. 98--107.

\bibitem{Kazemi2020Formal}
M.~Kazemi, S.~Soudjani, Formal policy synthesis for continuous-state systems
  via reinforcement learning, in: International Conference on Integrated Formal
  Methods, Springer, 2020, pp. 3--21.

\bibitem{Biyik2020Learning}
E.~B{\i}y{\i}k, D.~P. Losey, M.~Palan, N.~C. Landolfi, G.~Shevchuk, D.~Sadigh,
  Learning reward functions from diverse sources of human feedback: Optimally
  integrating demonstrations and preferences, arXiv:2006.14091(2020) .

\bibitem{Haesaert2020Robust}
S.~Haesaert, S.~Soudjani, Robust dynamic programming for temporal logic control
  of stochastic systems, IEEE Transactions on Automatic Control (2020) 1--16.

\bibitem{Shreve1978Stochastic}
S.~E. Shreve, Stochastic optimal control: The discrete time case, Academic
  Press, 1978.

\bibitem{Faruq2018Simultaneous}
F.~Faruq, D.~Parker, B.~Laccrda, N.~Hawes, Simultaneous task allocation and
  planning under uncertainty, in: 2018 IEEE/RSJ International Conference on
  Intelligent Robots and Systems (IROS), IEEE, 2018, pp. 3559--3564.

\bibitem{Saha2014Automated}
I.~Saha, R.~Ramaithitima, V.~Kumar, G.~J. Pappas, S.~A. Seshia, Automated
  composition of motion primitives for multi-robot systems from safe {LTL}
  specifications, in: 2014 IEEE/RSJ International Conference on Intelligent
  Robots and Systems, IEEE, 2014, pp. 1525--1532.

\bibitem{Girard2009Hierarchical}
A.~Girard, G.~J. Pappas, Hierarchical control system design using approximate
  simulation, Automatica 45~(2) (2009) 566--571.

\bibitem{Zhong2021Automata}
B.~Zhong, A.~Lavaei, M.~Zamani, M.~Caccamo, Automata-based controller synthesis
  for discrete-time stochastic systems: a stochastic game framework via
  approximate probabilistic relations, arXiv:2104.11803(2021) .

\bibitem{Tkachev2013Quantitative}
I.~Tkachev, A.~Mereacre, J.-P. Katoen, A.~Abate, Quantitative automata-based
  controller synthesis for non-autonomous stochastic hybrid systems, in:
  Proceedings of the 16th international conference on Hybrid systems:
  computation and control, ACM, 2013, pp. 293--302.

\bibitem{Lavaei2019Compositionalc}
A.~Lavaei, S.~Soudjani, M.~Zamani, Compositional construction of infinite
  abstractions for networks of stochastic control systems, Automatica 107
  (2019) 125--137.

\bibitem{Touze2021Model}
C.~Touz{\'e}, A.~Vizzaccaro, O.~Thomas, Model order reduction methods for
  geometrically nonlinear structures: a review of nonlinear techniques,
  Nonlinear Dynamics  1--50.

\bibitem{Baur2014Model}
U.~Baur, P.~Benner, L.~Feng, Model order reduction for linear and nonlinear
  systems: a system-theoretic perspective, Archives of Computational Methods in
  Engineering 21~(4)  331--358.

\bibitem{lavaei2021automated}
A.~Lavaei, S.~Soudjani, A.~Abate, M.~Zamani, Automated verification and
  synthesis of stochastic hybrid systems: A survey, arXiv:2101.07491(2021) .

\bibitem{Zhu2015Game}
Q.~Zhu, T.~Basar, Game-theoretic methods for robustness, security, and
  resilience of cyberphysical control systems: games-in-games principle for
  optimal cross-layer resilient control systems, IEEE Control Systems Magazine
  35~(1) (2015) 46--65.

\bibitem{Borkar2012Probability}
V.~S. Borkar, Probability theory: an advanced course, Springer Science \&
  Business Media, 2012.

\bibitem{Lavaei2019Compositionala}
A.~Lavaei, S.~Soudjani, M.~Zamani, Compositional abstraction-based synthesis of
  general mdps via approximate probabilistic relations, Nonlinear Analysis:
  Hybrid Systems 39(2021) .

\bibitem{Huijgevoort2020Similarity}
B.~C. van Huijgevoort, S.~Haesaert, Similarity quantification for linear
  stochastic systems as a set-theoretic control problem, arXiv:2007.09052(2020)
  .

\bibitem{Zaccarian2012DC}
L.~Zaccarian, {DC} motors: dynamic model and control techniques (2012).

\bibitem{Lillicrap2016Continuous}
T.~P. Lillicrap, J.~J. Hunt, A.~Pritzel, N.~Heess, T.~Erez, Y.~Tassa,
  D.~Silver, D.~Wierstra, Continuous control with deep reinforcement learning,
  in: International Conference on Learning Representations(Poster), 2016.

\end{thebibliography}
% and a bib file to produce the 
% bibliography (preferred). The
% correct style is generated by
% Elsevier at the time of printing.

\appendix
\section{Proof for Theorem~\ref{theorem:guarantee for rs} and Theorem~\ref{theorem:guarantee for wcv}}\label{proof}
To show Theorem~\ref{theorem:guarantee for rs} and Theorem~\ref{theorem:guarantee for wcv}, we first need to define the \emph{n-steps reachable state set} of DFA as follows.
\begin{definition}
	\emph{(n-steps Reachable State Set)}
	Given a gMDP $\mathfrak{D} =(X,U,x_0,T,Y,h)$ and a DFA $\mathcal{A} = (Q, q_0, \Pi,$ $ \tau, F)$ with a labelling function $L:Y\rightarrow\Pi$, an n-step reachable state set $\tilde{Q}_n(x_0)$ of $\mathcal{A}$ is recursively defined as
	\begin{align*}
	\tilde{Q}_0(x_0) &= \Big\{q\in Q\,\big|\,q=\tau(q_0,L\circ h(x_0))\Big\},\\
	\tilde{Q}_n(x_0)&= \Big\{q\in Q\,\big|\,\exists q'\in \tilde{Q}_{n-1}(x_0), \sigma \in \Pi\ s.t.\ q=\tau(q',\sigma)\Big\},
	\end{align*}
	with $n\in \N_{>0}$.
\end{definition} 
We also need the following lemma for the proof.

\begin{lemma}\label{lem:cosafeLTL1minus}
	Consider a gMDP $\mathfrak{D} =(X,U,x_0,T,Y,h)$ and its finite abstraction $\widehat{\mathfrak{D}}= (\hat X,\hat U,\hat{x}_0, \hat T, Y,\hat h)$ with $\widehat{\mathfrak{D}}\preceq^{\delta}_{\epsilon}\mathfrak{D}$, and a DFA $\mathcal{A}=(Q, q_0, \Pi,$ $\tau, F)$ that characterizes the desired safety specification.
	Given a Markov policy $\rho\,=\,(\rho_0, \rho_1,\ldots,$ $\rho_{H-1})$ over the product gMDP $\widehat{\mathfrak{D}}\otimes \mathcal{A}$ within the time horizon $[0,H]$, one has
	\begin{equation}\label{eq:help1}
	1-\bar{V}_{n+1}^{\rho}(\hat{x},q)=(1-\delta)\!\!\!\sum_{\tilde{x}\in\hat{X}'_{\epsilon}(q)}\!\!\!\Big(1-\bar{V}_{n}^{\rho}(\tilde{x},\underline{q}(\tilde{x},q))\Big)\hat{T}(\tilde{x}\,\big|\,\hat{x},\hat{u})+\delta,
	\end{equation}
	and 
	\begin{equation}\label{eq:help2}
	1-\ul{V}_{n+1}^{\rho}(\hat{x},q)= (1-\delta)\!\!\!\sum_{\tilde{x}\in\hat{X}'_{-\epsilon}(q)}\!\!\!\Big(1-\ul{V}_{n}^{\rho}(\tilde{x},\bar{q}(\tilde{x},q))\Big)\hat{T}(\tilde{x}\,\big|\,\hat{x},\hat{u}),\\
	\end{equation}
	with $q\notin F$, $\hat{u}=\rho_{H-n-1}(\hat{x},q)$,  $\bar{V}_{n+1}^{\rho}(\hat{x},q)$ as in \eqref{eq:P_general}, $\underline{q}$ as in \eqref{eq:underline_p}, $\hat{X}'_{\epsilon}(q)$ as in \eqref{eq:xeps},
	$\ul{V}_{n+1}^{\rho}(\hat{x},q)$ as in~\eqref{eq:T_general}, $\bar{q}$ as in~\eqref{eq:overline_p}, and $\hat{X}'_{-\epsilon}(q)$ as in \eqref{eq:xeps-}.
\end{lemma}	
The proof of Lemmas~\ref{lem:cosafeLTL1minus} can readily be derived based on \eqref{eq:P_general}, \eqref{eq:underline_p}, \eqref{eq:T_general} and \eqref{eq:overline_p}. 
Now, we are ready to show the results of both theorems.

{\bf Proof of Theorem~\ref{theorem:guarantee for rs}}
For the sake of clarity of the proof, we define
\begin{align*}
f(\hat{x}(k)&,q(k))=1-\!\!\!\!\!\!\sum_{\hat{x}(k+1)\in \hat{X}}\!\!\!\bar{V}^*_{H-k-1}\Big(\hat{x}(k+1),\underline{q}^*\big(\hat{x}(k+1),q(k)\big)\Big)\hat{T}\big(\hat{x}(k+1)\,\big|\,\hat{x}(k),\hat{u}(k)\big),
\end{align*}
for $k\in[0,H-1]$ with $\bar{V}^*_{H-k-1}$ as in~\eqref{eq:P_opt_sac}, $\underline{q}^*$ as in~\eqref{eq:underline_p_star}, and $\hat{u}(k)=\rho'_k(\hat{x}(k),q(k))$, and
\begin{align*}
g(\hat{x}(z-1),q(z-1))=\hat{T}\big(\hat{x}(z)\,\big|\,\hat{x}(z-1),\rho'_{z-1}(\hat{x}(z-1),q(z-1))\big),
\end{align*}	
for $z\in[0,k]$.
First, consider an initial state $(\hat{x}_0,\bar{q}_0)$.
By expanding out $\bar{V}^{\rho'}_H(\hat{x}_0,\bar{q}_0)$ up to the time instant $k\in[0,H-1]$ with the help of~\eqref{eq:help1}, we have $1-\bar{V}^{\rho'}_H(\hat{x}_0,\bar{q}_0) = \theta_1(k)+\theta_2(k)$, with
\begin{small}
	\begin{align}
	&\theta_1(k)=\,(1-\delta)^{k+1}\!\!\!\!\!\!\!\!\!\!\!\sum_{\hat{x}(1)\in \hat{X}'_{\epsilon}(q(0))}\!\!\!\!\Big( \sum_{\hat{x}(2)\in \hat{X}'_{\epsilon}(q(1))}\Big(\ldots\Big(\!\!\!\!\!\!\!\!\!\!\!\!\!\!\!\!\sum_{\hat{x}(k-1)\in \hat{X}'_{\epsilon}(q(k-2))}\Big(\!\!\!\!\sum_{\hat{x}(k)\in \hat{X}'_{\epsilon}(q(k-1))}\!\!\!\!\!\!\!\!\!\!f\big(\hat{x}(k),q(k)\big)\nonumber\\
	&\times\! g\big(\hat{x}(k-1),q(k-1)\big)\Big)g\big(\hat{x}(k-2),q(k-2)\big)\Big)\!\ldots\!\Big)g\big(\hat{x}(1),q(1)\big)\Big)g\big(\hat{x}(0),q(0)\big),\label{eq:proof1theta1}	
	\end{align}
\end{small}
and
\begin{small}
	\begin{align}
	&\theta_2(k)=\delta\Big(1+(1-\delta)\Big(\!\!\!\!\!\!\!\!\!\sum_{\hat{x}(1)\in \hat{X}'_{\epsilon}(q(0))}\!\!\!\!\!\!\!\!\!\!g\big(\hat{x}(0),q(0)\big)\Big)+(1-\delta)^2\Big(\!\!\!\!\!\!\!\!\!\!\sum_{\hat{x}(1)\in \hat{X}'_{\epsilon}(q(0))}\Big(\!\!\!\!\sum_{\hat{x}(2)\in \hat{X}'_{\epsilon}(q(1))}\!\!\!\!\!\!\!\!\!\!g\big(\hat{x}(1),q(1)\big)\Big)\nonumber\\
	&\!\!\times\! g\big(\hat{x}(0),q(0)\big)\Big)\!+\!\ldots\!+\!(1-\delta)^k\!\Big(\!\!\!\!\!\!\!\!\!\!\sum_{\hat{x}(1)\in \hat{X}'_{\epsilon}(q(0))}\Big(\!\!\!\sum_{\hat{x}(2)\in \hat{X}'_{\epsilon}(q(1))}\!\!\!\!\!\!\Big(\ldots\Big(\!\!\!\!\!\!\!\!\!\!\sum_{\hat{x}(k-1)\in \hat{X}'_{\epsilon}(q(k-2))}\Big(\!\!\!\!\sum_{\hat{x}(k)\in \hat{X}'_{\epsilon}(q(k-1))}\nonumber\\
	&\!g(\hat{x}\big(k-1),q(k-1)\big)\Big)g\big(\hat{x}(k-2),q(k-2)\big)\Big)\!\ldots\!\Big)g\big(\hat{x}(1),q(1)\big)\Big)g\big(\hat{x}(0),q(0)\big)\Big)\Big).\label{eq:proof1theta2}
	\end{align}
\end{small}
Let us choose $\hat{x}^*(k) := \argmax_{\hat{x}(k)\in \hat{X}'_{\epsilon}(q(k-1))}f(\hat{x}(k),q(k))$ with $q(k-1)\in\tilde{Q}_{k-1}(x_0)$, $\hat{X}'_{\epsilon}(q(k-1))$ as in~\eqref{eq:xeps}, and $q^*(k)=\underline{q}(q(k-1),\hat{x}^*(k))$ with $\underline{q}$ as in \eqref{eq:underline_p}.
Then, we have
\begin{small}
	\begin{align*}
	\sum_{\hat{x}(k)\in \hat{X}'_{\epsilon}(q(k-1))}\!\!\!\!\!\!\!\!\!\!\!\!\!\!\!f\big(\hat{x}(k),q(k)\big)g\big(\hat{x}(k-1),q(k-1)\big)\leq f\big(\hat{x}^*(k),q^*(k)\big)\!\!\!\!\!\!\!\!\!\!\!\!\!\!\!\sum_{\hat{x}(k)\in \hat{X}'_{\epsilon}(q(k-1))}\!\!\!\!\!\!\!\!\!\!\!\!\!\!\!g\big(\hat{x}(k-1),q(k-1)\big).
	\end{align*}
\end{small}
Thus, proceed from~\eqref{eq:proof1theta1}, we have 
\begin{small}
	\begin{align}
	&\theta_1(k)\leq(1-\delta)^{k+1}\!\!\!\!\!\!\!\!\!\!\!\!\sum_{\hat{x}(1)\in \hat{X}'_{\epsilon}(q(0))}\!\!\!\!\Big( \sum_{\hat{x}(2)\in \hat{X}'_{\epsilon}(q(1))}\Big(\ldots\Big(\!\!\!\!\!\!\!\!\!\!\!\!\!\!\!\!\sum_{\hat{x}(k-1)\in \hat{X}'_{\epsilon}(q(k-2))}\Big(\!\!\!\!\sum_{\hat{x}(k)\in \hat{X}'_{\epsilon}(q(k-1))}\!\!\!\!\!\!\!\!\!\!\!\!\!\!g\big(\hat{x}(k-1), \nonumber\\
	&q(k-1)\big)\Big)g\big(\hat{x}(k-2),q(k-2)\big)\Big)\!\ldots\!\Big)g\big(\hat{x}(1),q(1)\big)\Big)g\big(\hat{x}(0),q(0)\big)f\big(\hat{x}^*(k),q^*(k)\big).\label{eq:proof1step1}
	\end{align}
\end{small}
Next, let us select 
\begin{equation}\label{proof1opt1}
\hat{x}^*(k-1) = \mathop{\arg\max}_{\hat{x}(k-1)\in \hat{X}'_{\epsilon}(q(k-2))}\sum_{\hat{x}(k)\in \hat{X}'_{\epsilon}(q(k-1))}g\big(\hat{x}(k-1), q(k-1)\big),
\end{equation}
with $q(k-2)\in \tilde{Q}_{k-2}(x_0)$, and 
\begin{equation}\label{proof1opt2}
q^*(k-1)=\underline{q}(q(k-2),\hat{x}^*(k-1)),
\end{equation}
with $\underline{q}$ as in \eqref{eq:underline_p}.
With~\eqref{proof1opt1} and~\eqref{proof1opt2}, we have
\begin{small}
	\begin{align*}
	\sum_{\hat{x}(k-1)\in \hat{X}'_{\epsilon}(q(k-2))}&\Big(\!\!\!\!\sum_{\hat{x}(k)\in \hat{X}'_{\epsilon}(q(k-1))}\!\!\!\!\!\!\!\!\!\!\!g\big(\hat{x}(k-1),q(k-1)\big)\Big) g\big(\hat{x}(k-2),q(k-2)\big)\nonumber\\
	\leq&\quad\Big(\!\!\!\!\sum_{\hat{x}(k)\in \hat{X}'_{\epsilon}(q^*(k-1))}\!\!\!\!\!\!\!\!\!\!g\big(\hat{x}^*(k-1),q^*(k-1)\big)\Big)\!\!\!\!\!\!\!\!\!\!\!\!\!\!\sum_{\hat{x}(k-1)\in \hat{X}'_{\epsilon}(q(k-2))}\!\!\!\!\!\!\!\!\!\!\!\!\!\!g\big(\hat{x}(k-2),q(k-2)\big).
	\end{align*}
\end{small}
Therefore, from~\eqref{eq:proof1step1}, we have
\begin{small}
	\begin{align}
	&\theta_1(k)\leq(1-\delta)^{k+1}\!\!\!\!\!\!\!\!\!\!\!\sum_{\hat{x}(1)\in \hat{X}'_{\epsilon}(q(0))}\!\!\!\!\Big( \sum_{\hat{x}(2)\in \hat{X}'_{\epsilon}(q(1))}\Big(\ldots\Big(\!\!\!\!\!\!\!\!\!\!\!\!\!\!\!\!\sum_{\hat{x}(k-1)\in \hat{X}'_{\epsilon}(q(k-2))}\!\!\!\!\!\!\!\!\!\!\!\!\!\!g\big(\hat{x}(k-2),q(k-2)\big)\Big)\!\ldots\!\Big) \nonumber\\
	&\times g\big(\hat{x}(1),q(1)\big)\Big)g\big(\hat{x}(0),q(0)\big)f\big(\hat{x}^*(k),q^*(k)\big)\Big(\!\!\!\!\!\!\!\!\!\!\!\sum_{\hat{x}(k)\in \hat{X}'_{\epsilon}(q^*(k-1))}\!\!\!\!\!\!\!\!\!\!\!\!g\big(\hat{x}^*(k-1),q^*(k-1)\big)\Big).\label{eq:proof1step2}
	\end{align}
\end{small}
For all $z\in[2,k-1]$, one can choose $x^*(z-1)$ similar to~\eqref{proof1opt1} and $q^*(z-1)$ analogously to~\eqref{proof1opt2}.
Then, we have
\begin{small}
	\begin{align}
	&\sum_{\hat{x}(z-1)\in \hat{X}'_{\epsilon}(q(z-2))}\Big(\!\!\!\!\sum_{\hat{x}(z)\in \hat{X}'_{\epsilon}(q(z-1))}\!\!\!\!\!\!\!\!g\big(\hat{x}(z-1),q(z-1)\big)\Big) g\big(\hat{x}(z-2),q(z-2)\big)\nonumber\\
	\leq&\quad \Big(\!\!\!\!\sum_{\hat{x}(z)\in \hat{X}'_{\epsilon}(q^*(z-1))}\!\!\!\!\!\!\!\!\!\!g\big(\hat{x}^*(z-1),q^*(z-1)\big)\Big)\!\!\!\!\!\!\!\!\!\!\!\!\!\!\!\sum_{\hat{x}(z-1)\in \hat{X}'_{\epsilon}(q(z-2))}\!\!\!\!\!\!\!\!\!\!\!\!\!\!\!g\big(\hat{x}(z-2),q(z-2)\big).\label{eq:proof1final}
	\end{align}
\end{small}
Therefore, continuing from~\eqref{eq:proof1step2} with~\eqref{eq:proof1final} for all $z\in[2,k-1]$, we have 
\begin{small}
	\begin{align}
	\theta_1(k)\leq\,&(1-\delta)^{k+1}\Big(\!\!\!\!\!\!\!\!\!\!\!\sum_{\hat{x}(1)\in \hat{X}'_{\epsilon}(q(0))}\!\!\!\!\!g\big(\hat{x}(0),q(0)\big)\Big)\prod_{z=2}^{k}\Big(\!\!\!\!\sum_{\hat{x}(z)\in \hat{X}'_{\epsilon}(q^*(z-1))}\!\!\!\!\!\!\!\!\!\!\!\!\!\!g\big(\hat{x}^*(z-1),q^*(z-1)\big)\Big)f\big(\hat{x}^*(k),q^*(k)\big).\label{prrof1theta1final}
	\end{align}
\end{small}
Similar to the idea of going from~\eqref{eq:proof1step1} to~\eqref{prrof1theta1final} with $x^*(z-1)$ and $q^*(z-1)$ for all $z\in[2,k-1]$, starting from~\eqref{eq:proof1theta2}, we have
\begin{small}
	\begin{align}
	\theta_2(k)\leq&\ \delta\Big(1+(1-\delta)\Big(\!\!\!\!\!\!\!\!\!\!\sum_{\hat{x}(1)\in \hat{X}'_{\epsilon}(q(0))}\!\!\!\!\!\!\!\!\!\!g(\hat{x}(0),q(0))\Big)+(1-\delta)^2\Big(\!\!\!\!\!\!\!\!\!\!\sum_{\hat{x}(1)\in \hat{X}'_{\epsilon}(q(0))}\!\!\!\!\!\!\!\!\!\!g(\hat{x}(0),q(0))\Big)\Big(\!\!\!\!\!\!\!\sum_{\hat{x}(2)\in \hat{X}'_{\epsilon}(q^*(1))}\!\!\!\!\!\!\!\!\!\!g(\hat{x}^*(1),q^*(1))\Big)\nonumber\\
	& +\ldots+(1-\delta)^k\Big(\!\!\!\!\!\!\!\!\!\sum_{\hat{x}(1)\in \hat{X}'_{\epsilon}(q(0))}\!\!\!\!\!\!\!\!\!\!g(\hat{x}(0),q(0))\Big)\prod_{z=2}^{k}\Big(\!\!\!\!\!\!\!\sum_{\hat{x}(z)\in \hat{X}'_{\epsilon}(q^*(z-1))}\!\!\!\!\!\!\!\!\!\!g(\hat{x}^*(z-1),q^*(z-1))\Big)\Big).\label{prrof1theta2final}
	\end{align}
\end{small}
Finally. combining~\eqref{prrof1theta1final} and~\eqref{prrof1theta2final}, we have
\begin{small}
	\begin{align*}
	1-\bar{V}^{\rho'}_H(\hat{x}_0,\bar{q}_0)\leq&(1-\delta)^{k+1}\Big(\!\!\!\!\!\!\!\!\!\!\sum_{\hat{x}(1)\in \hat{X}'_{\epsilon}(q(0))}\!\!\!\!\!\!\!g(\hat{x}(0),q(0))\Big)\prod_{z=2}^{k}\Big(\!\!\!\!\!\sum_{\hat{x}(z)\in \hat{X}'_{\epsilon}(q^*(z-1))}\!\!\!\!\!\!\!\!\!\!g(\hat{x}^*(z-1),q^*(z-1))\Big)f(\hat{x}^*(k),q^*(k))\\
	&+\delta\Big( 1+\sum_{j=1}^{k}(1-\delta)^j\!\!\!\!\!\!\!\!\!\!\sum_{\hat{x}(1)\in \hat{X}'_{\epsilon}(q(0))}\!\!\!\!\!\!\!\!\!\!g(\hat{x}(0),q(0))\prod_{z=2}^{j}\Big(\sum_{\hat{x}(z)\in \hat{X}'_{\epsilon}(q^*(z-1))}\!\!\!\!\!\!\!\!\!\!g(\hat{x}^*(z-1),q^*(z-1))\Big)\Big).
	\end{align*}
\end{small}
Note that $\bar{\omega}'_k \!=\! \big(\hat{x}(0), q(0),\rho_0(\hat{x}(0),q(0)),\hat{x}^*(1), q^*(1),\rho_1(\hat{x}^*(1),q^*(1)),\ldots,$ $\hat{x}^*(k),\\q^*(k)\big)$ is one of the history paths as in Definition~\ref{def:path_new} up to the time instant $k$.
By applying the history-based supervisor as in Definition~\ref{def:History-based Supervisor_rs}, one can ensure that for an arbitrary path $\bar{\omega}_k$, one has
\begin{align*}
&(1-\delta)^{k+1}\prod_{z=1}^{k}\Big(\sum_{\bar{\omega}_{\hat{x}k}(z)\in \hat{X}'_{\epsilon}(\bar{\omega}_{qk}(z-1))}\!\!\!\!\!\!\!\!g(\bar{\omega}_{\hat{x}k}(z-1),\bar{\omega}_{qk}(z-1))\Big)f(\bar{\omega}_{\hat{x}k}(k),\bar{\omega}_{qk}(k))\\
&+\delta\Big(1+\sum_{j=1}^{k}(1-\delta)^j\prod_{z=1}^{j}\Big(\sum_{\bar{\omega}_{\hat{x}k}(z)\in \hat{X}'_{\epsilon}(\bar{\omega}_{qk}(z-1))}\!\!\!\!\!\!\!\!g(\bar{\omega}_{\hat{x}k}(z-1),\bar{\omega}_{qk}(z-1))\Big)\Big)\leq \eta.
\end{align*}
Therefore, we have $\bar{V}^{\rho'}_{H}(\hat{x}_0,\bar{q}_0) \geq 1-\eta$ with the supervisor as in	Definition~\ref{def:History-based Supervisor_rs}.
According to Theorem~\ref{thm:gua_prmax}, one has $\mathbb{P}_{\tilde{\mathbf{C}}_{\rho'}\times \mathfrak{D}}\{\exists k\leq H, y_{\omega k}\models \mathcal{A}\} \geq  \bar{V}^{\rho'}_{H}(\hat{x}_0,\bar{q}_0)$ so that $\mathbb{P}_{\mathfrak{D}}\Big\{y_{\omega H}\models \mathcal{A}\Big\}\geq 1-\eta$, which completes the proof.

{\bf Proof of Theorem~\ref{theorem:guarantee for wcv}}
For the sake of clarity of the proof, we define
\begin{align*}
&r(\hat{x}(k),q(k))=1-\!\!\!\!\!\!\sum_{\hat{x}(k+1)\in \hat{X}}\!\!\!\ul{V}_{*,H-k-1}\Big(\hat{x}(k+1),\bar{q}_*\big(\hat{x}(k+1),q(k)\big)\Big)\hat{T}\big(\hat{x}(k+1)\,\big|\,\hat{x}(k),\hat{u}(k)\big),
\end{align*}
for $k\in[0,H-1]$ with $\ul{V}^*_{H-k-1}$ as in~\eqref{eq:P_opt_vio}, $\underline{q}^*$ as in~\eqref{eq:overline_p_star}, and $\hat{u}(k)=\rho'_k(\hat{x}(k),q(k))$, and
\begin{align*}
g(\hat{x}(z-1),q(z-1))=\hat{T}\big(\hat{x}(z)\,\big|\,\hat{x}(z-1),\rho'_{z-1}(\hat{x}(z-1),q(z-1))\big),
\end{align*}	
for $z\in[0,k]$.
Consider an initial state $(\hat{x}_0,\bar{q}_0)$.
Then, with the help of~\eqref{eq:help2}, we expand out $\ul{V}^{\rho'}_H(\hat{x}_0,\bar{q}_0)$ up to the time instant $k\in[0,H-1]$ as 
\begin{small}
	\begin{align}
	1-\ul{V}^{\rho'}_H(\hat{x}_0,\bar{q}_0)=&(1-\delta)^{k+1} \!\!\!\!\!\!\!\!\!\!\!\!\sum_{\hat{x}(1)\in \hat{X}'_{-\epsilon}(q(0))}\!\!\!\!\Big( \sum_{\hat{x}(2)\in \hat{X}'_{-\epsilon}(q(1))}\!\!\!\!\!\!\!\!\Big(\ldots\Big(\!\!\!\!\sum_{\hat{x}(k-1)\in \hat{X}'_{-\epsilon}(q(k-2))}\Big(\!\!\!\!\sum_{\hat{x}(k)\in \hat{X}'_{-\epsilon}(q(k-1))}\!\!\!\!\!\!\!\!\!\!r\big(\hat{x}(k),q(k)\big)\nonumber\\
	&\times g\big(\hat{x}(k-1),q(k-1)\big)\Big)g\big(\hat{x}(k-2),q(k-2)\big)\Big)\!\ldots\!\Big)g\big(\hat{x}(1),q(1)\big)\Big)g\big(\hat{x}(0),q(0)\big).\label{eq:proof2step1}
	\end{align}
\end{small}
Let us select $\hat{x}_*(k) := \mathop{\arg\min}_{\hat{x}(k)\in \hat{X}'_{-\epsilon}(q(k-1))}r(\hat{x}(k),q(k))$ with $q(k-1)\in\tilde{Q}_{k-1}(x_0)$, $\hat{X}'_{-\epsilon}(q(k-1))$ as in~\eqref{eq:xeps-}, and $q_*(k)=\bar{q}(q(k-1),\hat{x}_*(k))$ with $\bar{q}$ as in \eqref{eq:overline_p}.
Then, we have
\begin{small}
	\begin{align*}
	\sum_{\hat{x}(k)\in \hat{X}'_{-\epsilon}(q(k-1))}\!\!\!\!\!\!\!\!\!\!\!\!\!\!\!r(\hat{x}(k),q(k))g(\hat{x}(k-1),q(k-1))\geq r(\hat{x}_*(k),q_*(k))\!\!\!\!\!\!\!\!\!\!\!\!\!\!\!\sum_{\hat{x}(k)\in \hat{X}'_{-\epsilon}(q(k-1))}\!\!\!\!\!\!\!\!\!\!\!\!\!\!\!g(\hat{x}(k-1),q(k-1)).
	\end{align*}
\end{small}
Therefore, from~\eqref{eq:proof2step1}, we have 
\begin{small}
	\begin{align}
	1-\ul{V}^{\rho'}_H(\hat{x}_0,\bar{q}_0)\geq&(1-\delta)^{k+1}\!\!\!\!\!\!\!\!\!\!\!\!\! \sum_{\hat{x}(1)\in \hat{X}'_{-\epsilon}(q(0))}\!\!\!\!\Big( \sum_{\hat{x}(2)\in \hat{X}'_{-\epsilon}(q(1))}\!\!\!\!\Big(\ldots\Big(\!\!\!\!\!\!\!\!\!\!\!\!\sum_{\hat{x}(k-1)\in \hat{X}'_{-\epsilon}(q(k-2))}\Big(\!\!\!\!\sum_{\hat{x}(k)\in \hat{X}'_{-\epsilon}(q(k-1))}\!\!\!\!\!\!\!\!\!\!\!\!\!\!\!\!g\big(\hat{x}(k-1),q(k-1)\big)\Big)\nonumber\\
	&\times g\big(\hat{x}(k-2),q(k-2)\big)\Big)\ldots\Big)g\big(\hat{x}(1),q(1)\big)\Big)g\big(\hat{x}(0),q(0)\big)r\big(\hat{x}_*(k),q_*(k)\big).\label{eq:proof2step2}
	\end{align}
\end{small}
Next, let us select 
\begin{equation}\label{proof2opt1}
\hat{x}_*(k-1) = \mathop{\arg\min}_{\hat{x}(k-1)\in \hat{X}'_{-\epsilon}(q(k-2))}\sum_{\hat{x}(k)\in \hat{X}'_{-\epsilon}(q(k-1))}g\big(\hat{x}(k-1), q(k-1)\big),
\end{equation}
with $q(k-2)\in \tilde{Q}_{k-2}(x_0)$, and 
\begin{equation}\label{proof2opt2}
q_*(k-1)=\bar{q}(q(k-2),\hat{x}_*(k-1)),
\end{equation}
with $\bar{q}$ as in \eqref{eq:overline_p}.
Then, one has
\begin{small}
	\begin{align*}
	&\sum_{\hat{x}(k-1)\in \hat{X}'_{-\epsilon}(q(k-2))}\Big(\!\!\!\!\sum_{\hat{x}(k)\in \hat{X}'_{-\epsilon}(q(k-1))}\!\!\!\!\!\!\!\!\!\!\!\!\!\!\!\!g\big(\hat{x}(k-1),q(k-1)\big)\Big) g\big(\hat{x}(k-2),q(k-2)\big)\nonumber\\
	\geq&\quad \Big(\!\!\!\!\sum_{\hat{x}(k)\in \hat{X}'_{-\epsilon}(q_*(k-1))}\!\!\!\!\!\!\!\!\!\!\!\!\!\!\!\!g\big(\hat{x}_*(k-1),q_*(k-1)\big)\Big) \sum_{\hat{x}(k-1)\in \hat{X}'_{-\epsilon}(q(k-2))}g\big(\hat{x}(k-2),q(k-2)\big).
	\end{align*}
\end{small}
Thus, proceed from~\eqref{eq:proof2step2}, we have
\begin{small}
	\begin{align}
	1-\ul{V}^{\rho'}_H(\hat{x}_0,\bar{q}_0)&\geq(1-\delta)^{k+1}\!\!\!\!\!\!\!\!\sum_{\hat{x}(1)\in \hat{X}'_{-\epsilon}(q(0))}\!\!\!\!\Big( \sum_{\hat{x}(2)\in \hat{X}'_{-\epsilon}(q(1))}\!\!\!\!\Big(\ldots\Big(\!\!\!\!\!\!\!\!\!\!\sum_{\hat{x}(k-1)\in \hat{X}'_{-\epsilon}(q(k-2))}g\big(\hat{x}(k-2),q(k-2)\big)\Big)\ldots\Big)\nonumber\\
	&\times\, g\big(\hat{x}(1),q(1)\big)\Big)g\big(\hat{x}(0),q(0)\big)\Big(\!\!\!\!\!\!\!\!\!\!\!\!\!\!\!\!\sum_{\hat{x}(k)\in \hat{X}'_{-\epsilon}(q_*(k-1))}\!\!\!\!\!\!\!\!\!\!\!\!\!\!\!\!g\big(\hat{x}_*(k-1),q_*(k-1)\big)\Big)r\big(\hat{x}_*(k),q_*(k)\big).\label{eq:proof2step3}
	\end{align}
\end{small}
For all $z\in[2,k-1]$, we can select $x_*(z-1)$ similar to~\eqref{proof2opt1} and $q_*(z-1)$ similar to~\eqref{proof2opt2}.
Accordingly, we have
\begin{small}
	\begin{align}
	&\sum_{\hat{x}(z-1)\in \hat{X}'_{-\epsilon}(q(z-2))}\Big(\!\!\!\!\sum_{\hat{x}(z)\in \hat{X}'_{-\epsilon}(q(z-1))}\!\!\!\!\!\!\!\!\!\!\!\!\!\!\!\!g\big(\hat{x}(z-1),q(z-1)\big)\Big) g\big(\hat{x}(z-2),q(z-2)\big)\nonumber\\
	\geq&\quad \Big(\!\!\!\!\sum_{\hat{x}(z)\in \hat{X}'_{-\epsilon}(q_*(z-1))}\!\!\!\!\!\!\!\!\!\!\!\!\!\!\!\!g\big(\hat{x}_*(z-1),q_*(z-1)\big)\Big)\!\!\!\!\sum_{\hat{x}(z-1)\in \hat{X}'_{-\epsilon}(q(z-2))}\!\!\!\!g\big(\hat{x}(z-2),q(z-2)\big).\label{eq:proof2final}
	\end{align}
\end{small}
Then, with~\eqref{eq:proof2final} for all $z\in[2,k-1]$ and~\eqref{eq:proof2step3}, we have
\begin{small}
	\begin{align*}
	1-\ul{V}^{\rho'}_H(\hat{x}_0,\bar{q}_0)\geq (1-\delta)^{k+1}\Big(\!\!\!\!\!\!\!\!\!\!\!\!\sum_{\hat{x}(1)\in \hat{X}'_{-\epsilon}(q(0))}\!\!\!\!\!\!\!\!\!\!\!\!\!g(\hat{x}(0),q(0))\Big)\prod_{z=2}^{k}\Big(\sum_{\hat{x}(z)\in \hat{X}'_{-\epsilon}(q_*(z-1))}\!\!\!\!\!\!\!\!\!\!\!\!\!\!\!g(\hat{x}_*(z-1),q_*(z-1))\Big)r(\hat{x}_*(k),q_*(k)).
	\end{align*}
\end{small}
Note that $\bar{\omega}'_k = \big(\hat{x}(0), q(0),\rho_0(\hat{x}(0),q(0)),\hat{x}_*(1), q_*(1),\rho_1(\hat{x}_*(1),q_*(1))\ldots\,\hat{x}_*(k),$ $q_*(k)\big)$ is one of history paths as in Definition~\ref{def:path_new} up to the time instant $k$, and the history-based supervisor as in Definition~\ref{def:History-based Supervisor_wcv} ensures that for all history paths $\bar{\omega}_k$, we have
\begin{equation*}
(1-\delta)^{k+1}\prod_{z=1}^{k}\Big(\!\!\!\!\!\sum_{\bar{\omega}_{\hat{x}k}(z)\in \hat{X}'_{-\epsilon}(\bar{\omega}_{qk}(z-1))}\!\!\!\!\!\!\!\!\!\!g(\bar{\omega}_{\hat{x}k}(z-1),\bar{\omega}_{qk}(z-1))\Big)r(\bar{\omega}_{\hat{x}k}(k),\bar{\omega}_{qk}(k))\geq 1-\eta.
\end{equation*}
Therefore, we have $\ul{V}^{\rho'}_{H}(\hat{x}_0,\bar{q}_0) \leq \eta$ when applying the supervisor as in Definition~\ref{def:History-based Supervisor_wcv}.	 
According to Theorem~\ref{thm:gua_prmin}, one has $\mathbb{P}_{\tilde{\mathbf{C}}_{\rho'}\times \mathfrak{D}}\{\exists k\leq H, y_{\omega k}\models \mathcal{A}\}\leq \ul{V}^{\rho'}_{H}(\hat{x}_0,\bar{q}_0)$ so that $\mathbb{P}_{\mathfrak{D}}\Big\{y_{\omega H}\models \mathcal{A}\Big\}\leq \eta$, which completes the proof.	

\end{document}